\documentclass[aps,prb,twocolumn,superscriptaddress,citeautoscript,floatfix,showpacs,reprint,longbibliography]{revtex4-1}

\pdfoutput=1
\usepackage[english]{babel}
\usepackage[utf8]{inputenc}
\usepackage{amsmath}
\usepackage{graphicx}
\usepackage[dvipsnames]{xcolor}
\usepackage{amssymb}
\usepackage{subfigure}
\usepackage{physics}
\usepackage{makeidx}
\usepackage{todonotes}

\definecolor{darkblue}{rgb}{0.1,0.2,0.6} \definecolor{darkred}{rgb}{0.8,0.1,0.2}
\usepackage[colorlinks,citecolor=darkblue,linkcolor=darkred,urlcolor=darkblue]{hyperref}	
\usepackage[all]{hypcap} 

\DeclareMathOperator{\argmax}{argmax}

\begin{document}

\title{Exploring one particle orbitals in large Many-Body Localized systems}
\author{Benjamin Villalonga}
\author{Xiongjie Yu}
\affiliation{Institute for Condensed Matter Theory and Department of Physics, University of Illinois at Urbana-Champaign, Urbana, IL 61801, USA}
\author{David J. Luitz}
\affiliation{Institute for Condensed Matter Theory and Department of Physics, University of Illinois at Urbana-Champaign, Urbana, IL 61801, USA}
\affiliation{Department of Physics, T42, Technische Universit\"at M\"unchen, James-Franck-Str. 1, 85748 Garching, Germany}
\author{Bryan K. Clark}
\affiliation{Institute for Condensed Matter Theory and Department of Physics, University of Illinois at Urbana-Champaign, Urbana, IL 61801, USA}
\date{\today}

\begin{abstract}
Strong disorder in interacting quantum systems can give rise to the phenomenon of Many-Body
Localization (MBL), which defies thermalization due to the formation of an extensive number of quasi
local integrals of motion.  The one particle operator content of these integrals of motion is
related to the one particle orbitals of the one particle density matrix and shows a strong signature
across the MBL transition as recently pointed out by Bera et al. [Phys. Rev. Lett. \textbf{115},
046603 (2015); Ann. Phys. \textbf{529}, 1600356 (2017)].  We study the properties of the one
particle orbitals
of many-body eigenstates of an MBL system in one dimension.
Using
shift-and-invert MPS (SIMPS), a matrix product state method to target highly excited many-body
eigenstates introduced in [Phys. Rev. Lett. \textbf{118}, 017201 (2017)], we are able to obtain accurate
results for large systems of sizes up to $L=64$.
We find that the one particle
orbitals drawn from eigenstates at different energy densities have high overlap and their
occupations are correlated with the energy of the eigenstates.  
Moreover, the standard deviation of the inverse participation ratio of these orbitals is maximal at the nose of the mobility edge.
Also, the one particle orbitals decay exponentially in real
space, with a correlation length that increases at low disorder. In addition, we find a ``$1/f$'' distribution of the coupling constants of a certain range of the number operators of the OPOs, which is related to their exponential decay.
\end{abstract}


\maketitle

\section{Introduction}
\label{sec:intro}

The eigenstate thermalization hypothesis (ETH) \cite{peres_ergodicity_1984,deutsch_quantum_1991,srednicki_chaos_1994, srednicki_approach_1999,rigol_thermalization_2008,dalessio_quantum_2016,borgonovi_quantum_2016}
provides a mechanism for the thermalization of generic isolated quantum systems.
A pure quantum state initially prepared to be sharply peaked in energy can relax to the
thermodynamic equilibrium in the sense that subsystems evolve such that their reduced density matrix
looks like a mixed thermal density matrix whose temperature is characterized by the energy of the
initial state.
In this way, a pure quantum state can behave locally like a mixed thermal state. The mechanism of
thermalization is provided by the special structure of local operators in the eigenbasis of the
Hamiltonian, where they become a smooth function of energy in very large systems.

In contrast, the phenomenon of Anderson localization
\cite{anderson_absence_1958}
describes the existence of an insulating phase that fails to thermalize in closed, non-interacting, quantum systems with quenched disorder.
In one dimension, any arbitrarily small amount of disorder leads to localization. 

\begin{figure}[t]
\includegraphics[width=1.00\columnwidth]{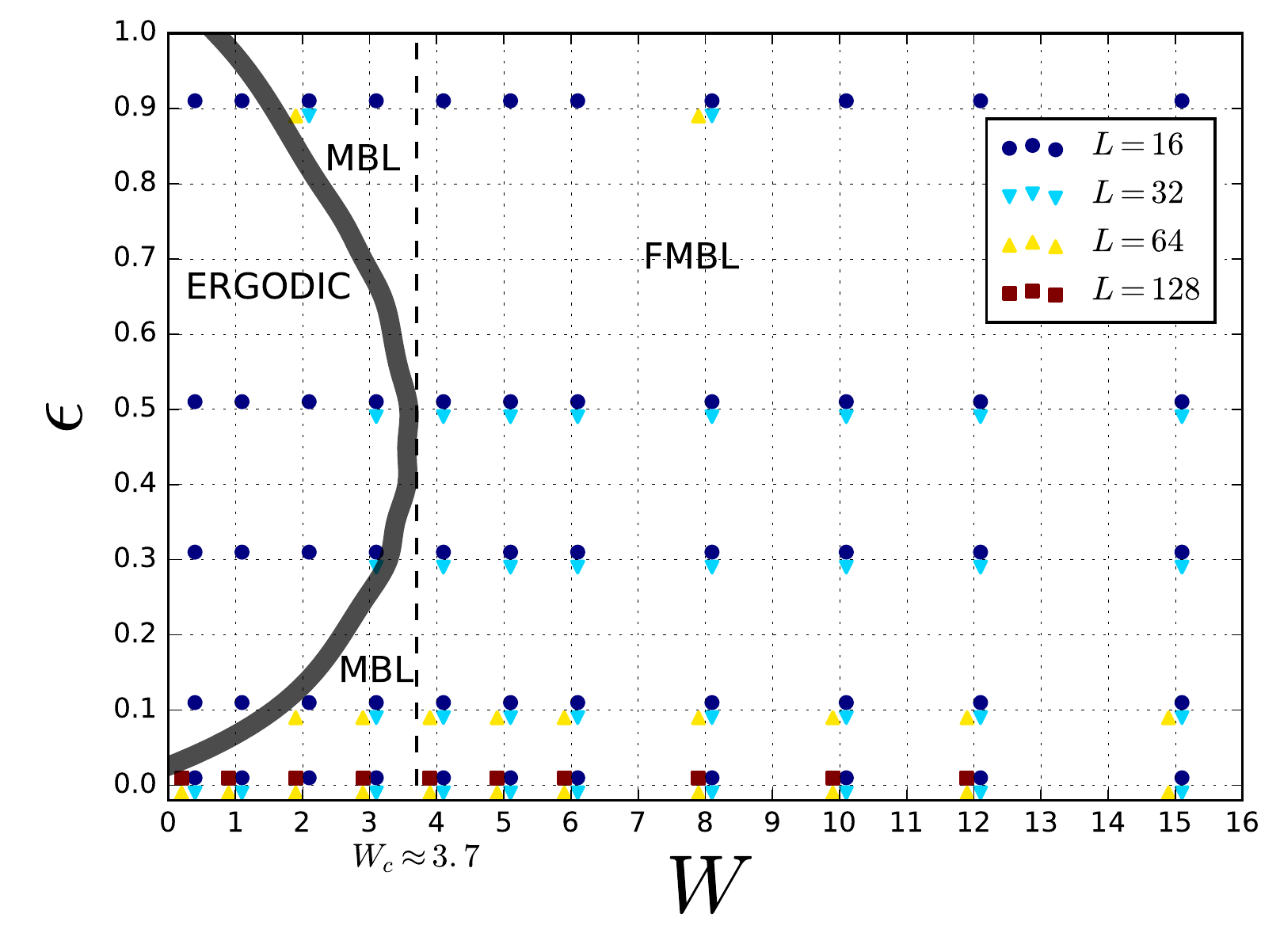}
\caption{\label{fig:phase_diagram} Phase diagram in the disorder strength $W$ and energy density $\epsilon$ plane
for the model in Eq.~\eqref{H_fermion} with $t=V=1$. The mobility edge is plotted from the results
of Ref.~\onlinecite{luitz_many-body_2015}.  In this paper, we numerically access eigenstates at the
depicted points.} 
\end{figure}

Surprisingly, the presence of strong interactions does not completely destroy this phenomenon.
Contrary to naive expectations that strongly interacting systems are always ergodic, a large number
of studies following pioneering works\cite{fleishman_interactions_1980,basko_metalinsulator_2006,gornyi_interacting_2005}
showed that usually interactions can stabilize an ergodic phase only at weak disorder, while at strong disorder the system many-body localizes (MBL)
(see
Refs.~\onlinecite{nandkishore_many-body_2015,altman_universal_2015,luitz_ergodic_2017,imbrie_local_2017,agarwal_rare-region_2017,abanin_recent_2017,laflorencie_many-body_2017}
 for recent reviews).
The MBL transition between the ergodic and localized phases has been the focus of many recent numerical studies
\cite{oganesyan_localization_2007,znidaric_many-body_2008,berkelbach_conductivity_2010,pal_many-body_2010,kjall_many-body_2014,luitz_many-body_2015,vosk_theory_2015,yu_bimodal_2016,khemani_critical_2017,
khemani_two_2017,pekker_hilbert-glass_2014,altman_universal_2015,potter_universal_2015,zhang_many-body_2016}
, and numerical evidence points to the
existence of a mobility edge (although the existence of a mobility edge is not settled~\cite{de_roeck_absence_2016}): for disorder strengths $W$
below a critical value, MBL eigenstates at low and high energy density are separated at a critical energy density $\epsilon$ from
extended eigenstates in the center of the spectrum \cite{luitz_many-body_2015} (see Fig.~\ref{fig:phase_diagram} for an illustration of the phase diagram). 
MBL can be seen as a novel eigenstate quantum phase transition
\cite{parameswaran_eigenstate_2017}
in which eigenstates radically change their nature as a function of disorder strength (or energy),
going from thermal eigenstates in the ergodic phase, which follow ETH and exhibit a volume law
scaling of the entanglement entropy, to MBL eigenstates in the MBL phase, which violate ETH and
exhibit an area law scaling of the entanglement entropy
\cite{bauer_area_2013,yu_bimodal_2016}.

For systems whose entire spectrum is MBL (fully MBL or FMBL), it is possible to find a complete set
of local integrals of motion or l-bits
\cite{serbyn_local_2013,huse_phenomenology_2014,chandran_constructing_2015,chandran_spectral_2015,ros_integrals_2015,pekker_fixed_2017,pollmann_efficient_2016,inglis_accessing_2016,pekker_encoding_2017,wahl_efficient_2017,imbrie_local_2017,monthus_many-body-localization_2017}
, which are responsible for a logarithmic growth of the entanglement entropy following a global quench in the MBL phase
\cite{chiara_entanglement_2006,znidaric_many-body_2008,bardarson_unbounded_2012,serbyn_universal_2013,luitz_extended_2016,singh_signatures_2016,zhou_operator_2017}.
The emergent integrability as signaled by a complete set of l-bits of an FMBL
system is lost below the critical disorder strength in the presence of a mobility edge, where the
existence of thermal eigenstates prevents any set of integrals of motion from containing only local
operators. Unfortunately, the numerical determination of the l-bit operators is very difficult and
does not scale favorably for large system sizes.
Therefore, a simplified proxy of l-bits is desirable and has been proposed earlier
\cite{bera_many-body_2015,bera_one-particle_2017,lin_many-body_2017}: the one particle density matrix (OPDM) and its eigenvectors, the one particle orbitals (OPOs).
The OPOs, which in the non-interacting limit become exact integrals of motion, have occupations in the MBL phase that are close to 0 and 1
\cite{bera_many-body_2015}.
They provide an effective first approximation to the l-bits and a well-defined, natural, continuous connection to the notion of integrability in the absence of interactions. Unlike the integrals of motion, the OPDM is defined over single eigenstates, which in the MBL phase are obtainable for large systems using DMRG-like methods
\cite{kennes_entanglement_2016,hu_excited-state_2015,khemani_obtaining_2016, lim_many-body_2016,
yu_finding_2017, devakul_obtaining_2017}.

\section{Summary of results}
\label{sec:summary}

\begin{figure}[t]
\includegraphics[width=0.8\columnwidth]{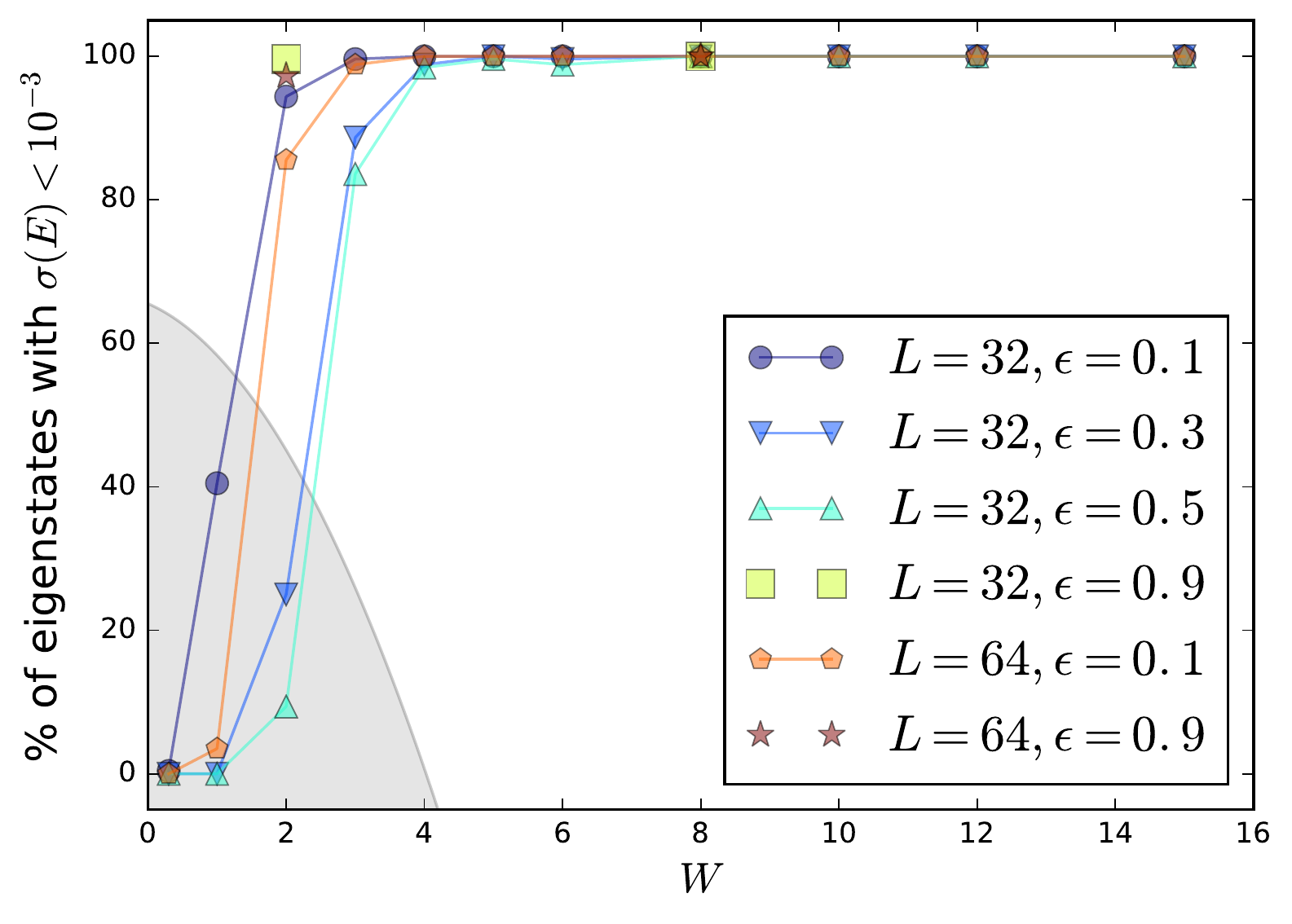}
\includegraphics[width=0.8\columnwidth]{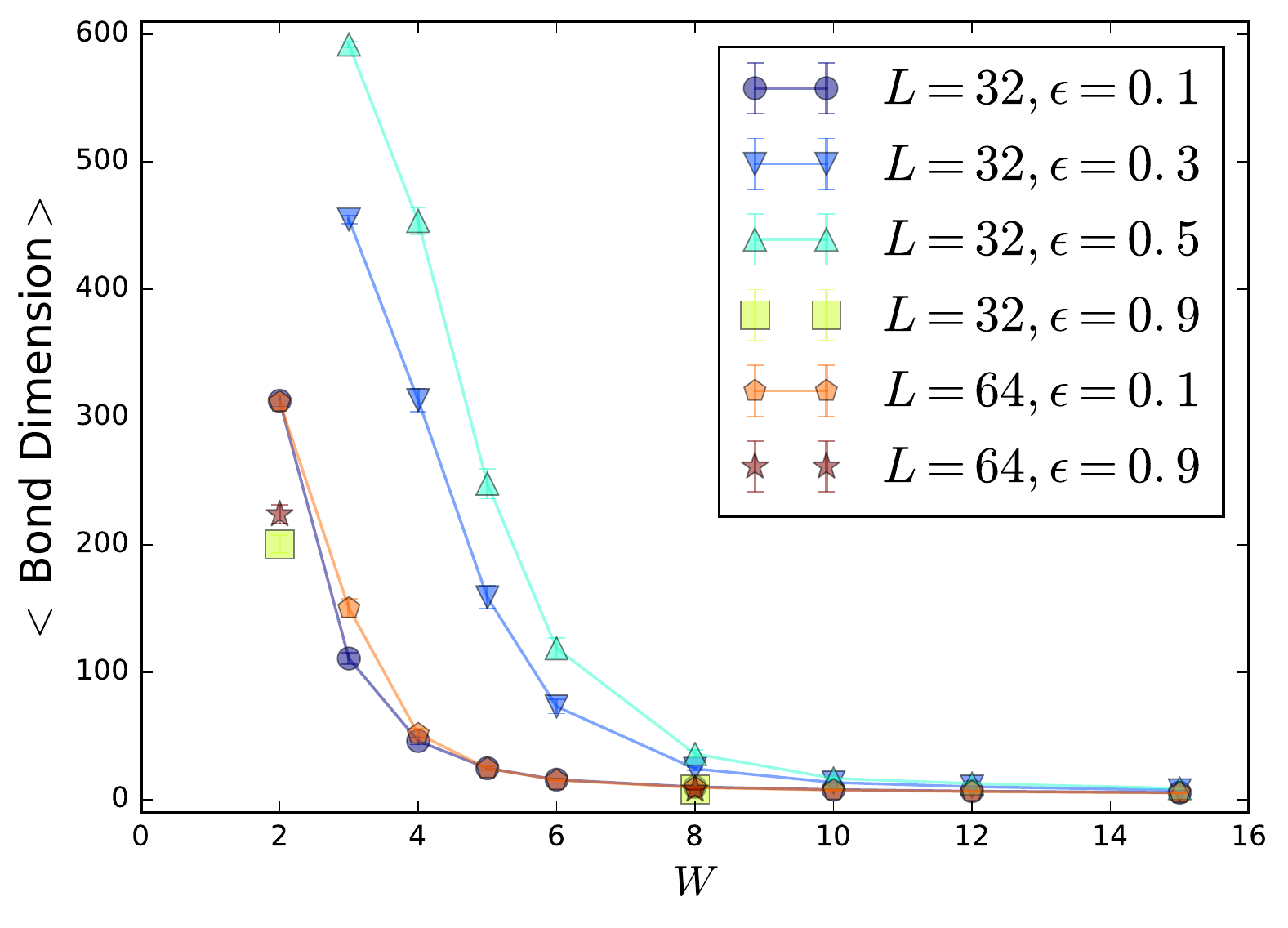}
\caption{\label{fig:survive} \emph{Top:} percentage of eigenstates accessed by SIMPS that pass our filter for the standard deviation of the energy, $\sigma(E)<10^{-3}$.  Eigenstates in the MBL region are accessed successfully through SIMPS with a low value of $\sigma(E)$, whereas eigenstates in the ergodic region (see Fig.~\ref{fig:phase_diagram}) fail to be represented accurately by the low bond dimension MPS ansatz. We neglect the eigenstates in the gray area due to the bias the strong filtering might introduce.
\emph{Bottom:} average bond dimension at half-cut of the eigenstates kept after filtering. As expected, the bond dimension diverges close to the transition, where it also becomes strongly system size dependent and it is eventually cutoff by the finite bond dimension used in SIMPS.}
\end{figure}

In this article we present a detailed numerical study of the behavior of the OPOs of the eigenstates of the model of Eq.~\eqref{H_fermion} using shift-and-invert MPS (SIMPS)\cite{yu_finding_2017}, an MPS-based algorithm  that allows us to access excited MBL eigenstates for 1D systems of size much larger than those studied using exact diagonalization (ED) techniques.
For systems of size up to at least $L=64$, SIMPS can access eigenstates at low energy densities at disorder $W<W_c$ (see Figs.~\ref{fig:phase_diagram} and \ref{fig:survive}), which supplies evidence for the existence of the mobility edge.

In Section~\ref{sec:support} we study the structure of the OPOs and their number operators as one particle approximations of the integrals of motion.
We find that the OPOs of MBL eigenstates decay exponentially in real space.
The OPOs' number operators, which encode the one particle content of the l-bits, have also an exponentially decaying weight in real space.
Their correlation length (same in both cases) increases monotonically as the disorder is lowered, but does not obviously diverge.
The correlation length is weakly system size dependent in the MBL phase and, as we can see for small systems, its dependence with energy density $\epsilon$ suggests the existence of the mobility edge.
The number operators of the OPOs are defined by string operators of different ranges in real space whose coupling constants approach a ``$1/f$'' distribution for a fixed range at strong disorder and large ranges, similar to the distribution seen in  Ref.~\onlinecite{pekker_fixed_2017}.
This distribution follows naturally from the exponential decay of the OPOs.
The OPOs and their number operators have a localized support at strong disorder.
The distribution of supports decays exponentially fast away from weak disorder, but becomes flat and extensive when the disorder is small.
The correlation length of this decay has several similarities with the one of the decay of the OPOs in real space.

In Section~\ref{sec:IPR} we analyze the inverse participation ratio of the OPOs as a measure of their localization.
Our results suggest that MBL eigenstates below a mobility edge in energy density $\epsilon$ reveal the presence of an ergodic phase at a higher $\epsilon$. This makes it possible to estimate the critical value of the disorder strength $W_c$, typically determined for the ergodic-MBL transition at $\epsilon=0.5$ (see Fig.~\ref{fig:phase_diagram}), from MBL eigenstates at $\epsilon\ll 0.5$.

In Section~\ref{sec:correspondence} we find that the OPOs extracted from eigenstates at different values of $\epsilon$ have high overlap, and their occupations are correlated to the energy of the eigenstate.
This provides the OPOs with a certain universality across the energy spectrum.

In Section~\ref{sec:occupations} we analyze the occupation spectrum of the eigenstates obtained for large systems.
Our results are in agreement with those of Refs.~\onlinecite{bera_many-body_2015,bera_one-particle_2017}: the occupations present a gap for MBL systems that becomes smaller as the ergodic phase is approached.
The $\epsilon$ dependence of the gap is the one expected in the presence of a mobility edge.
In addition, larger systems seem to have an ergodic region of the phase diagram that penetrates further into larger $W$ values.

Finally, we study the standard deviation of the entanglement entropy of the MBL eigenstates at half-cut in Section~\ref{sec:standard}.
As is discussed in Ref.~\onlinecite{kjall_many-body_2014}, it shows a peak at the critical disorder
strength.
Our results confirm that all eigenstates accessed by SIMPS are in the MBL region.
Also, the location of the peaks at different energy densities indicate once again the presence of a mobility edge.

\section{The model}
\label{sec:model}
We study spinless fermions with nearest neighbor repulsion $V$, a hopping matrix element $t$, subject to a random potential $\mu_i$ on an open chain with Hamiltonian:
\begin{align}
\label{H_fermion}
  \hat{H} = -\frac{t}{2}& \sum_{i=0}^{L-2} \left( \hat{c}^\dagger_i \hat{c}_{i+1} + \hat{c}^\dagger_{i+1} \hat{c}_i\right) \\ \nonumber
    + V& \sum_{i=0}^{L-2} \hat{n}_i \hat{n}_{i+1} + \sum_{i=0}^{L-1} \mu_i \hat{n}_i \text{.}
\end{align}
where $\hat{n}_i = \left( \hat{c}^\dagger_i \hat{c}_i - \frac{1}{2} \right)$ and the random
potential is sampled from a uniform distribution of width $2W$, \emph{i.e.} $\mu_i \in [-W,W]$,
where $W$ denotes the disorder strength. In this work, we let $t=V=1$. The model in
Eq.~\eqref{H_fermion} has been extensively studied in the context of 
MBL~\cite{oganesyan_localization_2007,znidaric_many-body_2008,berkelbach_conductivity_2010,pal_many-body_2010,bauer_area_2013,
nandkishore_many-body_2015,luitz_many-body_2015,
bar_lev_dynamics_2014,bar_lev_absence_2015,vosk_many-body_2013,agarwal_anomalous_2015,bera_many-body_2015,luitz_extended_2016,luitz_long_2016,luitz_anomalous_2016,luitz_information_2017,yu_bimodal_2016,khemani_critical_2017,
khemani_two_2017,serbyn_power-law_2016}. Among its characteristics, this model exhibits a mobility edge that separates
the MBL phase (at low and high values of $\epsilon$) from the delocalized phase (at intermediate values of $\epsilon$) for $W<W_c$, where $W_c \approx 3.7$ (see Fig.~\ref{fig:phase_diagram} and
Ref.~\onlinecite{luitz_many-body_2015}).
In addition, eigenstates in the delocalized phase obey a volume law for the 
entanglement entropy as a function of subsystem size, while MBL eigenstates follow an area
law\cite{bauer_area_2013}.
Close to the transition, the subsystem entanglement entropies are described by a bimodal distribution~\cite{yu_bimodal_2016}, and the standard deviation of the distribution of half-cut entanglement entropies peaks at the transition value of $W$ for each energy density~\cite{kjall_many-body_2014}.

Note that the Hamiltonian in Eq.~\eqref{H_fermion} is connected to the random field Heisenberg chain
through a Jordan-Wigner transformation and that the model is integrable at $W=0$.

\section{One Particle Density Matrix (OPDM)}
\label{sec:OPDM}
Given a pure state $\ket{\psi}$ of a system, the OPDM $\rho$ is defined as:
\begin{align}
\label{density_matrix}
  \rho_{ij} \equiv \bra{\psi} \hat{c}^\dagger_i \hat{c}_j \ket{\psi} \text{,}
\end{align}
which was introduced in the context of Bose-Einstein condensation\cite{penrose_bose-einstein_1956},
and was studied in Ref.~\onlinecite{bera_many-body_2015,bera_one-particle_2017} in the context of MBL. For a spinless, fermionic chain of length $L$, $\rho$ is a matrix of size $L\times L$, while $\ket{\psi}$ is a vector of size $2^L$.

We can diagonalize $\rho$ as:
\begin{align}
\label{occupations}
	\rho_{ij} = U_{ik} n_k U^\dagger_{kj} \text{,}
\end{align}
where the eigenvalues $n_k$ of $\rho$ are the occupations of the number operators $a^\dagger_k a_k$,
where $a^\dagger_k \equiv \sum_i U^\dagger_{ki} c^\dagger_i$. These rotated operators define the $L$
one particle orbitals (OPOs) $\ket{\phi_k} \equiv \sum_i U^\dagger_{ki} \ket{i}$, where $\ket{i}$ is the one particle wave function with a single fermion on site $i$. For convenience, we will order the OPOs by increasing value of their occupation $n_k$, unless otherwise specified.

For a non-interacting system and a particular eigenstate $\ket{\psi}$, the set of eigenvalues of the OPDM $\rho$ (or equivalently, the set of occupations of the OPOs) is highly degenerate, consisting only of the values $0$ and $1$.
Furthermore, there is a set of OPOs which simultaneously diagonalizes the OPDM of all eigenstates.
The number operators associated to these OPOs form a complete set of integrals of motion of the system and their occupations uniquely specify an energy eigenstate.
For an interacting system there is no such set of OPOs.
However, we show in Section~\ref{sec:correspondence} that the OPOs drawn from different eigenstates have a high overlap, and their occupations are correlated with the energy of the eigenstates.

In interacting systems, it has been shown that the spectrum of occupations $\{n_k\}$ of the OPDM contains a large gap for MBL eigenstates which gets smaller as the ergodic phase is entered, eventually closing for small values of the disorder strength\cite{bera_many-body_2015,bera_one-particle_2017}
When the gap is large, the spectrum of occupations is close to that of the non-interacting system; in the limit of infinite disorder the non-interacting picture is fully recovered.
This one particle picture provides thus not only a heuristic to characterize MBL and ergodic phases, but also a powerful point of view on the emergence of integrability in the MBL phase, since the one particle orbitals may be interpreted as the one particle operator content of the l-bits, which makes them a very good approximation for l-bits at strong disorder.

While the occupations indicate the nature of the dynamical phase at the energy density and $W$
corresponding to an eigenstate, we will see in  Section~\ref{sec:IPR} that the structure of the OPOs
allows us to discern between an MBL eigenstate of an FMBL Hamiltonian from one that is located in
energy below a many-body mobility edge.
The OPDM encodes therefore two distinct pieces of information: while the occupations of the OPOs
characterize the phase of an eigenstate, some properties the OPOs themselves can signal the presence of a mobility edge.

\section{Numerical simulations}
\label{sec:numerical}

We analyze the model from Eq.~\eqref{H_fermion} by obtaining eigenstates in the half-filled sector
at different energy densities, for systems of different sizes $L$ and disorder strengths $W$.  The energy density is defined as $\epsilon\equiv (E-E_{min})/(E_{max}-E_{min})$, where $E$ is the energy of the eigenstate and $E_{max}$ and $E_{min}$ are respectively the maximum and minimum energies in the energy spectrum (in all sectors for finite energy density data and in the half-filled sector for ground state results).
The phase diagram of this model (taken from Ref.~\onlinecite{luitz_many-body_2015}) and the points studied can be seen in Fig.~\ref{fig:phase_diagram}.
For each of the eigenstates accessed, the OPDM of Eq.~\eqref{density_matrix} is computed and diagonalized, which leaves us with its OPOs and their occupations.
Several disorder realizations are considered, and for each of them we obtain multiple
eigenstates for each value of $\epsilon$.

For the ground state ($\epsilon=0$) we use Lanczos ED ($L=16$) with 400 disorder realizations and
DMRG constrained to the half-filled sector ($L>16$) with 128 disorder realizations.
For each value of $\epsilon>0$ at finite energy density, we use shift invert ED ($L=16$) with 400 disorder realizations and SIMPS~\cite{yu_finding_2017} ($L>16$) with 128 disorder realizations.
Two eigenstates are generated for every pair $(W,\epsilon)$.
If SIMPS converges to the same eigenstate twice (which happens in less than $0.31\%$ of the cases), the duplicate is removed~\footnote{Less than $0.66\%$ of the runs were initialized on an eigenstate, preventing the algorithm from flowing to the desired energy density, and were therefore discarded. Less than $0.14\%$ of the eigenstates were also removed because of code failure. Therefore, in addition to the subsequent energy standard deviation filter, there was a total discard rate of less than $1.11\%$ due to technical reasons.}. 
SIMPS exploits the low entanglement of the MBL eigenstates to represent them efficiently using an MPS ansatz.
For eigenstates accessed by SIMPS in the strong disorder limit, the standard deviation of the energy is as low as machine precision ($~10^{-8}$); the increase in entanglement as we approach the transition makes the algorithm obtain eigenstates with a lower precision for fixed bond-dimension.
For this reason, we filter the ensemble of eigenstates by removing states whose standard deviation of the energy is higher than $10^{-3}$ (see Fig.~\ref{fig:survive}).
To minimize the possibility that either allowing states with a big standard deviation of the energy or restricting our results to only the eigenstates that have a very small standard deviation  biases our results, we have tested different thresholds and find that $10^{-3}$ gives robust results against large changes in the threshold.

\section{Results}
\label{sec:results}

\subsection{Correlation length and support of the OPOs}
\label{sec:support}

\begin{figure}[t]
\includegraphics[width=1.0\columnwidth]{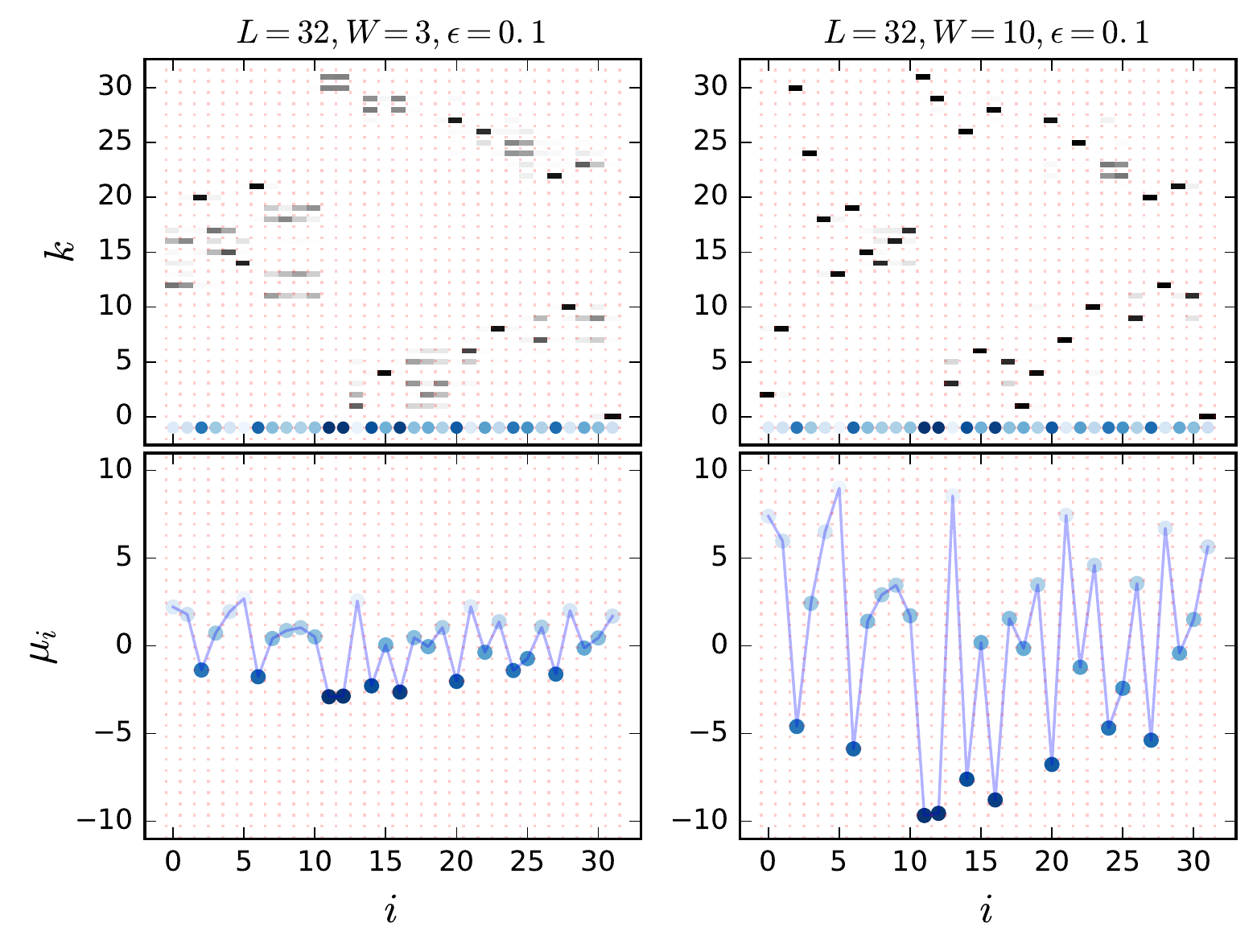}
\caption{\label{fig:opo_supports} \emph{Top:} Probability density $|U^\dagger_{ki}|^2$ in real space ($i$) of each OPO ($k$) of a particular eigenstate at $\epsilon=0.1$ of a system of size $L=32$ and $W=3.0, 10.0$.
\emph{Bottom:} profile of the random chemical potential $\mu_i$ at $W=3.0, 10.0$.
At strong disorder (right), the OPOs are highly localized on one site.
As the disorder is lowered, the OPOs start delocalizing, mixing over small non-overlapping subsystems of the chain.
There is a high probability of mixing along sites with a similar $\mu_i$, which occasionally gives rise to tunneling OPOs (see sites $14$ and $16$ at $W=3.0$ for an example).}
\end{figure}

\begin{figure}[t]
\includegraphics[width=1.00\columnwidth]{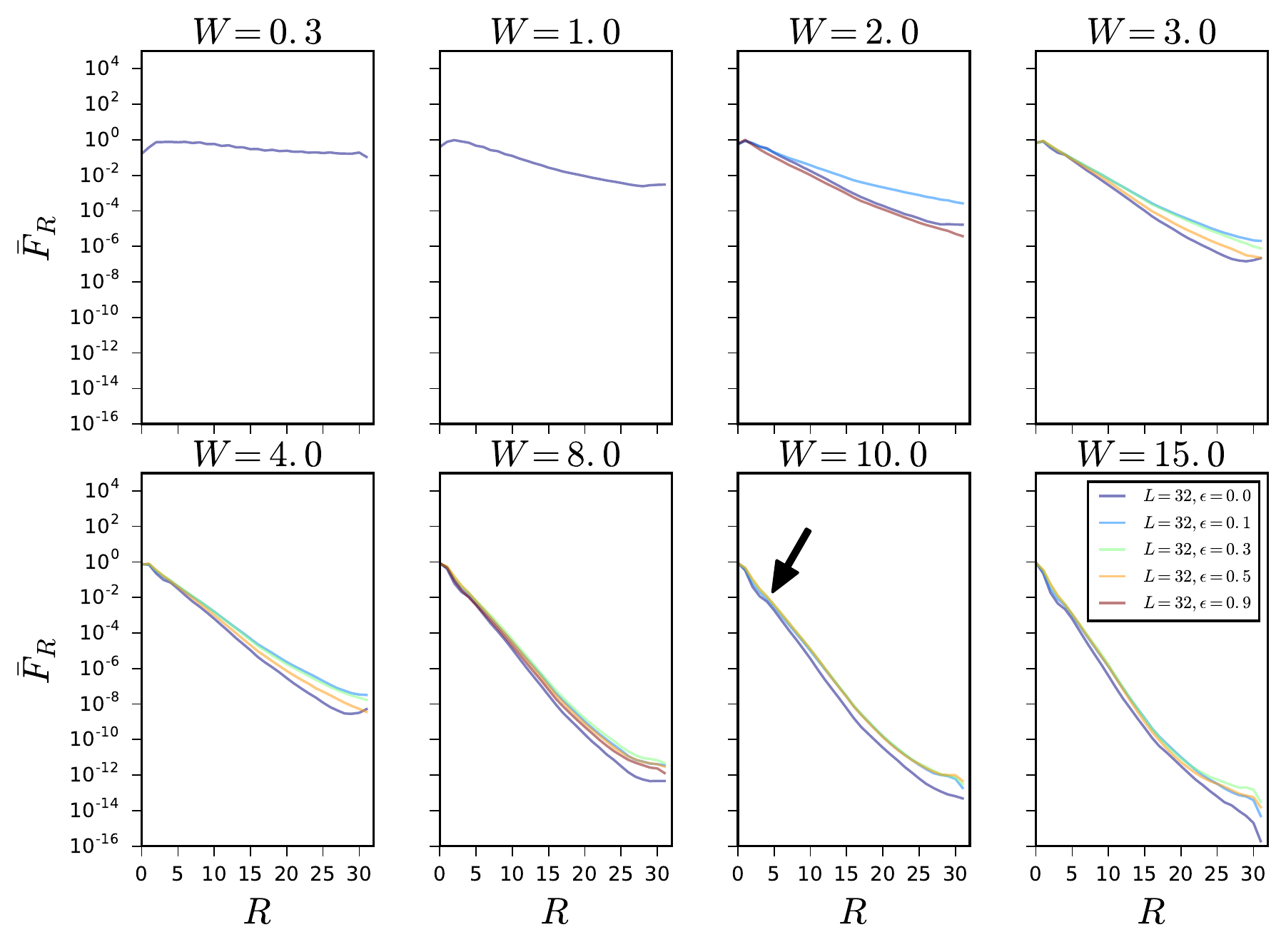}
\includegraphics[width=1.00\columnwidth]{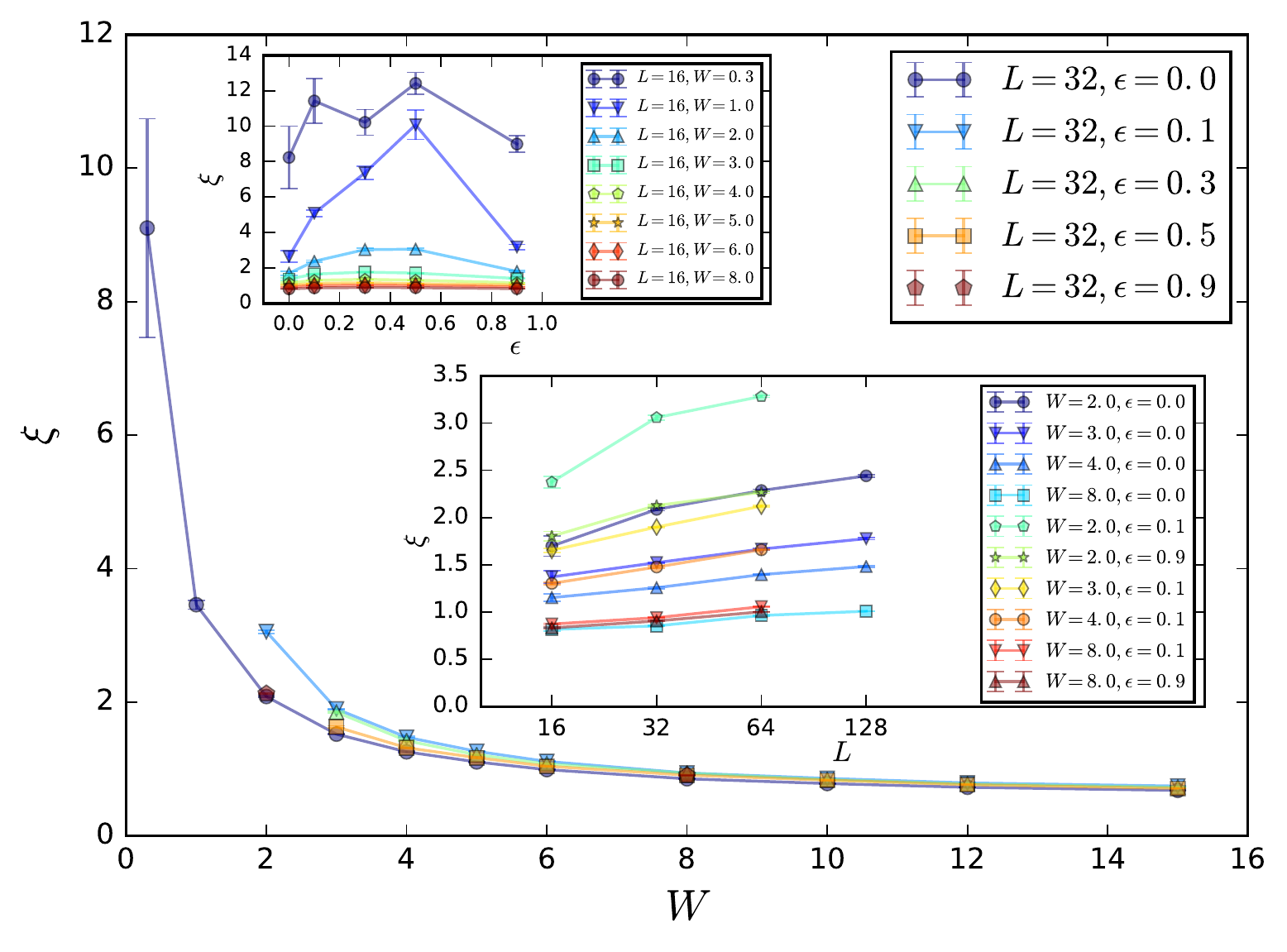}
\caption{\label{fig:FR_vs_R} \emph{Top:} total contribution $F_R$ from string operators of range $R$ to the definition of the OPO number operator $a^\dagger_k a_k$ (logarithmically) averaged over OPOs.
The average $\bar{F}_R$ decays exponentially with range $R$.
\emph{Bottom:} correlation length $\xi$ extracted from the exponential decay of $\bar{F}_R$.
\emph{Insets}: system size and $\epsilon$ dependence of $\xi$ }
\end{figure}

\begin{figure}[t]
\includegraphics[width=1.00\columnwidth]{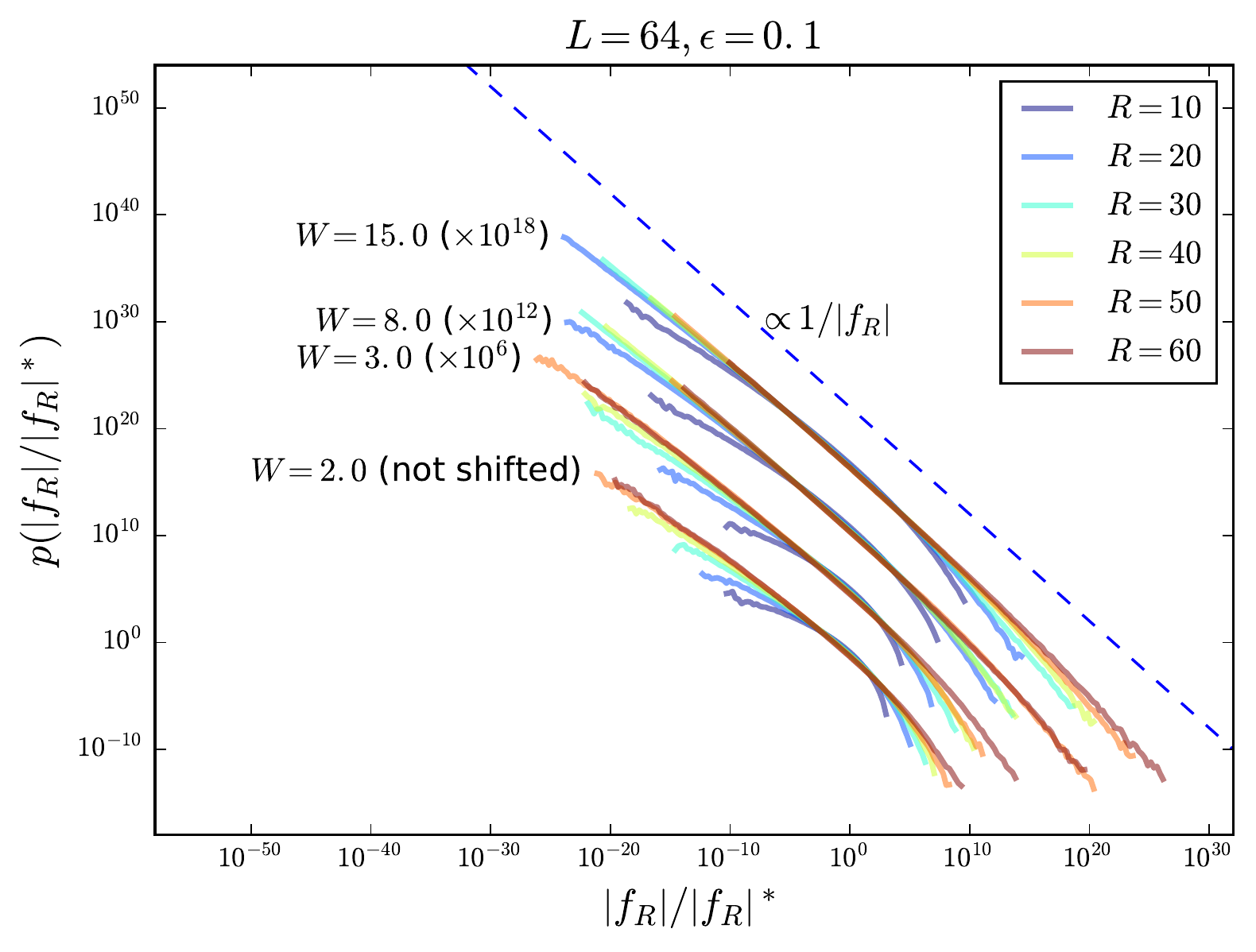}
\caption{\label{fig:pfR_vs_R} Probability distribution of the coupling constants $\left|f_R\right|$ divided by the typical coupling constant $\left|f_R\right|^* \equiv 10^{Mo\left(\log_{10}(|f_R|)\right)}$ (where the mode $Mo\left(\log_{10}(|f_R|)\right) \equiv \argmax\left[p\left(\log_{10}(|f_R|)\right)\right]$), $p\left(|f_R|/|f_R|^*\right)$, of the number operators of the OPOs at $\epsilon=0.1$ for fixed $W$ and range $R$, for systems of size $L=64$.
All curves, except for $W=2$, have been shifted in the $y$ axis for clarity; they would otherwise lay on top of each other and meet approximately at $|f_R|=|f_R|^*$ (where they are parallel to the $\propto 1/|f_R|$ reference line) and $p\left(|f_R|/|f_R|^*\right) \approx 10^{-1}$.}
\end{figure}

At strong disorder, each OPO is centered around a single site
with an exponentially fast decay (as will be discussed below).  As the disorder is lowered, we occasionally see more than one center and slower decay.
If we consider a single eigenstate, we can examine the probability density in real space of the set of OPOs (see Fig.~\ref{fig:opo_supports} for a generic example).
Notice that at moderate to large disorder, the OPOs are sharply localized at single sites.
At small disorder, the OPOs primarily mix in small groups (3-5 OPOs) over a local set of sites which
don't overlap each other.  Moreover, OPOs primarily mix with other OPOs which are at similar
occupation.  Take for example the four sites 7 through 10 for $W=3$ in Fig.~\ref{fig:opo_supports}, where all four OPOs which have
non-negligible amplitudes over these sites mix. We speculate that OPOs that tunnel a certain distance over the chain are related to resonances in the eigenstate.

\begin{figure}[t]
\includegraphics[width=1.00\columnwidth]{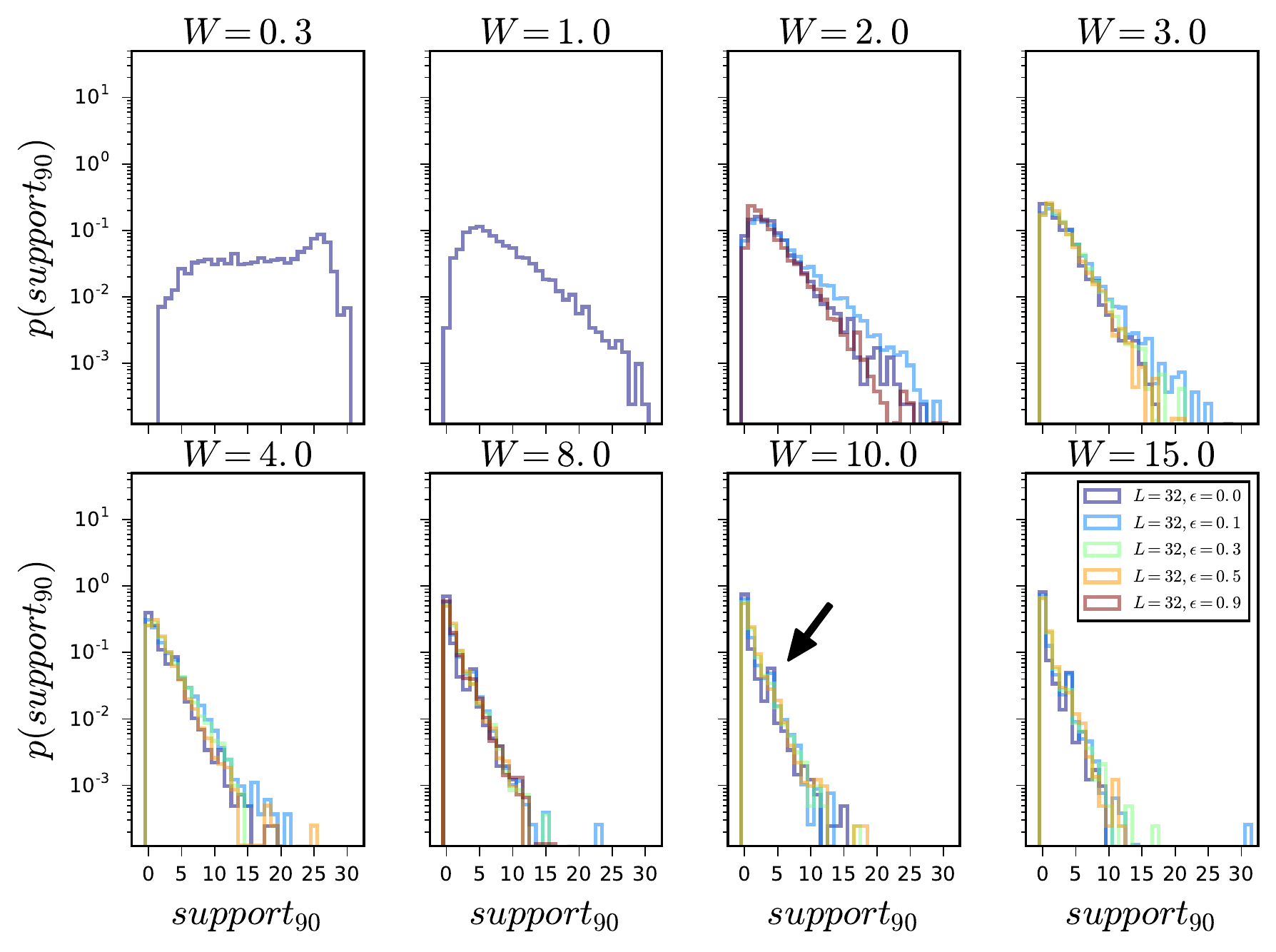}
\includegraphics[width=1.00\columnwidth]{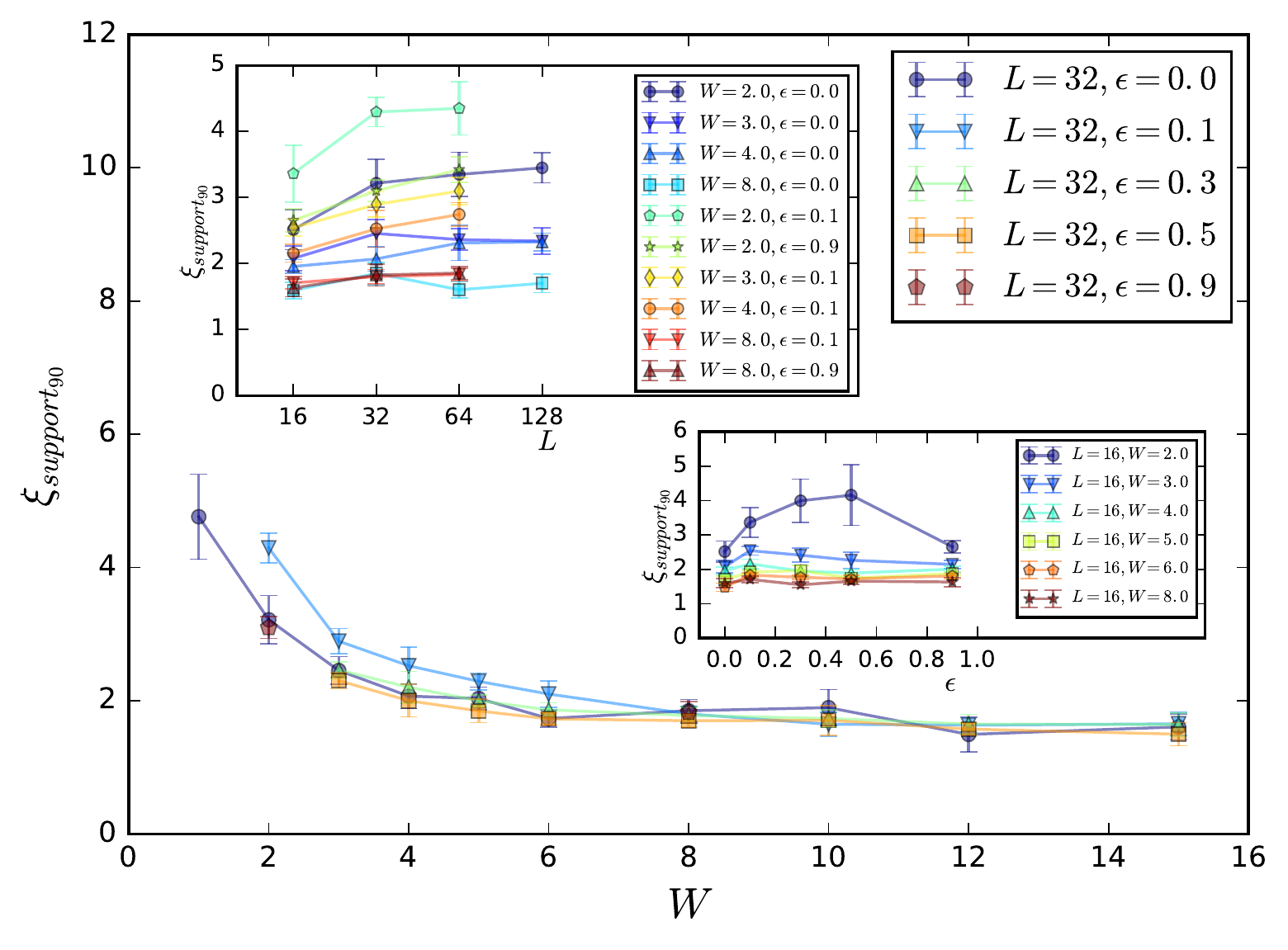}
\caption{\label{fig:hist_support_90} \emph{Top:} distribution of the support of the OPOs for different energy densities as a function of $W$ for $L=32$.
The $support_{90}$ is computed as the size of the region of that contains $90\%$ of the norm $L_2$ of the OPOs.
\emph{Bottom:} correlation length $\xi_{support_{90}}$ corresponding to the exponential decay of the distributions in the top panel.
\emph{Insets}: system size and energy density dependence of $\xi_{support_{90}}$.
}
\end{figure}

We now consider a  definition of the ``correlation length'' which applies to operators and is in the spirit of the correlation lengths 
used for  FMBL l-bits~\cite{huse_phenomenology_2014,abanin_recent_2017}.
Each OPO $k$ has its maximum amplitude at some site $m$ and has a number operator $a^\dagger_k a_k$  of the form:
\begin{align}
\label{number_operator}
a^\dagger_k a_k = \sum_{ij} f^k_{ij} c^\dagger_i c_j \text{,}
\end{align}
where $f^k_{ij} \equiv U^\dagger_{ki} U_{jk}$.
We define the range $R$ of the two-body strings $c^\dagger_i c_j$ relative to the localization
center $m$ as:
\begin{align}
\label{range}
	R\equiv\max\left( |i-m|, |j-m| \right)
\end{align}
(following the more general definition for l-bits of Ref.~\onlinecite{abanin_recent_2017}; a different choice of a definition for the range~\cite{huse_phenomenology_2014,pekker_fixed_2017} $R$ has few practical consequences, and is discussed in Appendix~\ref{sec:range}).
We expect the total contribution to $a^\dagger_k a_k$ from operators $c^\dagger_i c_j$ of different ranges to decay exponentially fast as a function of their range at strong disorder.
One way to quantify this is to define the contribution $F_R$ from range $R$ to OPO $k$ as the sum of all coefficients $\left| f^k_{ij} \right|$ of a particular range:
\begin{align}
\label{F_R}
F_R \equiv \sum_{\max\left( |i-m|, |j-m| \right)=R} \left|f^k_{ij}\right| \text{.}
\end{align}
Fig.~\ref{fig:FR_vs_R} presents the (logarithmic) average $\bar{F}_R$ across OPOs of $F_R$ as a function of $R$ for a system of size $L$=32 at different values of $\epsilon$ and $W$ (top panel).
Away from large $R$, where finite size effects are stronger, there is an exponential decay of $\bar{F}_R\propto e^{-R/\xi}$, with a characteristic correlation length $\xi$ that is shown in the bottom panel of Fig.~\ref{fig:FR_vs_R} (in fact, the exponential decay is not restricted to the average $\bar{F}_R$, but the raw distribution of $F_R$ also follows this form, as can be seen in Fig.~\ref{fig:hist_FR_vs_R} of Appendix~\ref{sec:sup_correlation}).
As $W$ gets smaller, $\xi$ increases monotonically; interestingly, in the ground state, the
correlation length seems to increase significantly at $W=0.3$. The lack of any clear divergence at
finite energy density is consistent with the fact that none of these points are in the ergodic phase.
While at large disorder $\xi$ is independent of energy density $\epsilon$, at smaller disorder ($W \approx 2, 3$) $\xi$ develops an
energy density dependence, with larger values towards the middle of the spectrum; this dependence becomes strong in the weak disorder limit (see upper inset of Fig.~\ref{fig:FR_vs_R}); this is clearly suggestive of the mobility edge.
The correlation length increases monotonically with system size (see lower inset of
Fig.~\ref{fig:FR_vs_R}); although the precise functional form of the scaling is unclear, it is
consistent with a logarithmically increasing correlation length within the MBL phase which might be
the result of exponentially rare regions.  
See Appendix~\ref{sec:sup_correlation} for additional information on the correlation length $\xi$.

The exponential decay of $\bar{F}_R$ can be related to the exponential decay of the tails of the OPOs.  
Assuming that the average exponential decay of $\bar{F}_R$ is representative of a typical case, it can be argued (see Appendix~\ref{sec:exponential}) that the decay of the tails of the OPOs is of the form $\left| U^\dagger_{ki} \right| \propto e^{-|i-m|/\xi} / \left(A+Bg(|i-m|)\right)$, where $A$ and $B$ are positive constants and $g(x)$ is a monotonically increasing function with limits $g(0)=0$ and $g(\infty)=1$.
The decay of the number operator $a^\dagger_k a_k$ and that of the OPOs' tails therefore have the same asymptotic exponential behavior, with the same correlation length $\xi$.  We verify this numerically (see Fig.~\ref{fig:tail} in Appendix~\ref{sec:exponential}).

Let $|f_R|$ be randomly sampled from the set of the magnitudes of the coefficients $f^k_{ij}$ (from Eq.~\eqref{number_operator}) for fixed range $R$ ($|f_R| \in \left\{\left|f^k_{ij}\right|\right\}_{R=const.}$) for fixed $L$, $W$ and $\epsilon$.
The probability that $|f_R|$ is of a given value, $p(|f_R|)$, decays as $\propto 1/|f_R|$ at large $W$ and $R$, as shown in Fig.~\ref{fig:pfR_vs_R}.
This is the same behavior found in Ref.~\onlinecite{pekker_fixed_2017} for l-bits (although for a slightly different definition of the range; see Appendix~\ref{sec:range}); the one particle approximation offers though a plausible explanation for this behavior, which arises directly from the exponential decay of the tails of the OPOs, and is discussed in Appendix~\ref{sec:range}.
In general, if the coupling constants of an l-bit decay exponentially at fixed range, in the sense that $p\left(\log(|f_R|)\right)=const.$, then $p(|f_R|)\propto 1/|f_R|$, due to the identity $d\left(\log(|f_R|)\right)/dp(|f_R|) = 1/|f_R|$.

\begin{figure}[t]
\includegraphics[width=1.00\columnwidth]{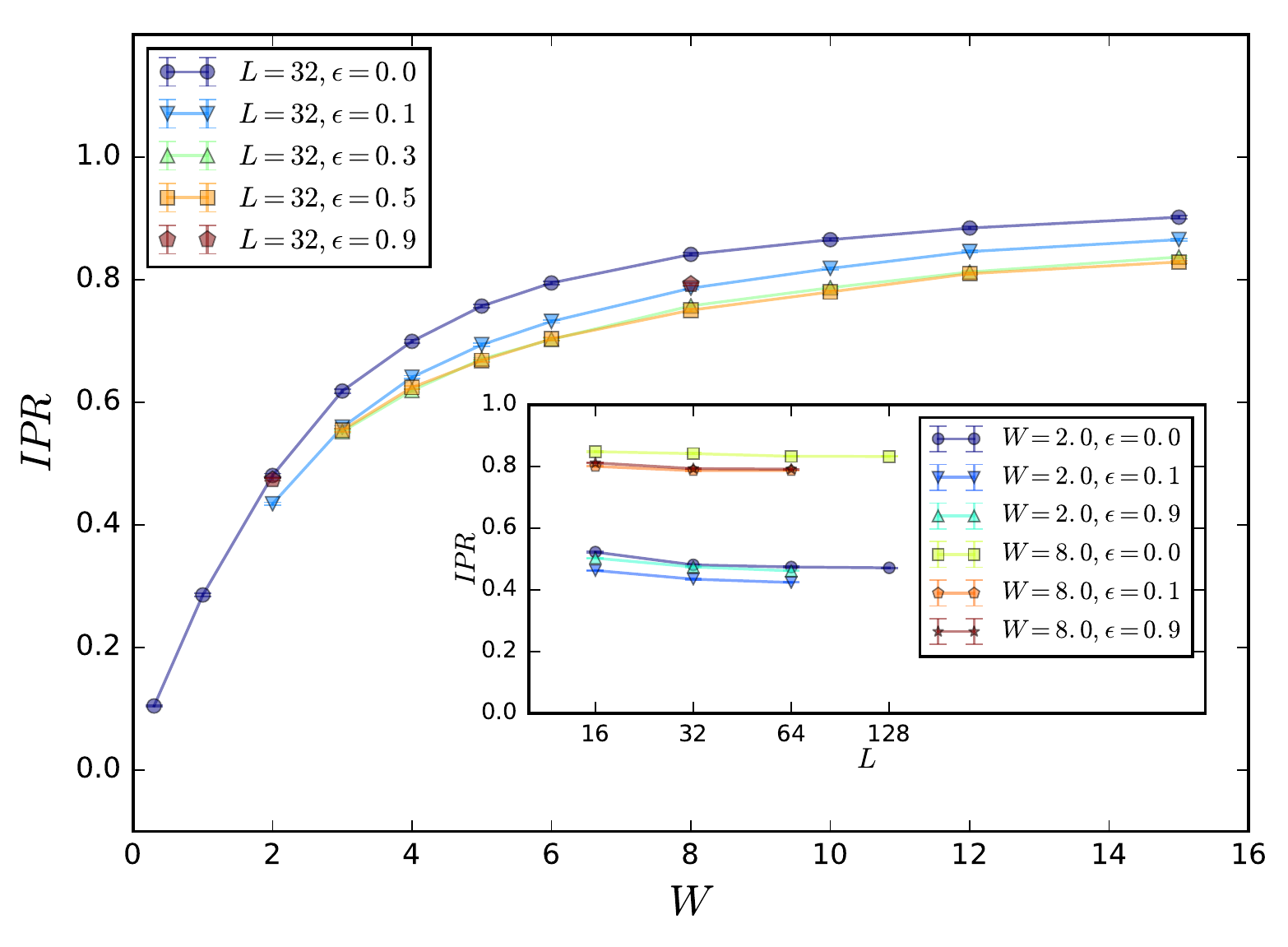}
\includegraphics[width=1.00\columnwidth]{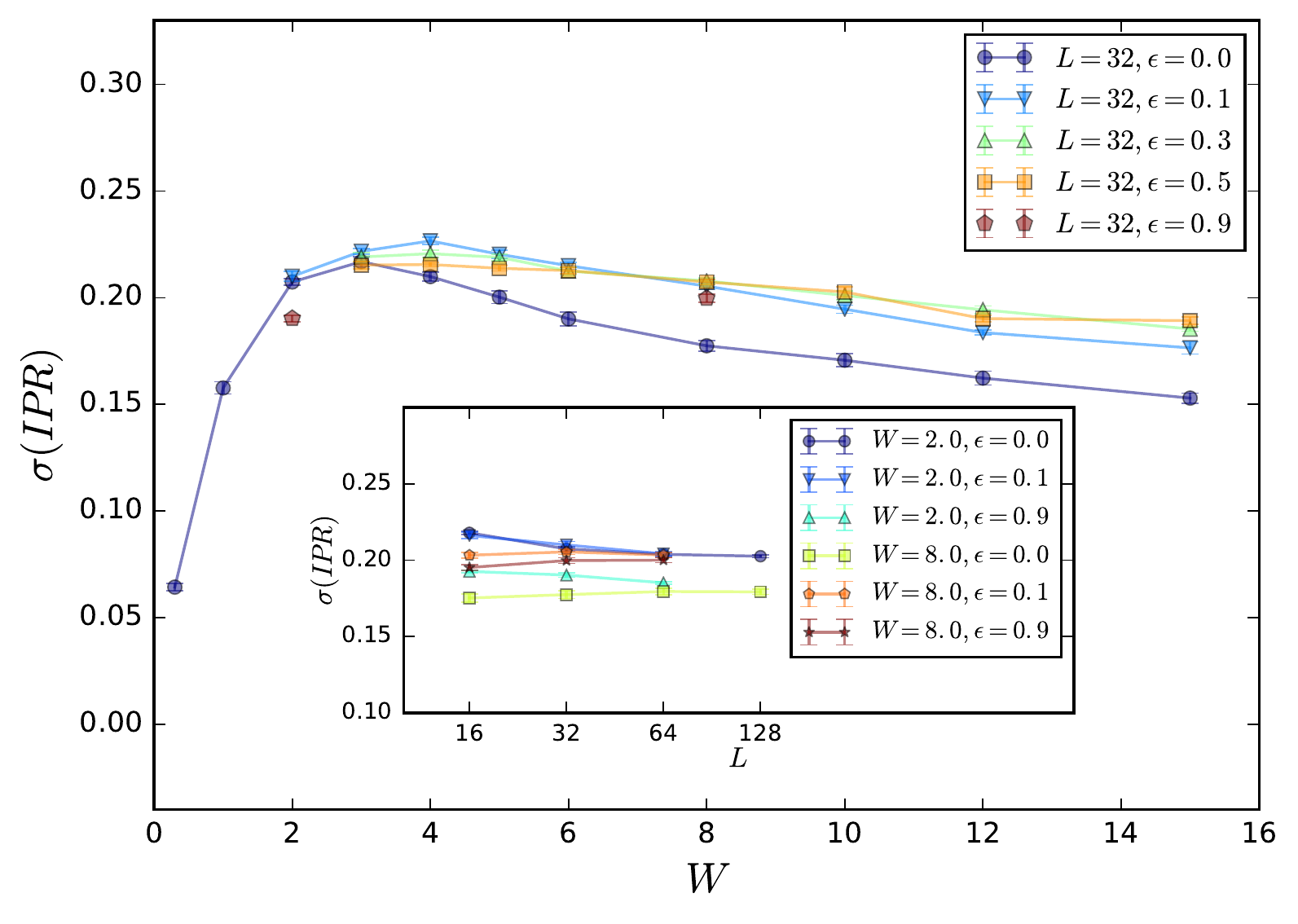}
\caption{\label{fig:mean_std_IPR} \emph{Top:} IPR as a function of $W$ for $L=32$ averaged over OPOs and disorder.  \emph{Inset:} average IPR as a function of $L$.  \emph{Bottom:} standard deviation of the IPR of the OPOs for $L=32$. \emph{Inset:} standard deviation of the IPR as a function of $L$. }
\end{figure}

\begin{figure}[t]
\includegraphics[width=1.00\columnwidth]{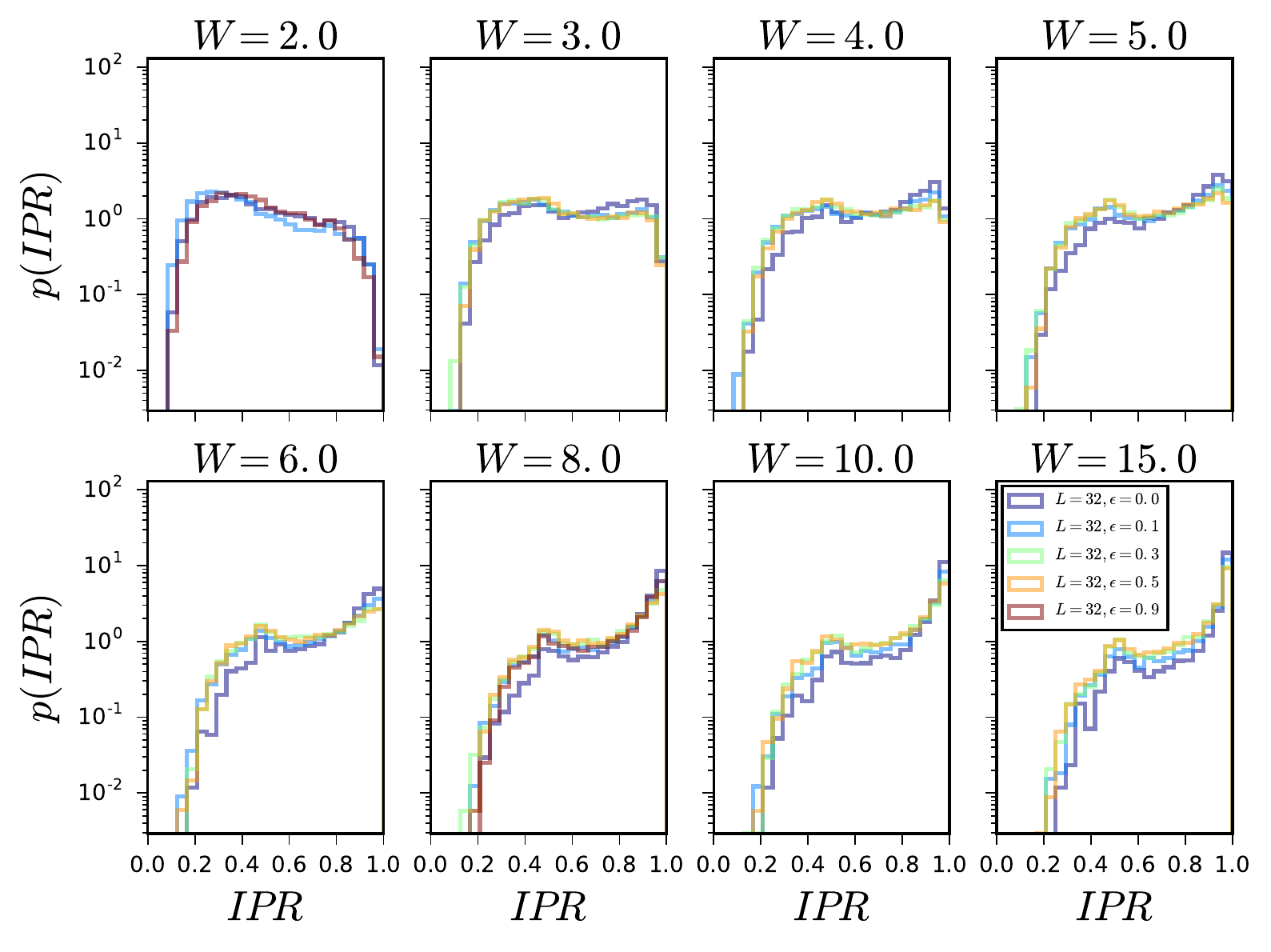}
\caption{\label{fig:bimodality} Distribution of IPR for $L=32$.
}
\end{figure}

\begin{figure}[t]
\includegraphics[width=1.0\columnwidth]{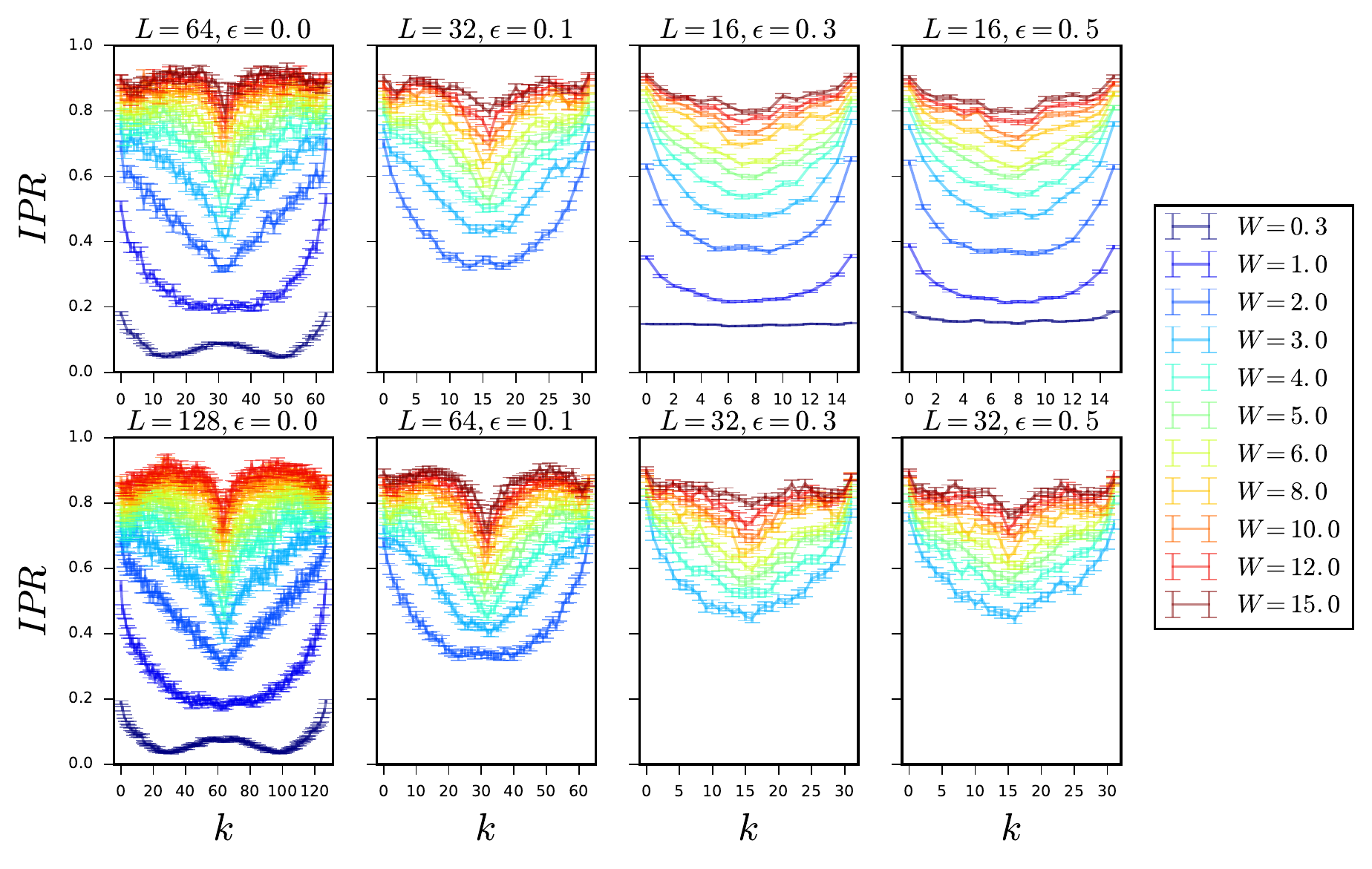}
\caption{\label{fig:IPR_vs_OPO} IPR as a function of $k$, \emph{i.e.} as a function of OPOs ordered by occupation, averaged over OPOs. 
The shapes of the curves are characteristic of, respectively, strong disorder eigenstates, eigenstates around the critical disorder strength $W_c$ and weak disorder, independent of $\epsilon$.
}
\end{figure}

\begin{figure}[t]
\includegraphics[width=1.00\columnwidth]{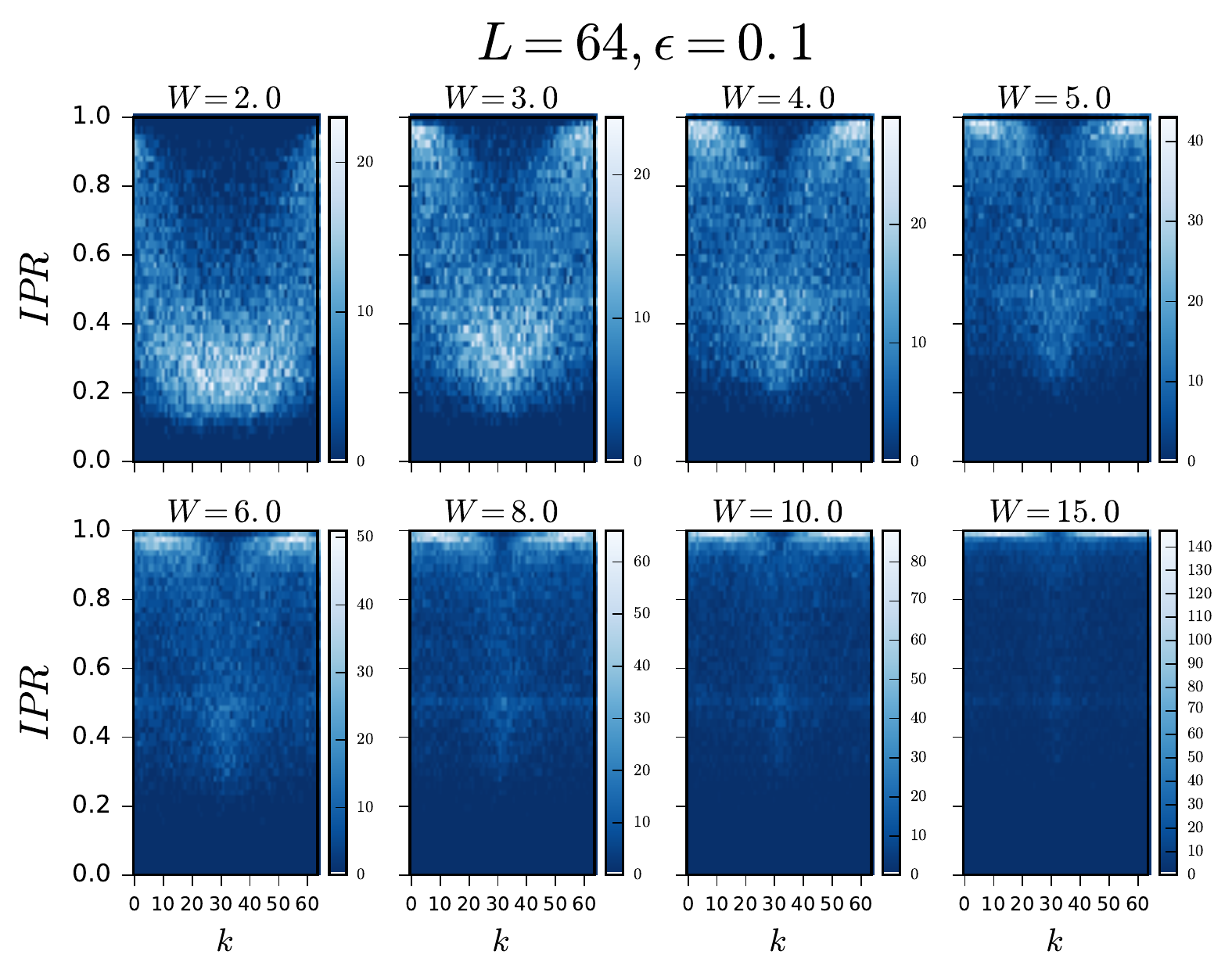}
\caption{\label{fig:corr_orbital_number} 2D histogram of the IPR of the OPOs vs. $k$ for $L=64$, $\epsilon=0.1$. It is easy to see the emergence of the characteristic curves presented in Fig.~\ref{fig:IPR_vs_OPO}. At strong disorder, intermediate values of $k$ have an IPR close to 0.5 (see Fig.~\ref{fig:bimodality}). }
\end{figure}

An alternative definition of the support of an OPO $k$ is to let it be the size of the smallest region of the chain that contains $90\%$ of the norm $\sum_i \left|U^\dagger_{ki}\right|^2$ of the OPO  
(the choice of a threshold of $90\%$ is arbitrary).
The effective support of the OPOs is representative of the localization of the system, and their distribution for several energy densities $\epsilon$ and disorder strengths $W$ for systems of $L=32$ is shown in the top panel of Fig.~\ref{fig:hist_support_90}.
The decay of the probability distribution is exponential at disorder strengths far from the weak disorder limit.
At small disorder the distribution becomes flat; a system size dependence arises because the extent of the OPOs becomes longer than the system length. (see Fig.~\ref{fig:hist_support_90_fixed_E} in Appendix~\ref{sec:sup_support}).
For exponentially decaying distributions, $p(support_{90}) \propto e^{support_{90}/\xi_{support_{90}}}$, we define a correlation length $\xi_{support_{90}}$  (bottom panel of Fig.~\ref{fig:hist_support_90})~\footnote{Due to the difficulty of extracting $\xi_{support_{90}}$ from a single linear fit over the distributions of $p(support_{90})$, which are rather noisy, we compute their slope as the average of several linear fits performed over different ranges of the x-axis, weighted by the inverse of their standard errors. The error in the estimation of the slope is computed as the standard deviation of the weighted samples}.
At strong disorder, $\xi_{support_{90}}$ is effectively independent of $\epsilon$ and of system size.
At $W < W_c$, an $\epsilon$ dependence arises, with higher values towards the middle of the energy spectrum (see lower inset of Fig.~\ref{fig:hist_support_90}).
Below $W\approx W_c$ the correlation length rises sharply (but does not obviously diverge) and might be weakly system size dependent.
Note that at large disorder and low energy density there exists a kink (see arrow for an example) in the distribution for a support of length $4$, which biases the probability of finding an OPO of $support_{90}=4$.
We think that this is related to the kink seen in Fig.~\ref{fig:FR_vs_R} for the same cases (see arrow). 
This same effect is barely seen in the distribution of $support$ of Fig.~\ref{fig:hist_support} of
Appendix~\ref{sec:sup_support}, but is visible in the distribution of $support_{90}$ of Fig.~\ref{fig:hist_support_90}.

\subsection{Inverse participation ratio of the OPOs}
\label{sec:IPR}

\begin{figure*}[t]
\includegraphics[width=0.58\columnwidth]{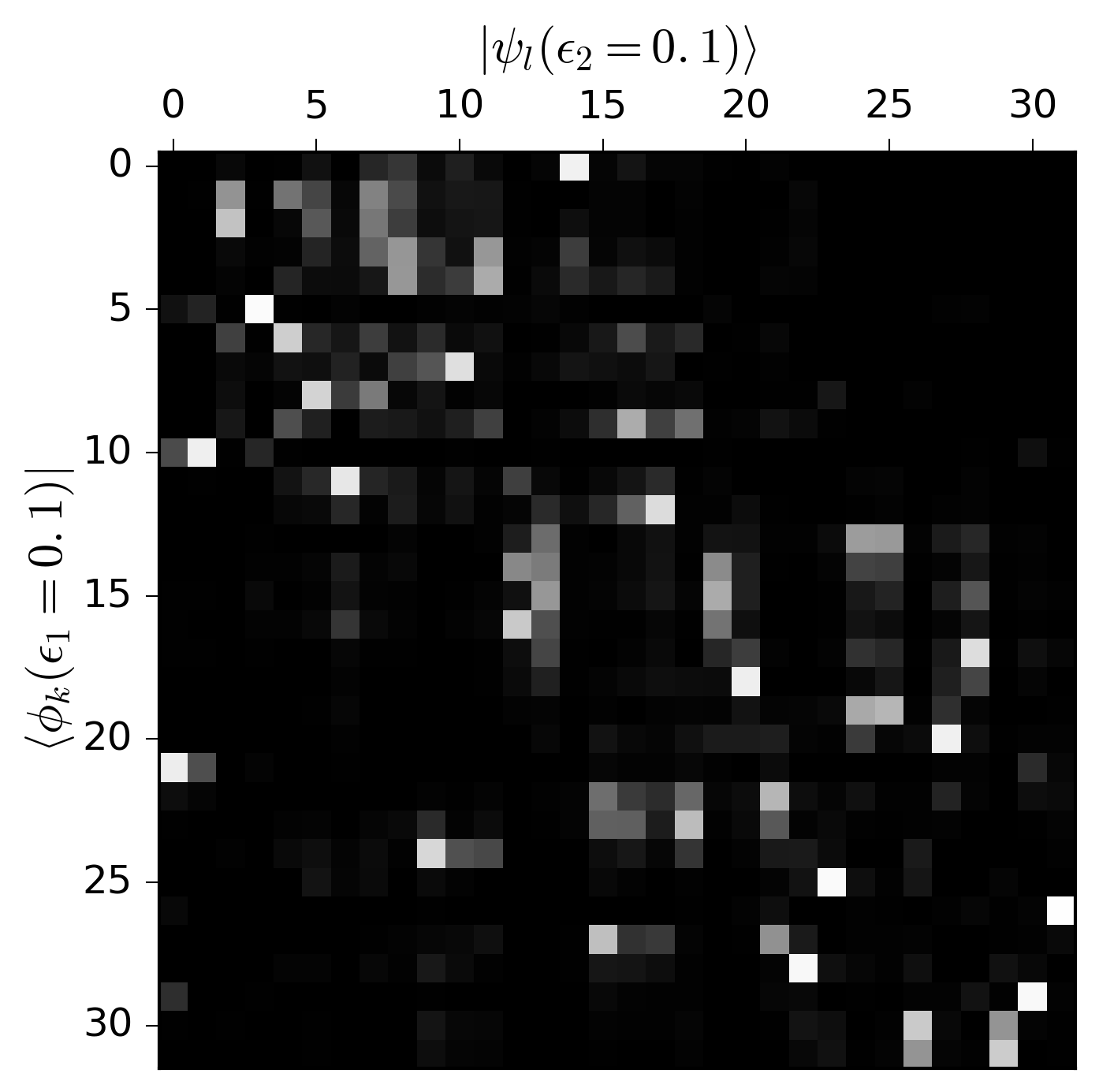}
\includegraphics[width=0.75\columnwidth]{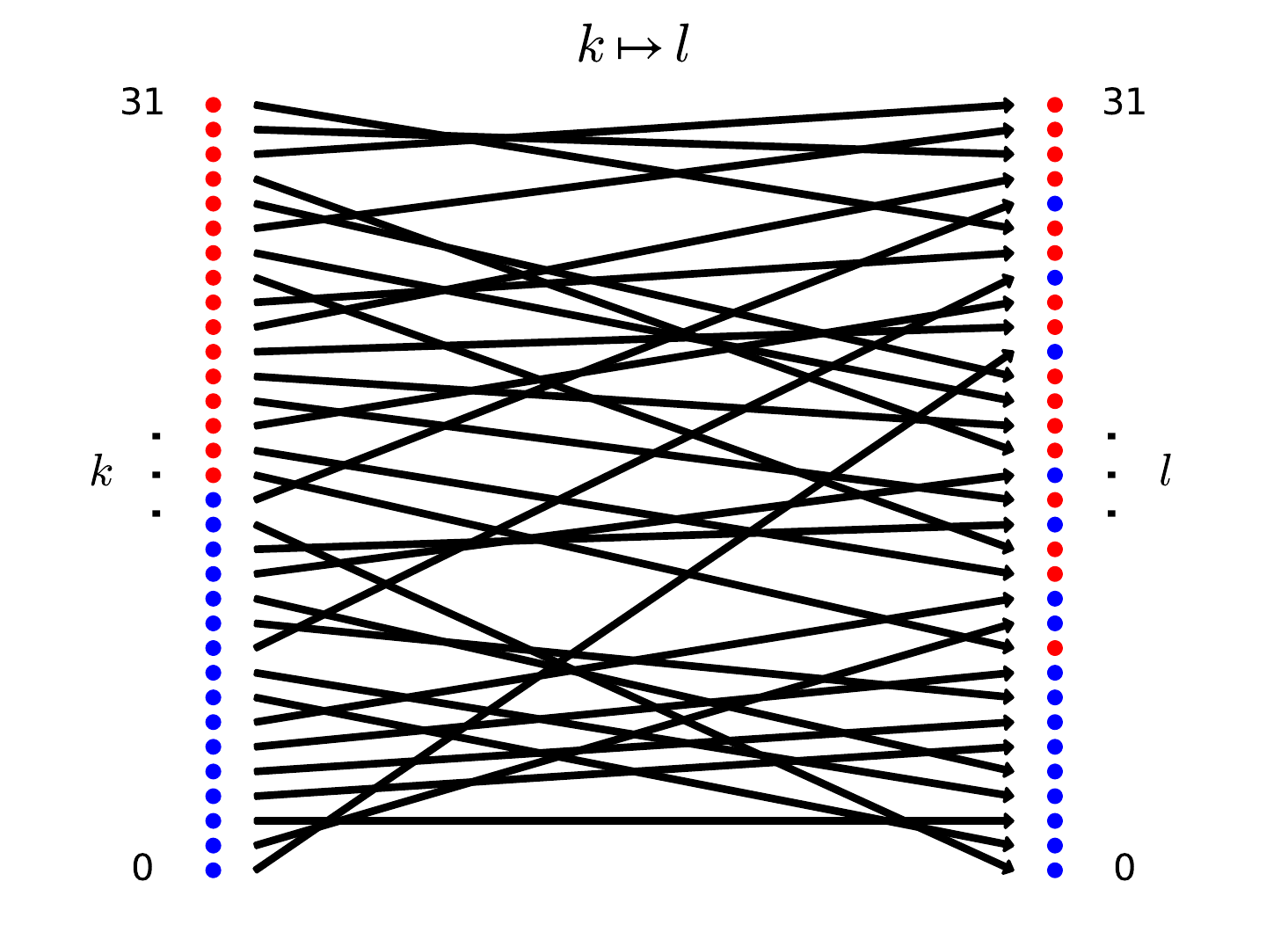}
\includegraphics[width=0.58\columnwidth]{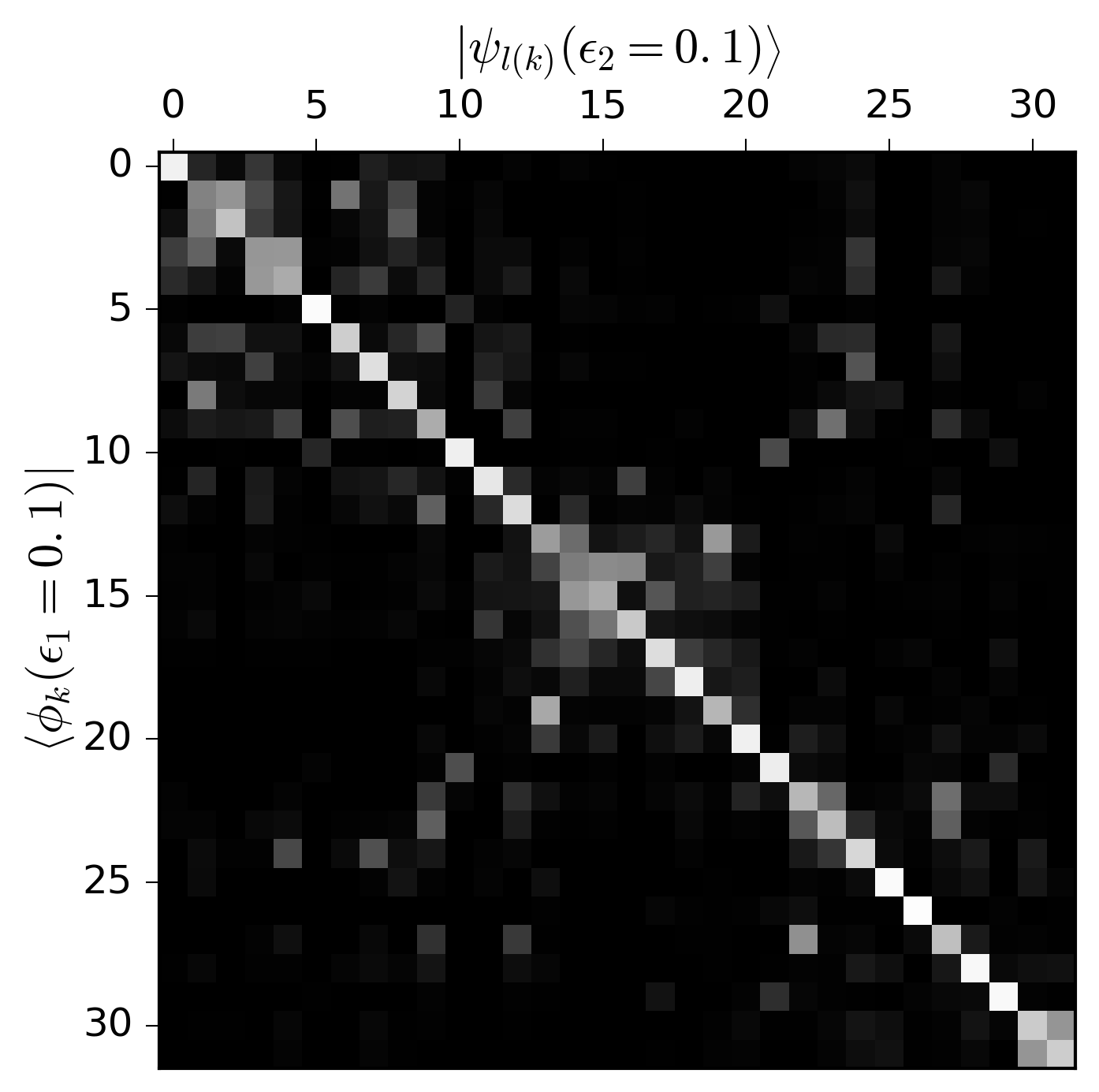}
\includegraphics[width=0.58\columnwidth]{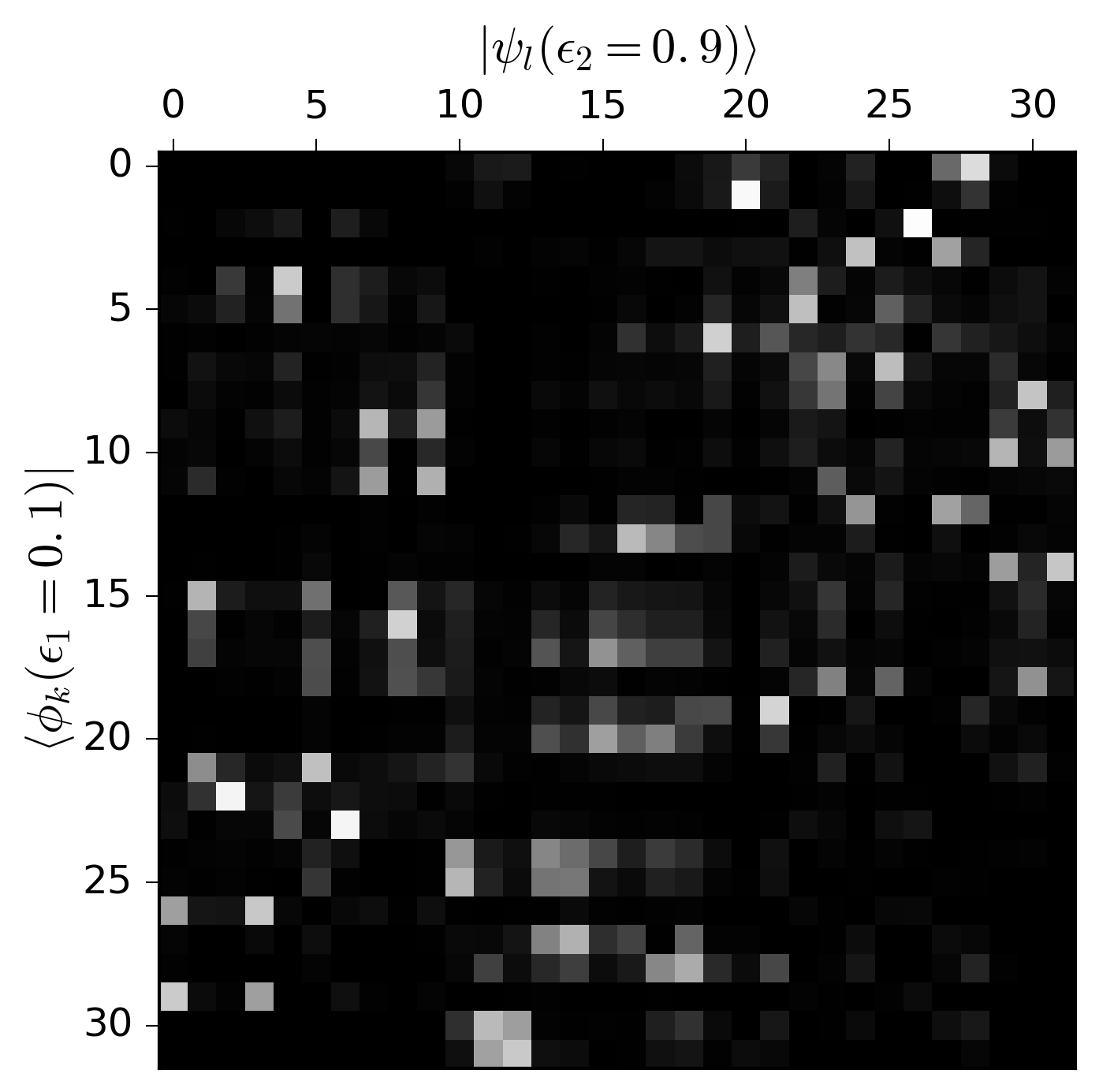}
\includegraphics[width=0.75\columnwidth]{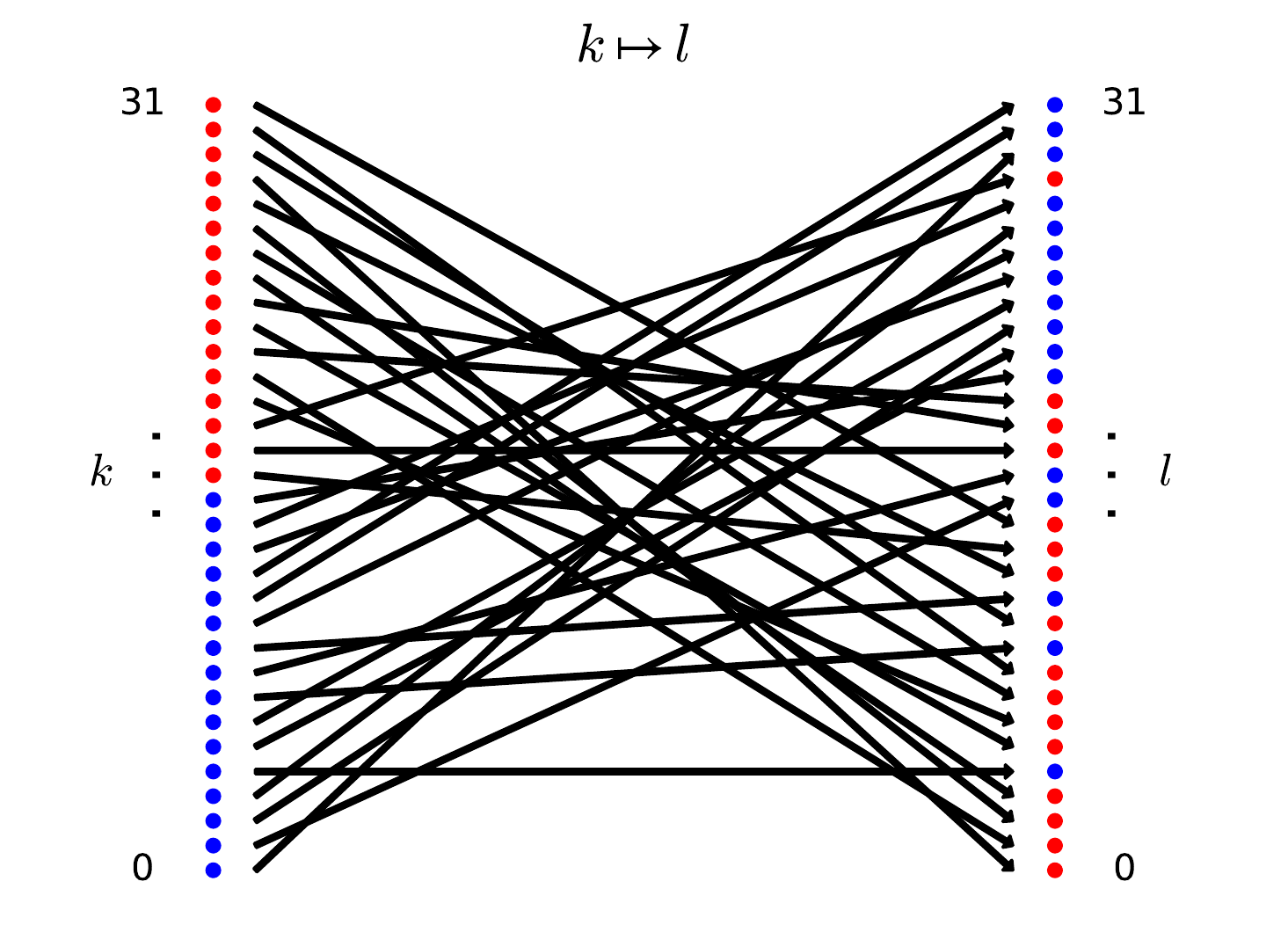}
\includegraphics[width=0.58\columnwidth]{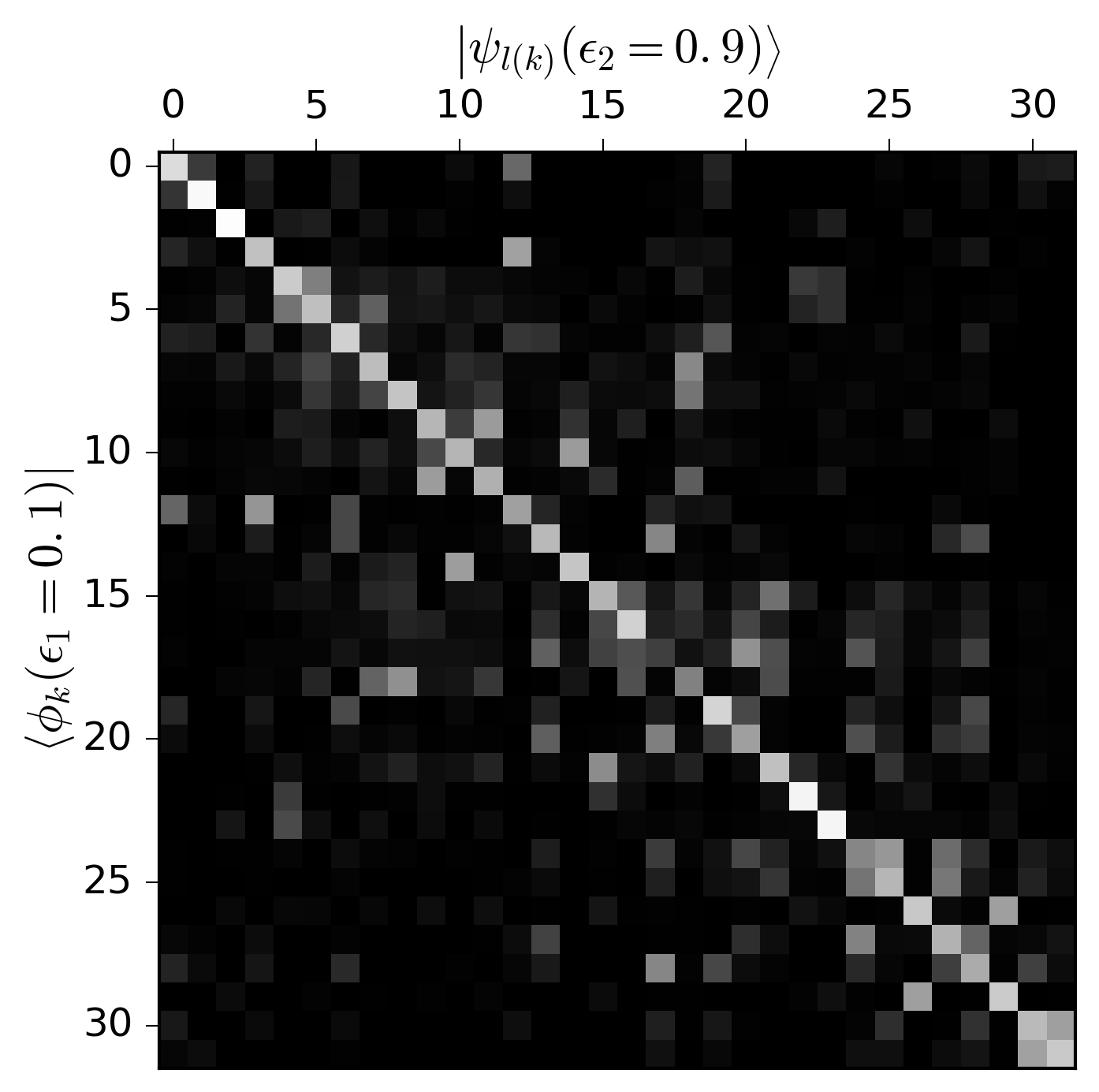}
\begin{flushleft}
\caption{\label{fig:matrix_overlaps} \emph{Left:} matrix of overlaps $M_{overlap}=|\langle \phi_k|\psi_l \rangle|$} between the OPOs of two different eigenstates with $W=2$ at  energy densities $\epsilon_1$ and $\epsilon_2$, for a system of size $L=32$. \emph{Middle:} permutation between OPOs of the two eigenstates, $k\mapsto l$. The coloring on the left column is red for the half of the OPOs that have highest occupation and blue for the half with lowest occupation.  The coloring on the right is inherited from the color of the OPO on the left to which it is mapped.  \emph{Right:} matrix of overlaps $M_{overlap}$ with the columns ordered after the permutation shown by the middle diagrams.
\end{flushleft}
\end{figure*}

In this section we consider the inverse participation ratio (IPR), a measure of localization commonly used in Anderson localization. The IPR of the $k$'th OPO is defined as:
\begin{align}
\label{IPR}
  \text{IPR} = \sum_{i=0}^{L-1} |U^\dagger_{ki}|^4 \text{.}
\end{align}
where $U^\dagger_{ki}$ is the matrix of OPOs that diagonalizes $\rho$, as defined in Section~\ref{sec:OPDM}, and $k$ labels the OPOs. The IPR of an OPO that is completely localized on one site is equal to $1$, while a delocalized OPO that is evenly distributed among all sites of the chain has an IPR of $1/L$.

We study the distribution of IPRs of the OPOs obtained for different points in the phase diagram.
Note that the average IPR increases monotonically with $W$ (see top panel of Fig.~\ref{fig:mean_std_IPR}), implying more localized orbitals at stronger disorder.
Although this behavior is common to all values of the energy density, the curves depend slightly on $\epsilon$, with lower values of the IPR towards the middle of the spectrum, and have a weak system size dependence at small disorder.

The standard deviation of the distribution of IPRs, $\sigma(IPR)$,  is presented in the bottom panel of Fig.~\ref{fig:mean_std_IPR}.
For all $\epsilon$ we find a peak of $\sigma(IPR)$.  Like the peak seen in the standard deviation of the entanglement entropy at half-cut at the transition\cite{kjall_many-body_2014,yu_bimodal_2016}, the peak in $\sigma(IPR)$ can be viewed as identifying a transition between the ergodic and MBL phase. Interestingly, while the eigenstates we consider at low energy density ($\epsilon=0.1,0.3$)  are in the MBL phase (see Fig.~\ref{fig:phase_diagram}), we find a peak at $W\approx 4$ near the critical disorder strength $W_c$ at the nose of the mobility edge.
This result suggests the possibility that MBL eigenstates \emph{know} whether they lie in the FMBL region  of the phase diagram or instead lie below a mobility edge.
The $\sigma(IPR)$ obtained from ground states also shows a peak, although at a lower value of $W$.
$\sigma(IPR)$ at $\epsilon=0.5$ is almost flat around the peak at $W_c$.
Note also that the curves of $\sigma(IPR)$ are only weakly system size dependent for the disorder strengths considered, \emph{i.e.} away from the $W=0$ limit.

To better understand this peak in $\sigma(IPR)$ we can consider the full probability distribution of the IPR of the OPOs.
As we see in Fig.~\ref{fig:bimodality}, it follows a bimodal distribution.  At large disorder, the distribution is peaked at 1.0 corresponding to most of the OPO's being highly localized;  the secondary peak at 0.5 at large disorders corresponds to OPO's with their amplitude evenly distributed between two sites.  At small disorder, for eigenstates in the MBL phase but deep below (or above) the mobility edge, there is a broad distribution of the OPO's with a maximum at small IPR;  this suggests some orbitals are localized but the plurality of them are extended.
The distribution presents its maximum spread (and most apparent bimodality) between $W=3$ and $W=4$, \emph{i.e.} around $W_c$, in agreement with the peak in $\sigma(IPR)$ (Fig.~\ref{fig:mean_std_IPR}).
As with the averaged IPR and $\sigma(IPR)$, the distribution's behavior is independent of $\epsilon$, although it slightly drifts towards higher values of the IPR for ground states.
In addition, there is system size independence (see Fig.~\ref{fig:bimodality_fixed_E} in Appendix~\ref{sec:size_independence}) at the $W$'s considered;  presumably though, in the $W=0$ limit, the IPR would collapse to $1/L$.
The bimodality observed here is similar to the bimodality of the distribution of the entanglement entropy at half-cut around the transition found in Ref.~\onlinecite{yu_bimodal_2016}.
Unlike in Ref~\onlinecite{yu_bimodal_2016}, where the bimodality of the entanglement entropy is only studied at $\epsilon=0.5$, here different values of $\epsilon$ are studied;  because the distribution of the IPR is independent of $\epsilon$, we can identify a transition from MBL eigenstates at small $\epsilon$ deep below the mobility edge and far from the transition (see Fig.~\ref{fig:phase_diagram}).

The bimodality of the distribution of the IPR of the OPOs of Fig.~\ref{fig:bimodality} is not visible in the distribution of the support of Fig.~\ref{fig:hist_support_90} of Section~\ref{sec:support}, although the distributions are broad in the transition region.
Indeed, the IPR and the support measure different things.
The IPR is very sensitive to the broadening of an OPO, but it can be insensitive to the size of its support.
Take for example an OPO with its amplitudes equally distributed between two nearest neighbor sites; while the support of this OPO is very small, its IPR is equal to $0.5$ (we attribute the bump found in the IPR at $0.5$ for strong $W$ in Fig.~\ref{fig:bimodality} and~\ref{fig:corr_orbital_number} to this).
At the same time, if the OPO's amplitudes are distributed evenly over two distant sites, its IPR is still $0.5$, but its support is large.
This explains why the bimodality found in the IPR does not imply a bimodal distribution of the support, however the broad distribution of the support confirms the coexistence of localized and extended OPOs in the transition region below the mobility edge.

We now analyze the correlation of the IPR of an OPO with its occupation.
In Fig.~\ref{fig:IPR_vs_OPO} we present the average IPR of the OPOs as a function of OPO number $k$, which are ordered by increasing occupation $n_k$.
We find curves have higher IPR at low and high occupations (close to 0 and 1) as compared to intermediate occupations (which are near the gap in the occupation spectrum).
Both the very strong and very weak disorder IPR curve is largely flat with an exception at occupation near the very middle of the spectrum in large systems where there is an inverted peak. 
These OPOs in the middle of the spectrum have occupations away from 0 or 1, even for fairly strong disorder, as can be seen in Refs.~\onlinecite{bera_many-body_2015,bera_one-particle_2017} and in Fig.~\ref{fig:occupations}.
For intermediate disorder strengths there is significant curvature around the critical disorder strength $W_c$.

Fig.~\ref{fig:corr_orbital_number} shows the distribution of the IPR vs. $k$ for a system of size $L=64$ at an energy density $\epsilon=0.1$.
The appearance of the inverted peak  at strong disorder results from the orbitals with an IPR of 0.5, which accounts for the secondary peak seen in Fig.~\ref{fig:bimodality} at strong disorder, and which correspond primarily to OPOs with amplitudes evenly distributed between two (usually nearby)
sites.
This correlation between the IPR and $k$ will be discussed further in Section~\ref{sec:correspondence}.

\subsection{OPOs at different energy densities}
\label{sec:correspondence}

\begin{figure}[t]
\includegraphics[width=1.0\columnwidth]{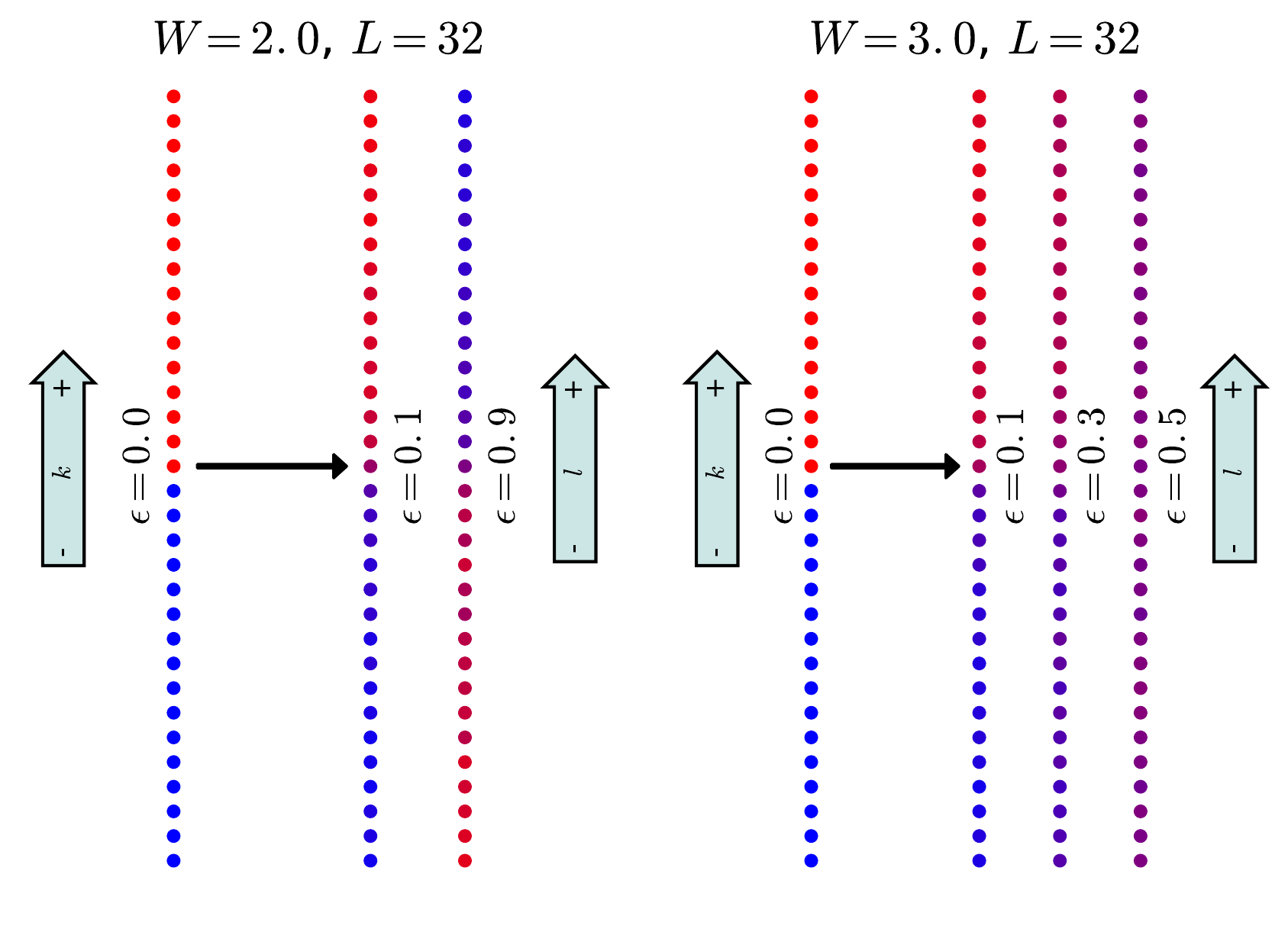}
\caption{\label{fig:average_arrows} The coloring on the left column is red for the half of the OPOs that have highest occupation and blue for the half with lowest occupation.  The coloring on the right is the disordered average of all colors inherited from the OPO on the left to which it is mapped (see  Fig.~\ref{fig:matrix_overlaps} (middle) for a non-averaged version of this).
The closer the eigenstate energies are to each other, the more likely the occupations of the OPOs of two eigenstates will be preserved, ranging from a few swaps in occupation when the energies are close in the spectrum to almost all swaps when the energies are in opposite sides of the spectrum.}
\end{figure}

\begin{figure}[t]
\includegraphics[width=1.0\columnwidth]{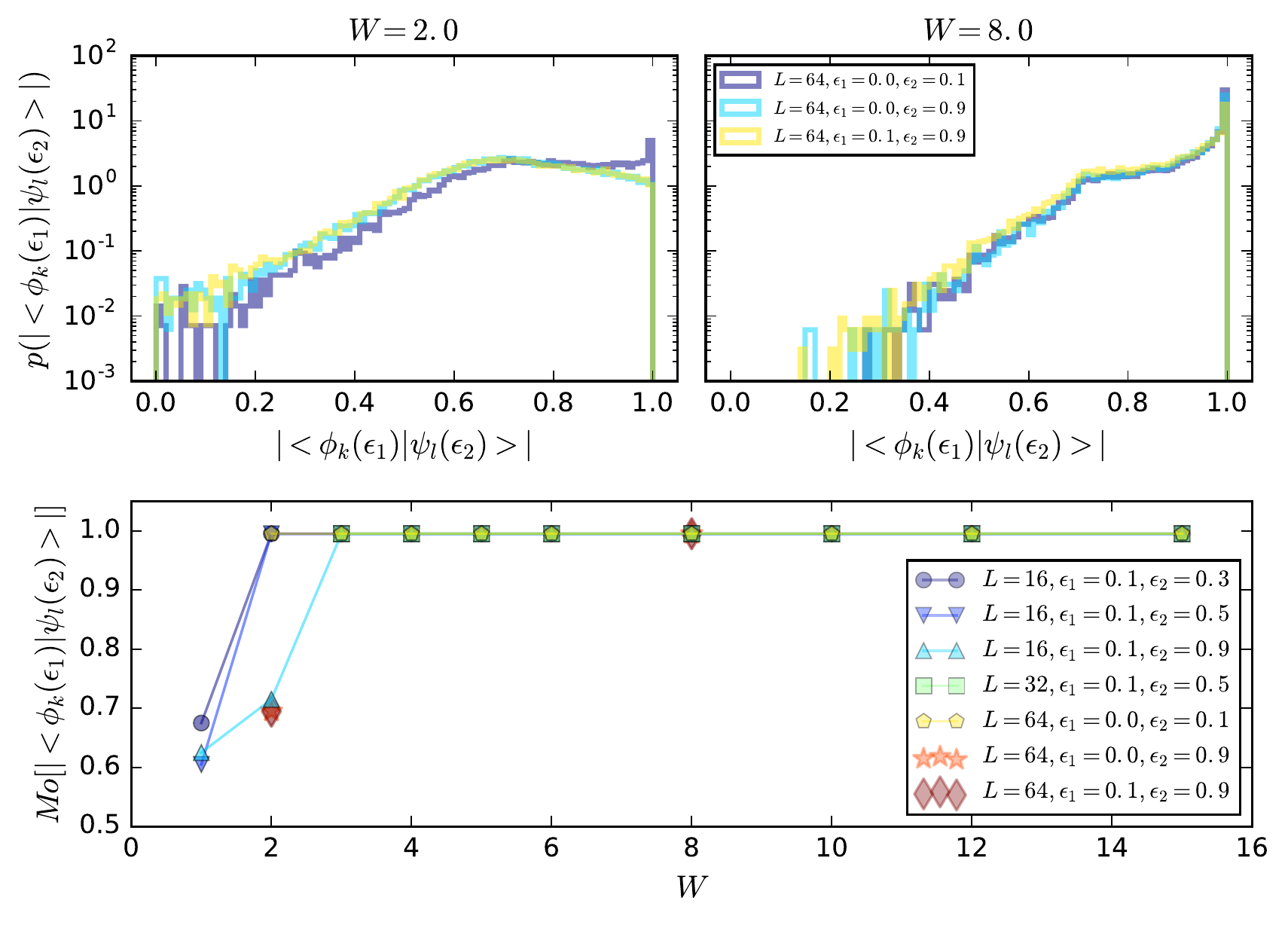}
\caption{\label{fig:hist_overlaps} \emph{Top:} distribution of overlaps  $|\langle \phi_{k}(\epsilon_1)|\psi_{l(k)}(\epsilon_2)\rangle |$ between corresponding OPOs of eigenstates at energy densities $\epsilon_1$ and $\epsilon_2$.
\emph{Bottom:} mode of the distribution of overlaps of corresponding OPOs.
}
\end{figure}

Motivated by the suggestive picture that OPOs represent approximately the one particle operator content of the l-bits,
 we expect that the OPOs of different eigenstates are very similar, since they originate from the same set of l-bit operators.
To test this simple picture, we compute the matrix of overlaps $M_\textrm{overlap}=\left|\left<\phi_k\left(\epsilon_1\right)|\psi_l\left(\epsilon_2\right)\right>\right|$ between the OPOs of two different eigenstates of the same Hamiltonian at different energy densities $\{\epsilon_1, \epsilon_2\}$ (see Fig.~\ref{fig:matrix_overlaps} for a prototypical example).
We find high overlap between OPOs drawn from different eigenstates.  After ordering OPOs by their occupation, we can consider which OPO's of one eigenstate map to OPO's of another eigenstate.
This is accomplished by finding the permutation of columns of $M_\textrm{overlap}$ which makes it maximally diagonal (see Fig.~\ref{fig:matrix_overlaps}).  Note that for two different eigenstates at a similar energy density, the permutation is close to the identity,  with highly occupied orbitals mapping to other highly occupied orbitals. However, for eigenstates at opposite sides of the energy spectrum the permutation essentially swaps highly occupied and unoccupied orbitals. Fig.~\ref{fig:average_arrows} shows a disordered average version of this behavior even from OPO's generated from the ground state.

The top panel of Fig.~\ref{fig:hist_overlaps} shows the distribution of overlaps $\left|\left<\phi_k\left(\epsilon_1\right)|\psi_{l(k)}\left(\epsilon_2\right)\right>\right|$ between matching pairs of OPOs for eigenstates at different pairs of energy densities $\{\epsilon_1, \epsilon_2\}$.
At moderate disorder ($W=8$) we find that the overlaps are extremely high and largely independent of $\epsilon_1$ and $\epsilon_2$.
Note that in the strong disorder limit all overlaps should be $1$.
For $W=2$, the magnitude of the overlaps decreases, but it is still surprisingly high; there is now a dependence on the energy densities, with better overlaps for $\epsilon_1\approx\epsilon_2$.
The typical overlap between matching pairs of OPOs is represented by the mode of the distribution, which is shown in the bottom panel of Fig.~\ref{fig:hist_overlaps} to be extremely close to $100\%$ at moderate and strong disorder as well as small disorder when $\epsilon_1\approx\epsilon_2$.  At $W=2$ it falls to $70\%$ for $\epsilon_1$ far from $\epsilon_2$ and at $W=1$ is drops below $70\%$ for all $\{\epsilon_1,\epsilon_2\}$ and $L=16$.
It should be noted that this strong overlap is not just caused by the fact that OPO's are generally centered on a site (see Appendix~\ref{sec:sup_overlaps} for further analysis and discussion).
Notice also that  the OPOs have high overlap even in the ergodic phase (for $L=16$).

The OPOs can be regarded as an approximate version of a set of integrals of motion of the system:
the high overlap between OPOs at different energy densities lets them acquire universality across the spectrum, and each eigenstate carries a particular permutation (correlated to its energy density) of the occupations of the OPOs.  It is interesting that this occupation dependence doesn't seem to be apparent in the results of Fig.~\ref{fig:IPR_vs_OPO} of Section~\ref{sec:IPR} where the behavior as a function of occupation order $k$ is independent of energy density. This suggests that both metrics are probing different aspects of the OPOs: the IPR is sensitive to small broadening of the OPOs to which the overlap is primarily insensitive.  Note also that those slightly broader OPOs are closer to the center of the occupation spectrum, and have therefore a less well defined occupation than the rest of the OPOs, contributing to the breakdown of the one particle approximation of the integrals of motion.

\subsection{Occupations of the OPOs}
\label{sec:occupations}

\begin{figure}[t]
\includegraphics[width=1.0\columnwidth]{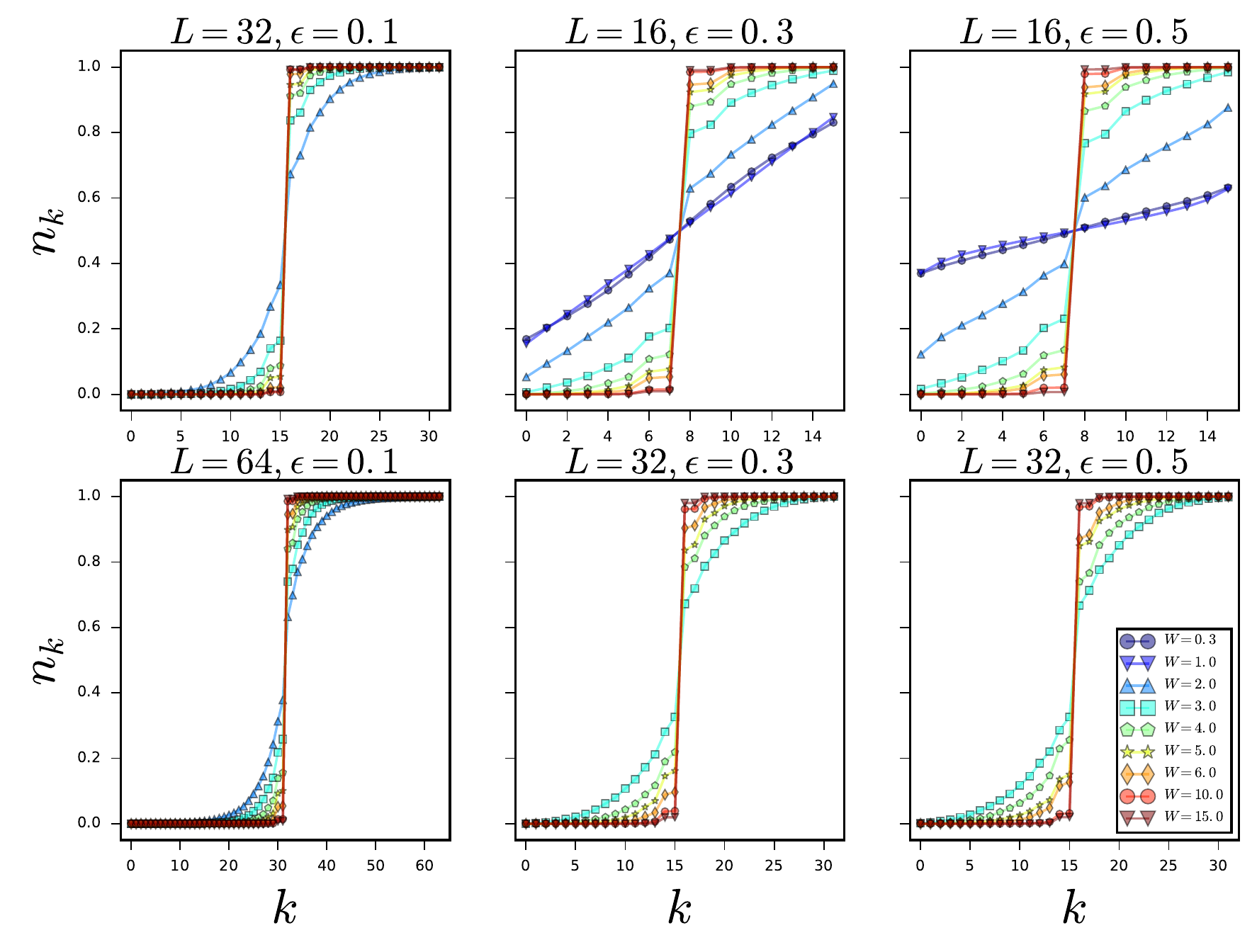}
\caption{\label{fig:occupations} Average occupations $n_k$ of the OPOs.
All eigenstates accessed by SIMPS are in the MBL phase, and so the occupation spectrum of the OPOs shows a finite gap~\cite{bera_many-body_2015,bera_one-particle_2017}.
}
\end{figure}

The gap in the occupations of the OPOs serves as a proxy for the characterization of the ergodic and the MBL phases, as shown in Refs.~\onlinecite{bera_many-body_2015}~and~\onlinecite{bera_one-particle_2017}.
An MBL system presents a large gap, which becomes smaller entering the ergodic phase and vanishes in the small disorder limit.
This is in agreement with our results for large systems (see Fig.~\ref{fig:occupations}) in the MBL phase.
Notice that for all values of $W$ the gap is smaller closer to the middle of the spectrum for fixed $L$, which agrees with the existence of a mobility edge.
In addition, for fixed $W$ and $\epsilon$, the gap decreases with system size, which is also in agreement with the usual numerical results, which point to the fact that the ergodic region of the phase diagram penetrates further into large disorder strengths for larger system sizes.

\subsection{Standard deviation of the entanglement entropy}
\label{sec:standard}
\begin{figure}[!tbp]
\includegraphics[width=1.0\columnwidth]{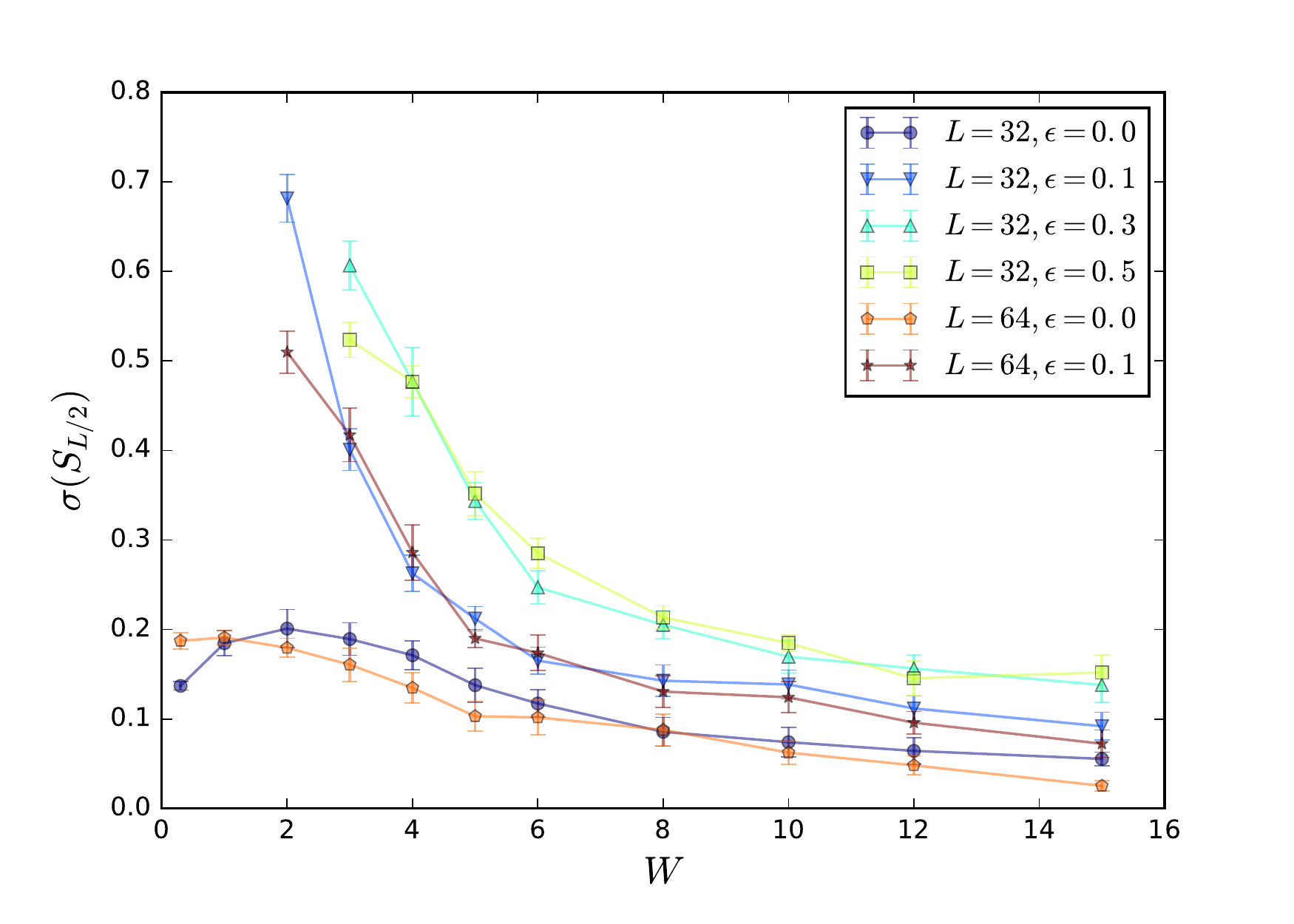}
\caption{\label{fig:std_EE} Standard deviation of the entanglement entropy of the half system
$\sigma\left(S_{L/2}\right)$ as a function of disorder strength $W$ for different system sizes and
energy densities.  $\sigma\left(S_{L/2}\right)$ exhibits a maximum at each energy density close to the transition
point~\cite{kjall_many-body_2014}. Our results for the finite energy eigenstates considered in this
article accessed by SIMPS only show the approach to this maximum. 
}
\end{figure}

At the MBL transition, the nature of many-body eigenstates changes radically, which is clearly
signaled in the different scaling behavior of the entanglement entropy: while in the MBL phase
almost all eigenstates have an area law entanglement entropy (EE), in the thermal phase, the EE is extensive.
 It has been demonstrated that the change of this behavior leads to a coexistence of
area law and volume law states at the
transition\cite{kjall_many-body_2014,luitz_extended_2016,yu_bimodal_2016,khemani_critical_2017},
which is signaled by a bimodal distribution of the entanglement entropy $S$ and, consequently by a peak
of the standard deviation $\sigma(S)$ at the critical point. Using SIMPS, we only have access to 
eigenstates at the MBL side of the transition and therefore can only observe the approach to the
peak in $\sigma(S_{L/2})$ (at half-cut) in Fig.~\ref{fig:std_EE}. In addition, the EE, even for states of small $\sigma(E)$, is likely to be much more sensitive than other observables to the finite bond dimension used in the SIMPS calculations; this probably explains the inverted system size and energy density dependence of $\sigma(S_{L/2})$ at low $W$. Our results are consistent with the existence of a
mobility edge, since it is apparent that the peak is located at different disorder strengths for
different energy densities.

\section{Conclusion}
\label{sec:conclusion}

In this work we study the properties of the eigenstates, particularly of their OPDM, of the model in Eq.~\eqref{H_fermion} deep into the mobility edge using the SIMPS algorithm~\cite{yu_finding_2017}.  

These SIMPS eigenstates give us various ways to probe the MBL transition. Interestingly enough, even the phase points at which
SIMPS (restricted to small bond-dimensions) succeeds or fails (see Fig.~\ref{fig:survive}) gives evidence for the location of the transition.  We can even identify the mobility edge by noting the $\epsilon$ dependence of the rate of failure of SIMPS and of the apparent divergence of the bond dimension of the eigenstates, as well as seeing that SIMPS successfully computes eigenstates at $W<W_c$ for small and large $\epsilon$ (see Fig.~\ref{fig:phase_diagram}).  The location of the transition can further be bounded using the approach to the peak in $\sigma(S_L/2)$ (see Fig.~\ref{fig:std_EE}) and the closing gap of the eigenvalues of the OPDM (see Fig.~\ref{fig:occupations}); both of these measures suggest that at low $\epsilon$ the transition happens at $W<W_c$.  Using the decay of either the OPOs or the number operators $a_k a_k^\dagger$ generated from them, we can define a correlation length.  As we approach the transition the correlation length gets larger but does not obviously diverge.  For small systems, we could probe this correlation length even within the ergodic phase; we find that deep in the MBL or ergodic phase the correlation length has a little $\epsilon$ dependence, while there is significant dependence on $\epsilon$ in the mobility edge.

Beyond probing physics near the transition, we can also use the OPDM to further elucidate properties about the MBL phase itself.  Within the MBL phase, we see a clear but small increase in the correlation length with system size (see inset of Fig.~\ref{fig:FR_vs_R}). Moreover, we consider the probability distribution of the magnitude of the coefficients $f^k_{ij}$ (from Eq.~\eqref{number_operator})  and find that deep within the MBL phase and at large range $R$ (defined in Eq.~\eqref{range}) it approaches a ``$1/f$'' distribution (see Fig.~\ref{fig:pfR_vs_R}). This is the same distribution seen in Ref.~\onlinecite{pekker_fixed_2017} for the l-bits.

Interestingly, we are also able to identify properties of the entire spectrum using MBL eigenstates at single points in the spectrum.  This is possible because, surprisingly, a single MBL eigenstate provides a `universal' set of OPOs (i.e. they have significant overlap with the OPOs generated from eigenstates at different energy densities (see  Fig.~\ref{fig:hist_overlaps})). While the OPOs at different energy densities have high overlap, the OPDMs are very different. This difference comes from a change in the occupations of the OPOs among the eigenstates.  There is correlation between the energy of the eigenstates and which OPOs have high occupation; for example, the set of high and low occupied OPOs at $\epsilon=0.1$ and $\epsilon=0.9$ are almost completely flipped (see Figs.~\ref{fig:matrix_overlaps}~and~\ref{fig:average_arrows}).

We show that the $\sigma(IPR)$ has a peak, for multiple $\epsilon$, at $W\approx 4$ (see Fig.~\ref{fig:mean_std_IPR}), suggesting that even MBL eigenstates deep under the mobility edge are aware of the presence or absence of an ergodic phase at a higher value of $\epsilon$.

The use of SIMPS allows us to access MBL eigenstates of systems of size beyond those accessible by other techniques, even deep into the mobility edge. By looking at the OPDM we are able to study the one particle approximation to the integrals of motion.  Despite its approximate nature, and the limitations of working with an MPS approach (with difficulty in probing the ergodic region of the phase diagram), our study leads to phenomenological conclusions that are not accessible from exact diagonalization techniques or an exact treatment of the integrals of motion. We think that the study of the MBL transition, as well as other problems, can benefit greatly from this promising approach.

\begin{acknowledgments}
We thank Fabian Heidrich-Meisner for useful discussions. DJL also would like to thank Jens Bardarson
and David Pekker for interesting discussions on one particle orbitals and l-bits. BKC would like to thank David Pekker for valuable discussions and collaboration on related projects involving correlation lengths of MBL phases.
This project has received support under SciDAC grant DE-FG02-12ER46875 and funding from the European Union's Horizon 2020 research and innovation
programme under the Marie Sk\l{}odowska-Curie grant agreement No 747914 (QMBDyn).  DJL acknowledges
PRACE for awarding access to HLRS's Hazel Hen computer based in Stuttgart, Germany under grant
number 2016153659. 
Our SIMPS code used in this work is built on top of the ITensor library~\cite{stoudenmire_itensor_nodate}.
This research is part of the Blue Waters
sustained-petascale computing project, which is supported by the National
Science Foundation (awards OCI-0725070 and ACI-1238993) and the state
of Illinois. Blue Waters is a joint effort of the University of Illinois
at Urbana-Champaign and its National Center for Supercomputing Applications.
\end{acknowledgments}

\appendix
\section{Supplementary data on the correlation length of the OPOs}
\label{sec:sup_correlation}
\begin{figure}[t]
\includegraphics[width=1.00\columnwidth]{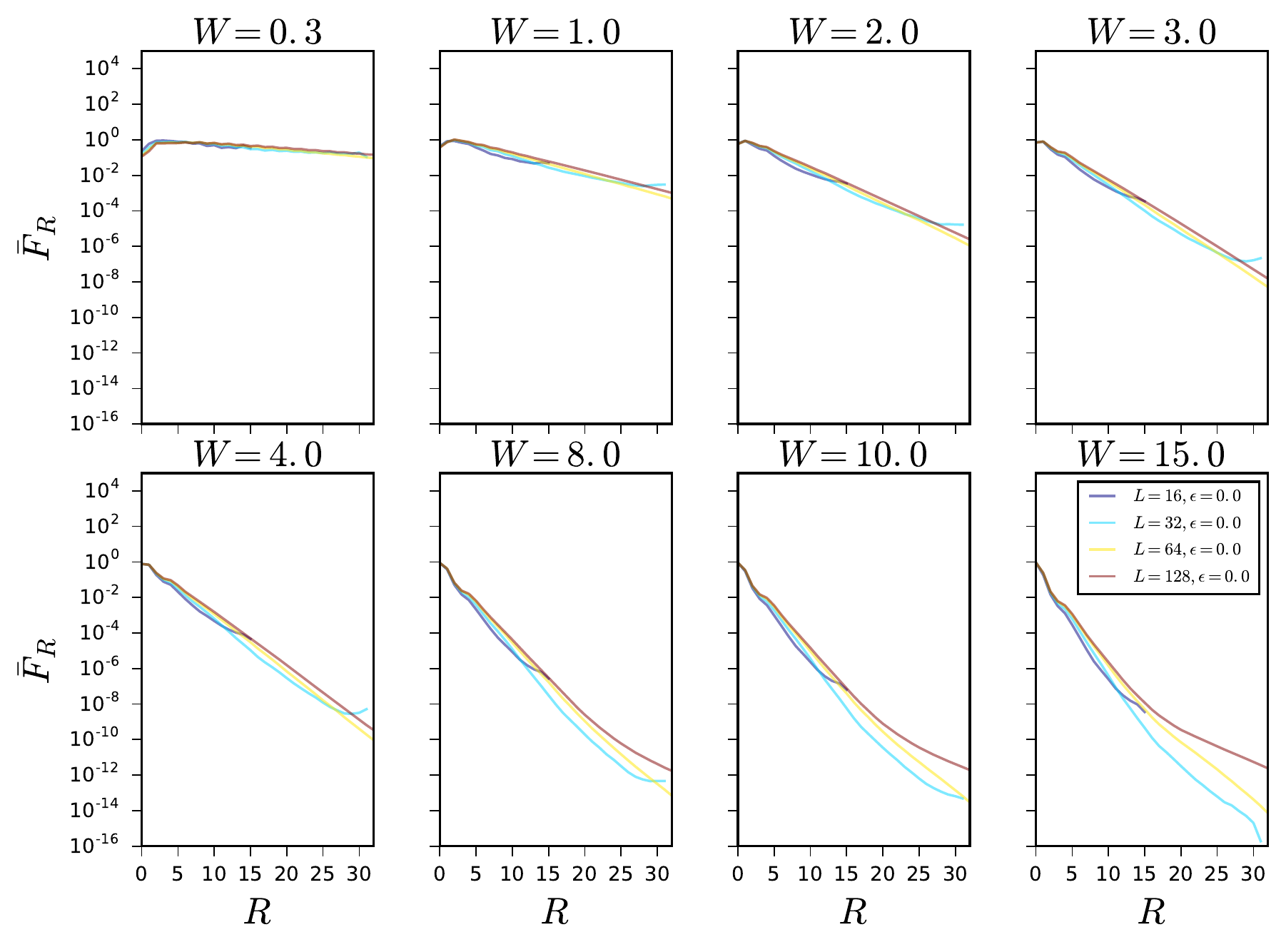}
\includegraphics[width=1.00\columnwidth]{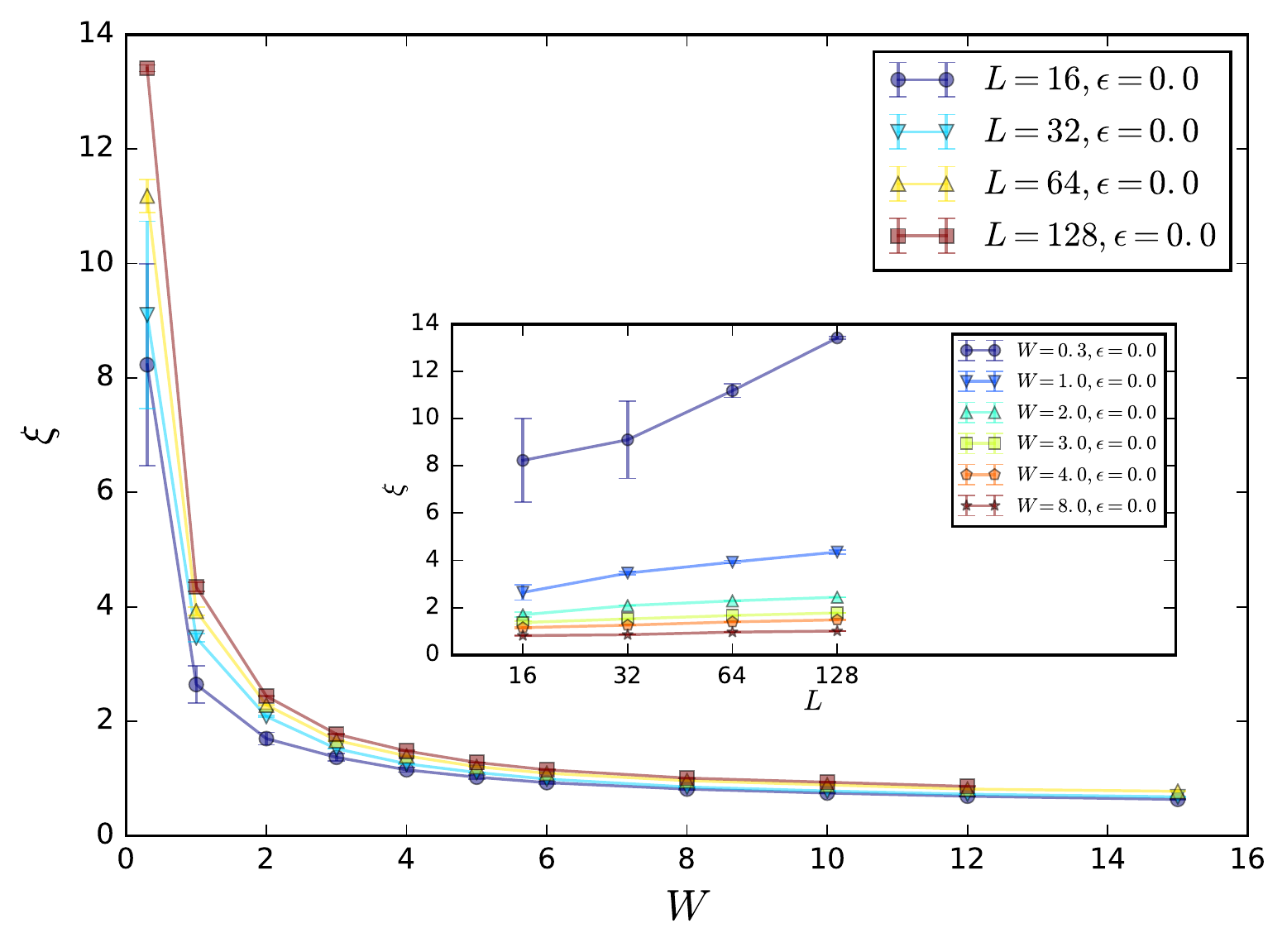}
\caption{\label{fig:FR_vs_R_GS} \emph{Top:} exponential decay of $\bar{F}_R$ for ground states. \emph{Bottom:} correlation length $\xi$ for ground states. $\xi$ is a monotonically increasing function of the system size $L$ at small disorder.} 
\end{figure}

The SIMPS algorithm does not allow us to access the weak disorder limit at finite energy density, due to the transition to an ergodic phase.
However, it is possible to access this limit at $\epsilon=0.0$ using DMRG, and it is interesting to see the system size dependence of the the decay of $\bar{F}_R$ and its associated $\xi$ for ground states.
We can see in Fig.~\ref{fig:FR_vs_R_GS} that the decay is seemingly exponential well into the weak disorder limit, where $\xi$ becomes large and increases strongly with $L$.

\begin{figure}[t]
\includegraphics[width=1.00\columnwidth]{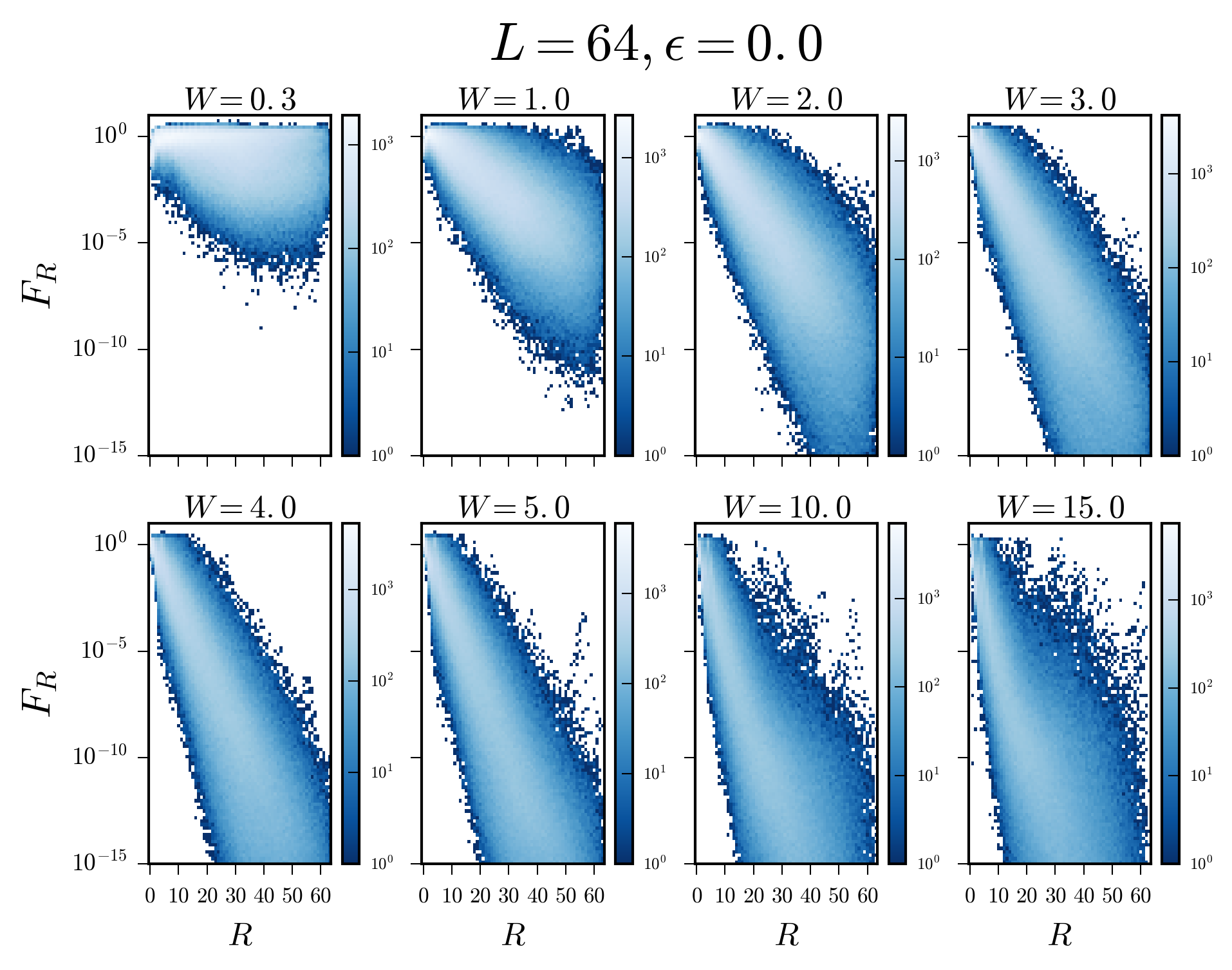}
\includegraphics[width=1.00\columnwidth]{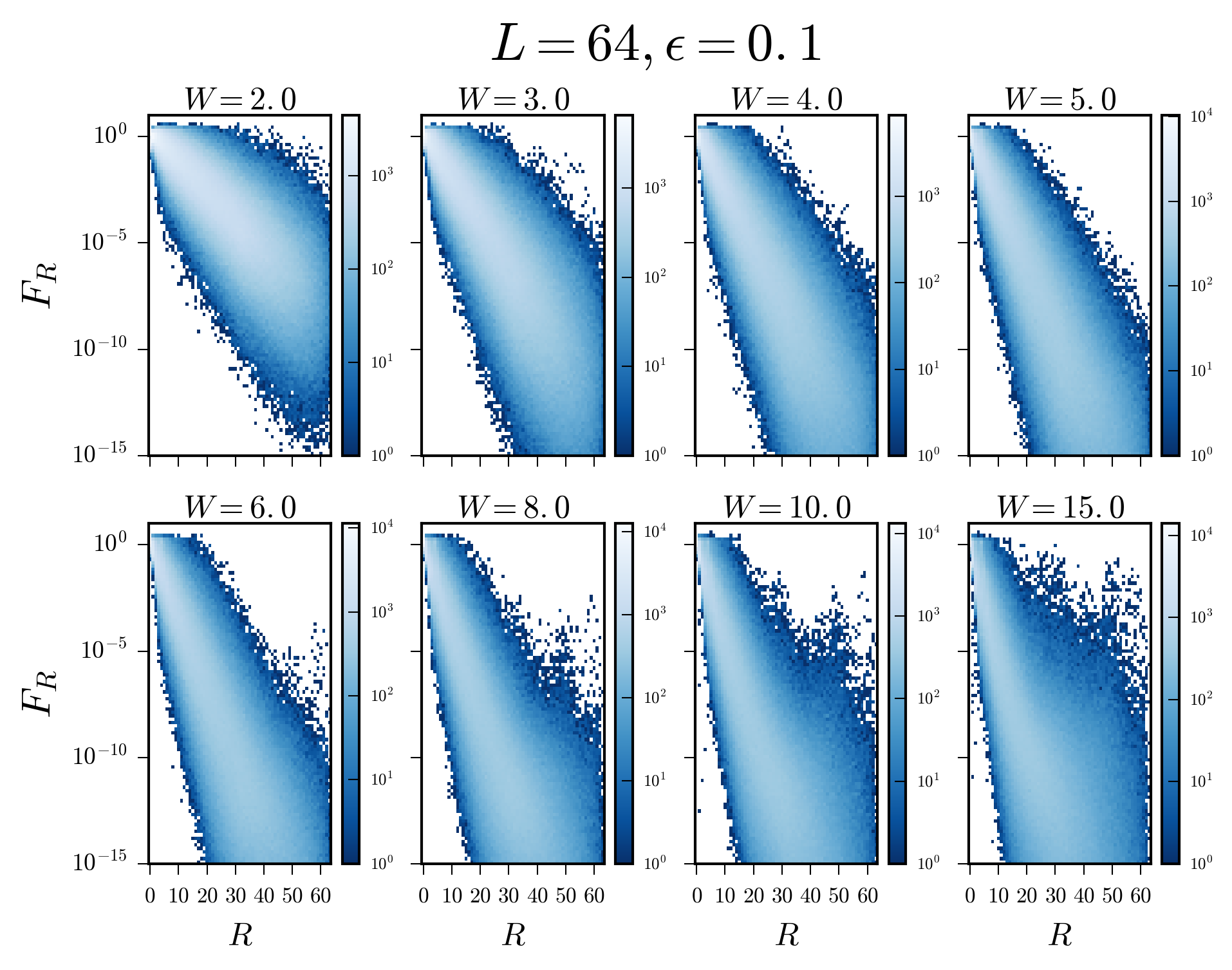}
\caption{\label{fig:hist_FR_vs_R} Histogram of $F_R$ vs. $R$ for systems of size $L=64$ at different values of $W$ and eigenstates at $\epsilon=0.0, 0.1$.
}
\end{figure}

\begin{figure}[t]
\includegraphics[width=0.7\columnwidth]{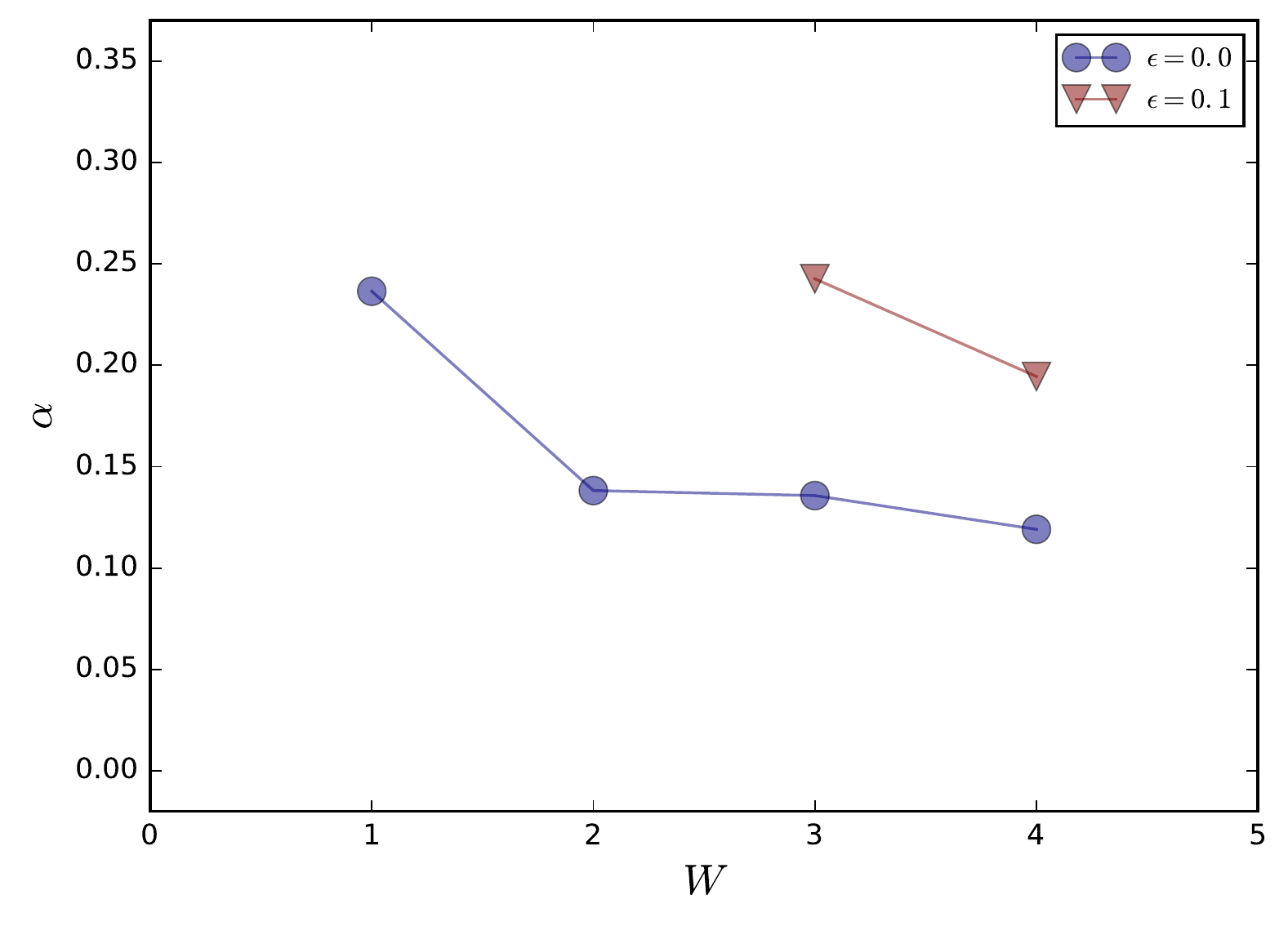}
\caption{\label{fig:alphas} Exponent $\alpha$ of the scaling law $\xi=\log\left(\beta L^\alpha\right)$ for the correlation length of the OPOs.
}
\end{figure}

\begin{figure}[t]
\includegraphics[width=1.00\columnwidth]{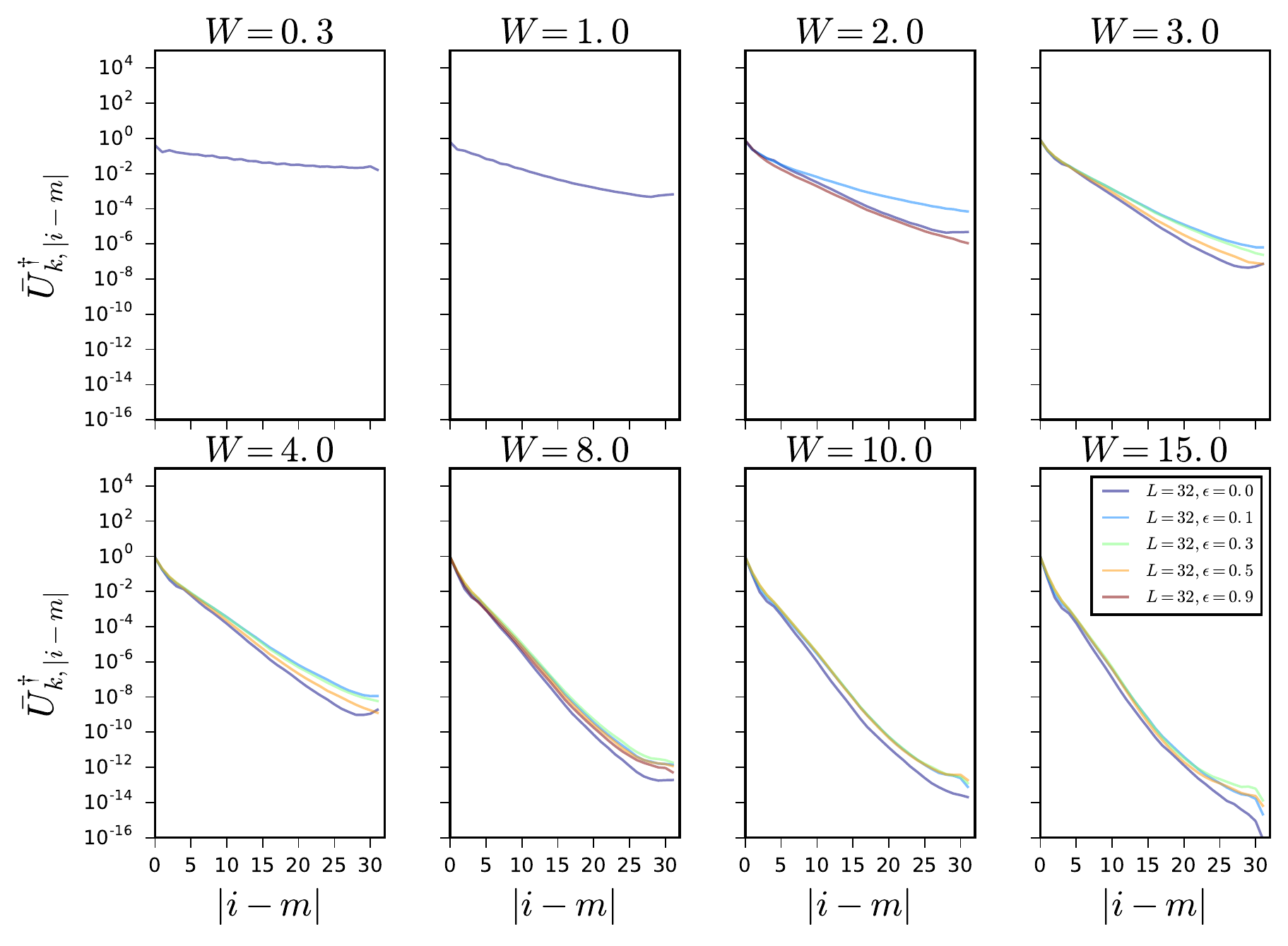}
\includegraphics[width=1.00\columnwidth]{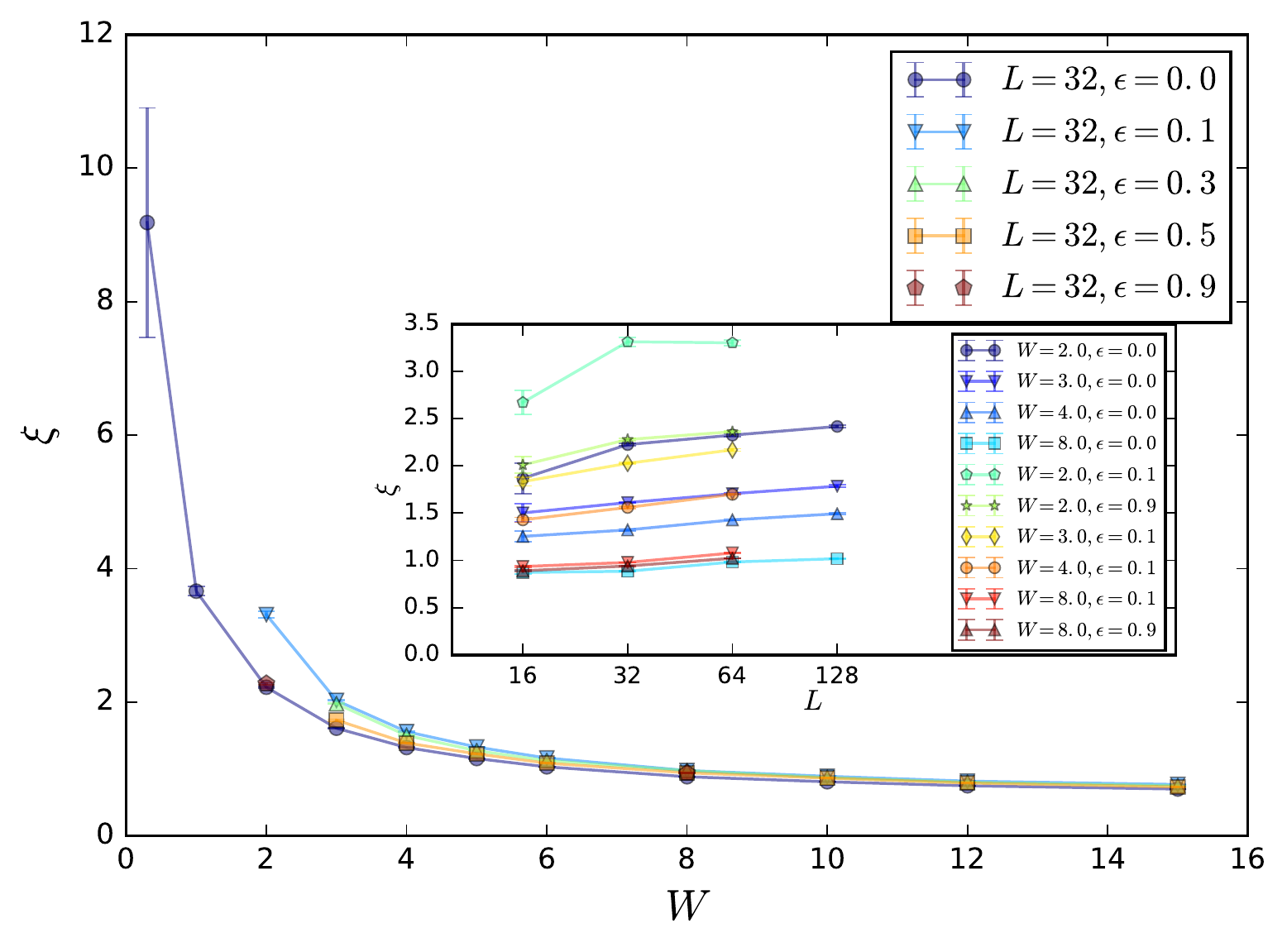}
\caption{\label{fig:tail} Average decay of the OPOs' tails. The asymptotic behavior of the tails is
equal to the one of $\bar{F}_R$ presented in Fig.~\ref{fig:FR_vs_R}.} 
\end{figure}

\begin{figure}[t]
\includegraphics[width=1.00\columnwidth]{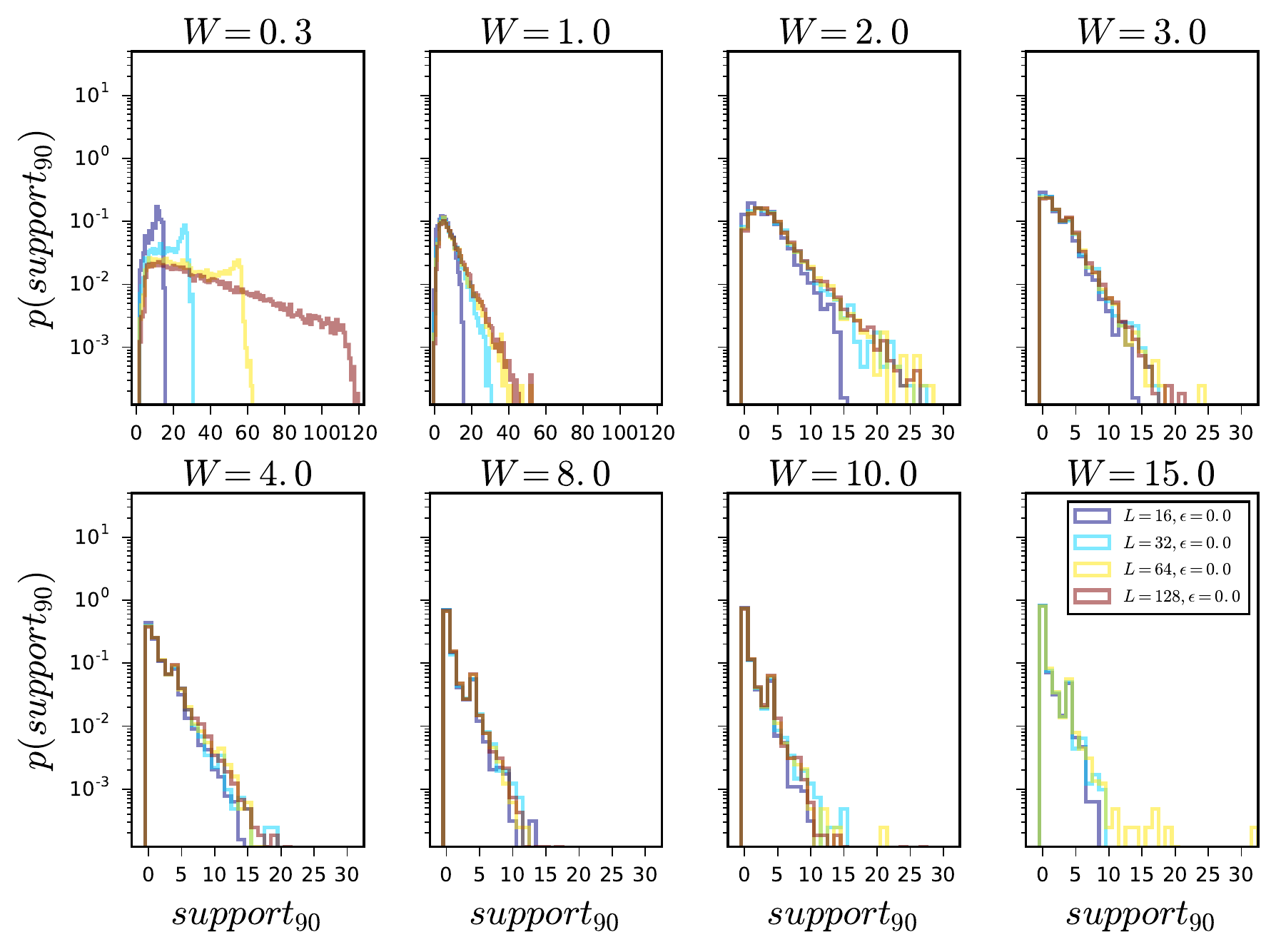}
\includegraphics[width=1.00\columnwidth]{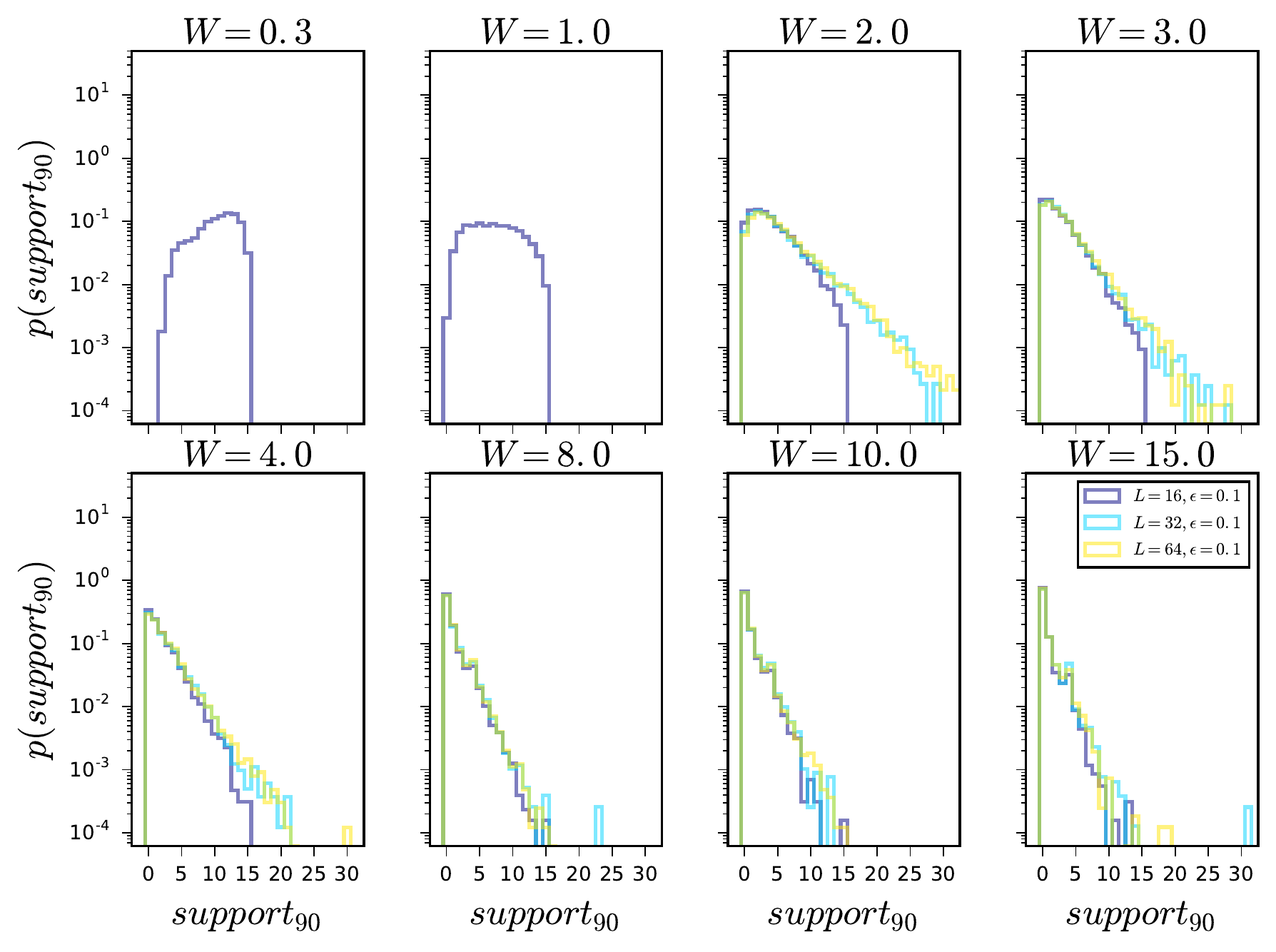}
\caption{\label{fig:hist_support_90_fixed_E} Distribution of the $support_{90}$ of the OPOs of eigenstates at $\epsilon=0.0, 0.1$ of systems of different sizes $L$.
At strong disorder the distribution decays exponentially and is largely system size independent, while it collapses to the system size at weak disorder, where the exponential decay is lost.} 
\end{figure}

\begin{figure}[t]
\includegraphics[width=1.00\columnwidth]{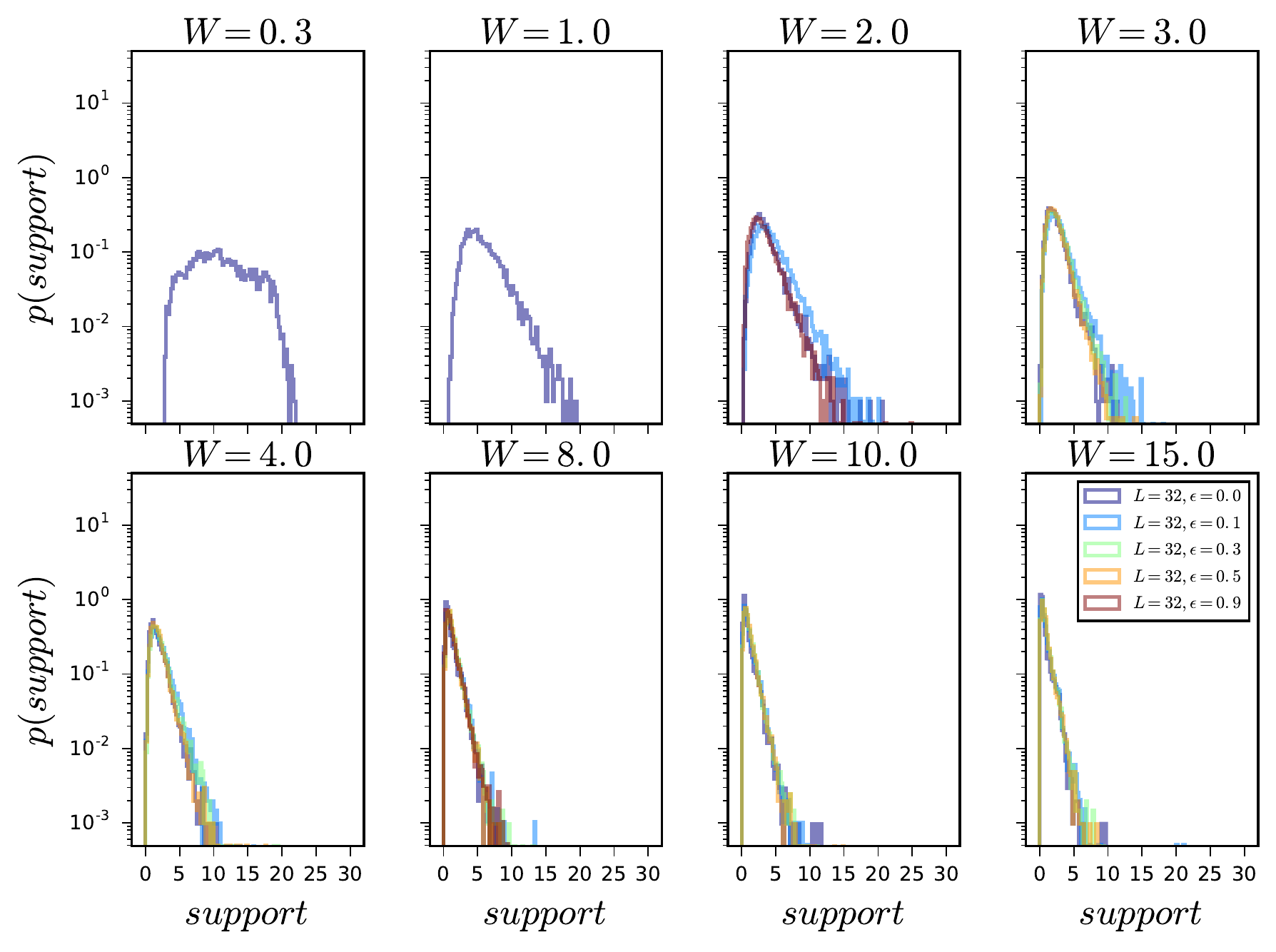}
\includegraphics[width=1.00\columnwidth]{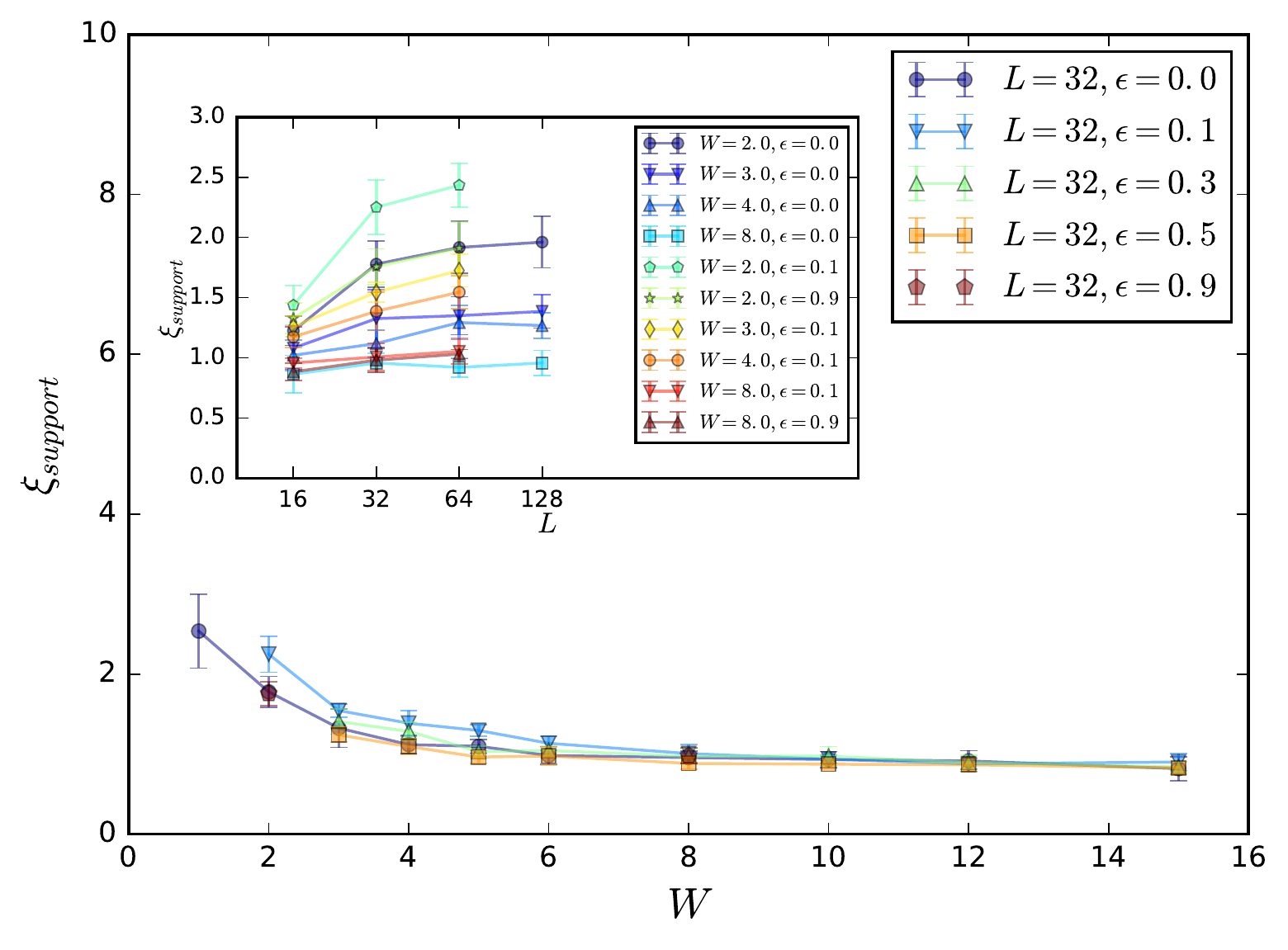}
\caption{\label{fig:hist_support} Equivalent to Fig.~\ref{fig:hist_support_90}.
The support ($support$) is now computed for the number operator of the OPO $a_k^\dagger a_k$ as the average range $R$ weighted by $F_R$.
The phenomenology is extremely similar to the one found for the $support_{90}$ in Fig.~\ref{fig:hist_support_90} in Section.~\ref{sec:support}.}
\end{figure}

\begin{figure}[t]
\includegraphics[width=1.00\columnwidth]{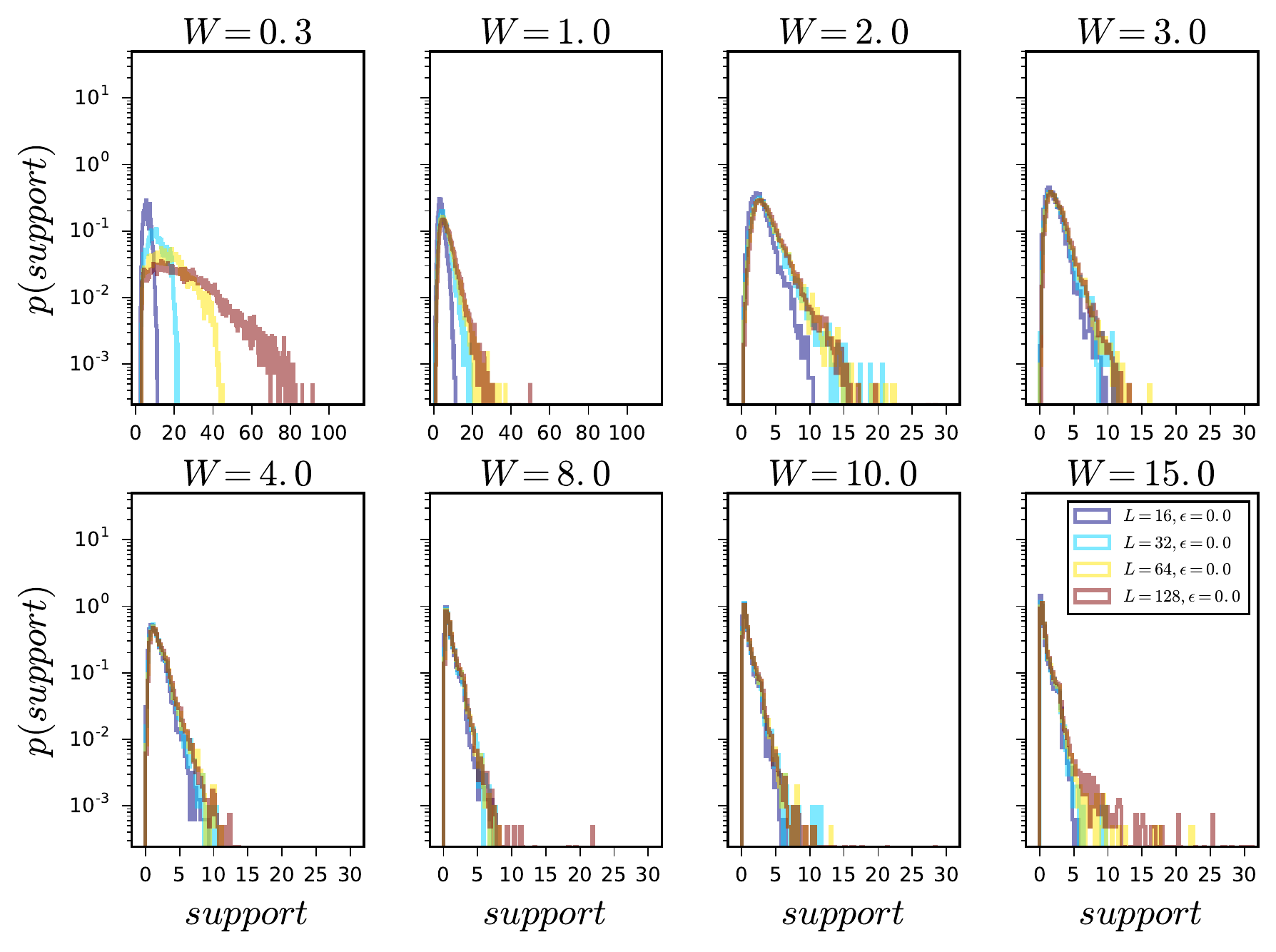}
\includegraphics[width=1.00\columnwidth]{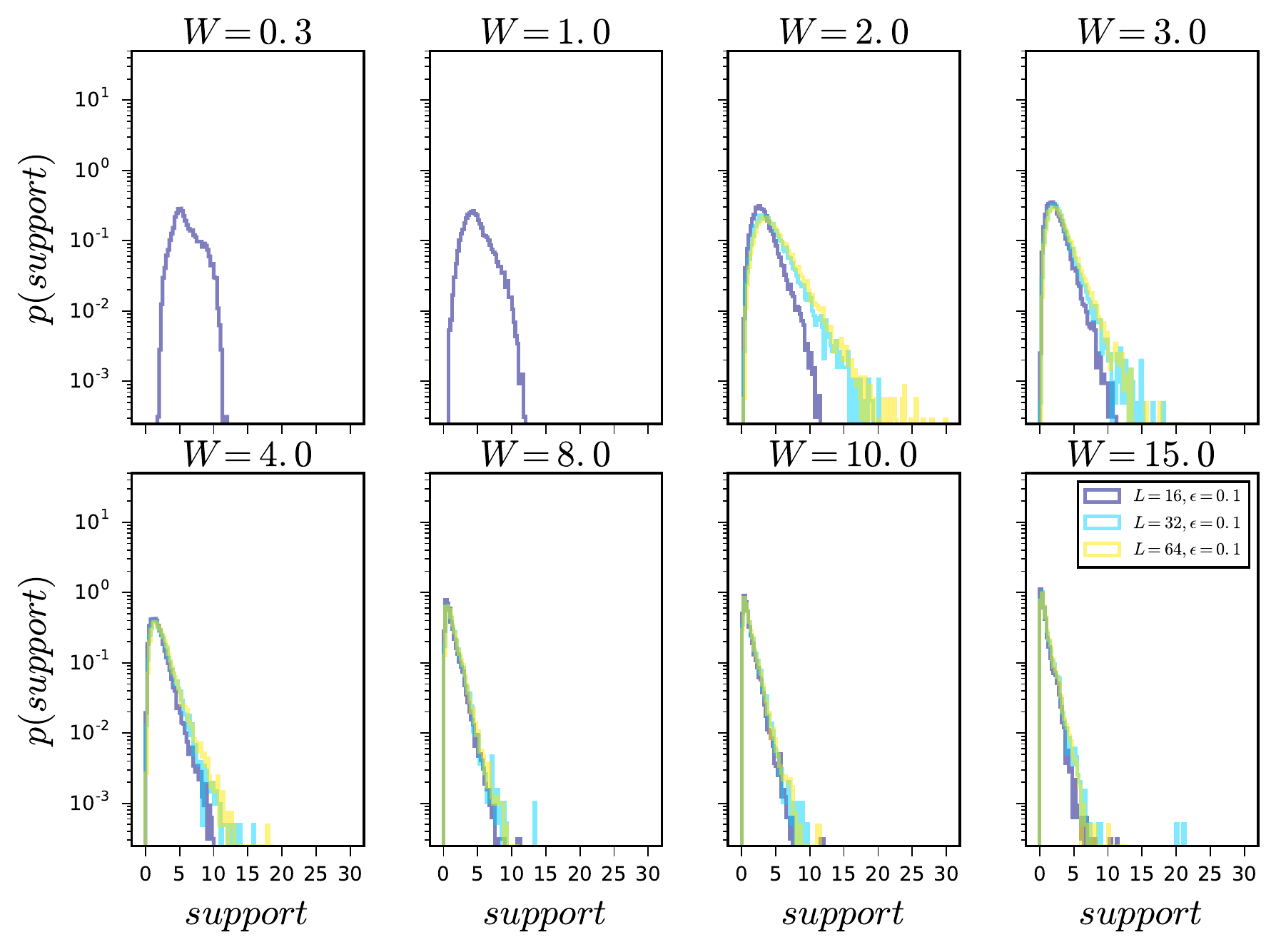}
\caption{\label{fig:hist_support_fixed_E} Distribution of the $support$ of the OPOs of systems of different size $L$ at $\epsilon=0.0, 0.1$.
The phenomenology is extremely similar to the one found in Fig.~\ref{fig:hist_support_90_fixed_E}.
}
\end{figure}

\begin{figure}[t]
\includegraphics[width=1.00\columnwidth]{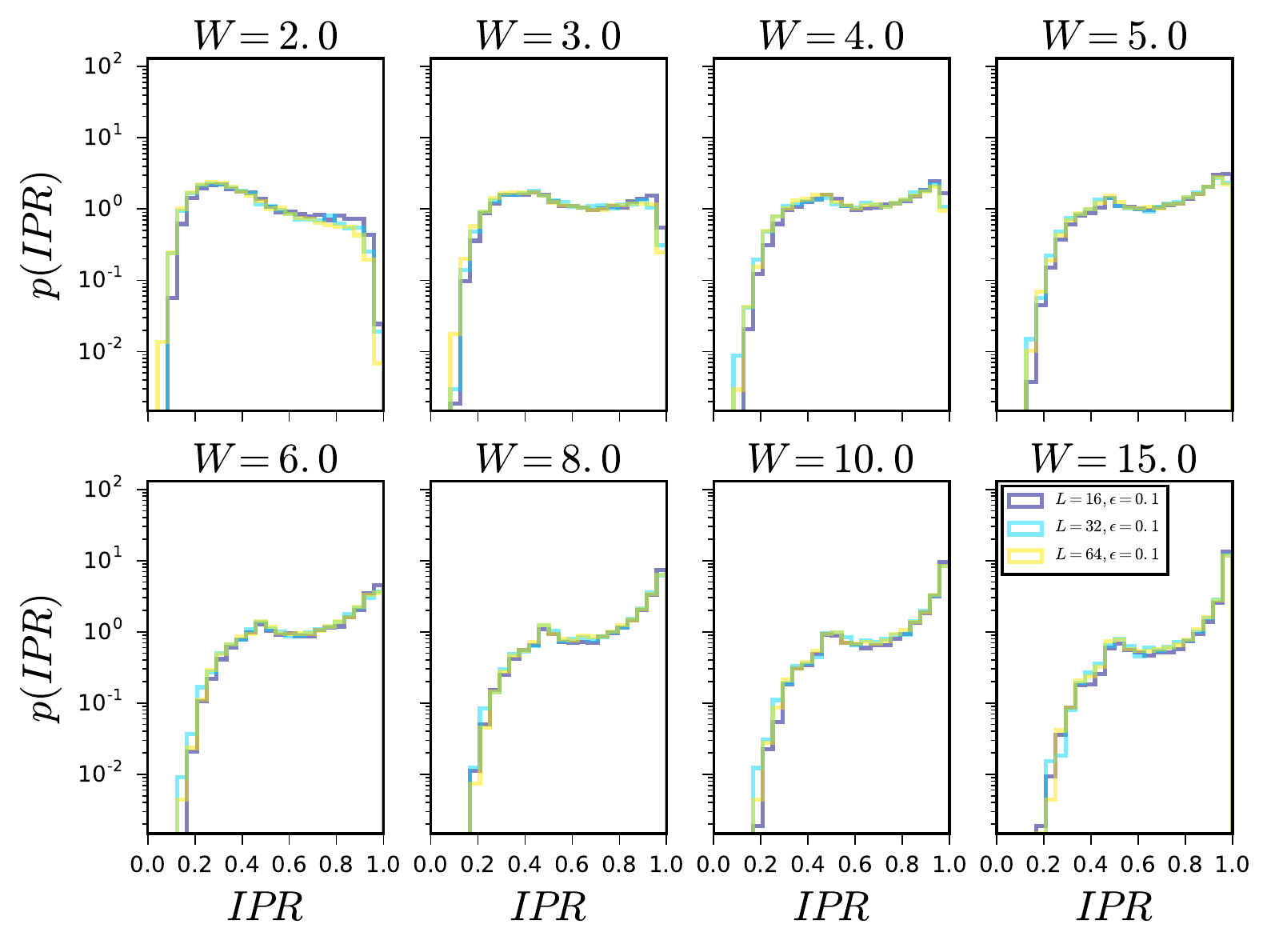}
\caption{\label{fig:bimodality_fixed_E} Histogram of the distribution of IPR for different system sizes at $\epsilon=0.1$. 
The distribution is system size independent for almost all values of $W$, with only a slight drift towards high IPR for small systems at $W=2$.}
\end{figure}

\begin{figure}[t]
\includegraphics[width=1.00\columnwidth]{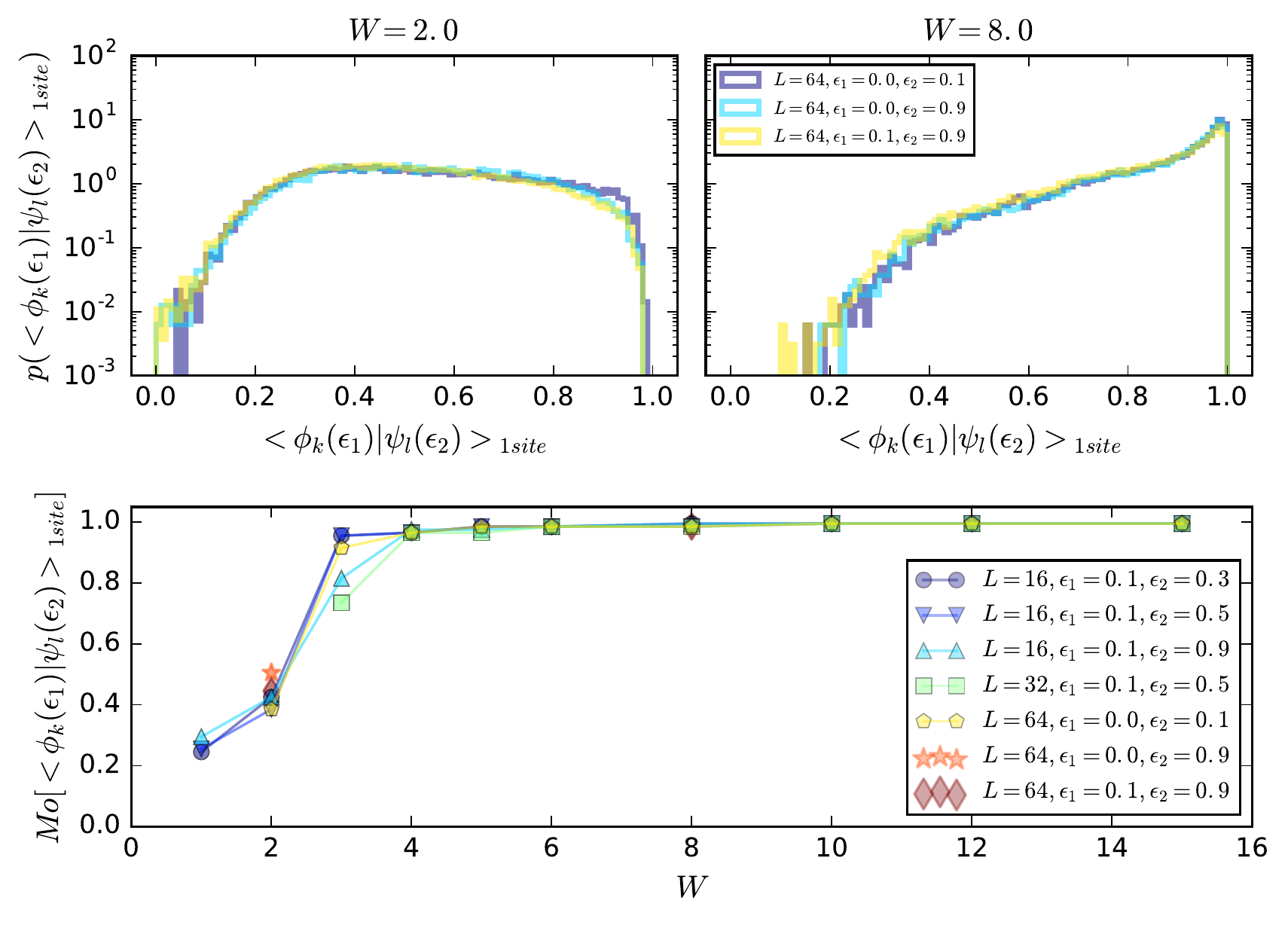}
\caption{\label{fig:hist_one_site_overlaps} \emph{Top:} distribution of the biggest one site contribution to the overlaps between corresponding OPOs ($k\leftrightarrow l(k)$) obtained from pairs of eigenstates of the same Hamiltonian at two particular energy densities $\epsilon_1$ and $\epsilon_2$.
\emph{Bottom:} mode of the distribution of the one site contributions to the overlaps for different pairs of energy densities as a function of $W$. The one site overlap is lower than the total overlap between the OPOs.}
\end{figure}

\begin{figure}[t]
\includegraphics[width=0.5\columnwidth]{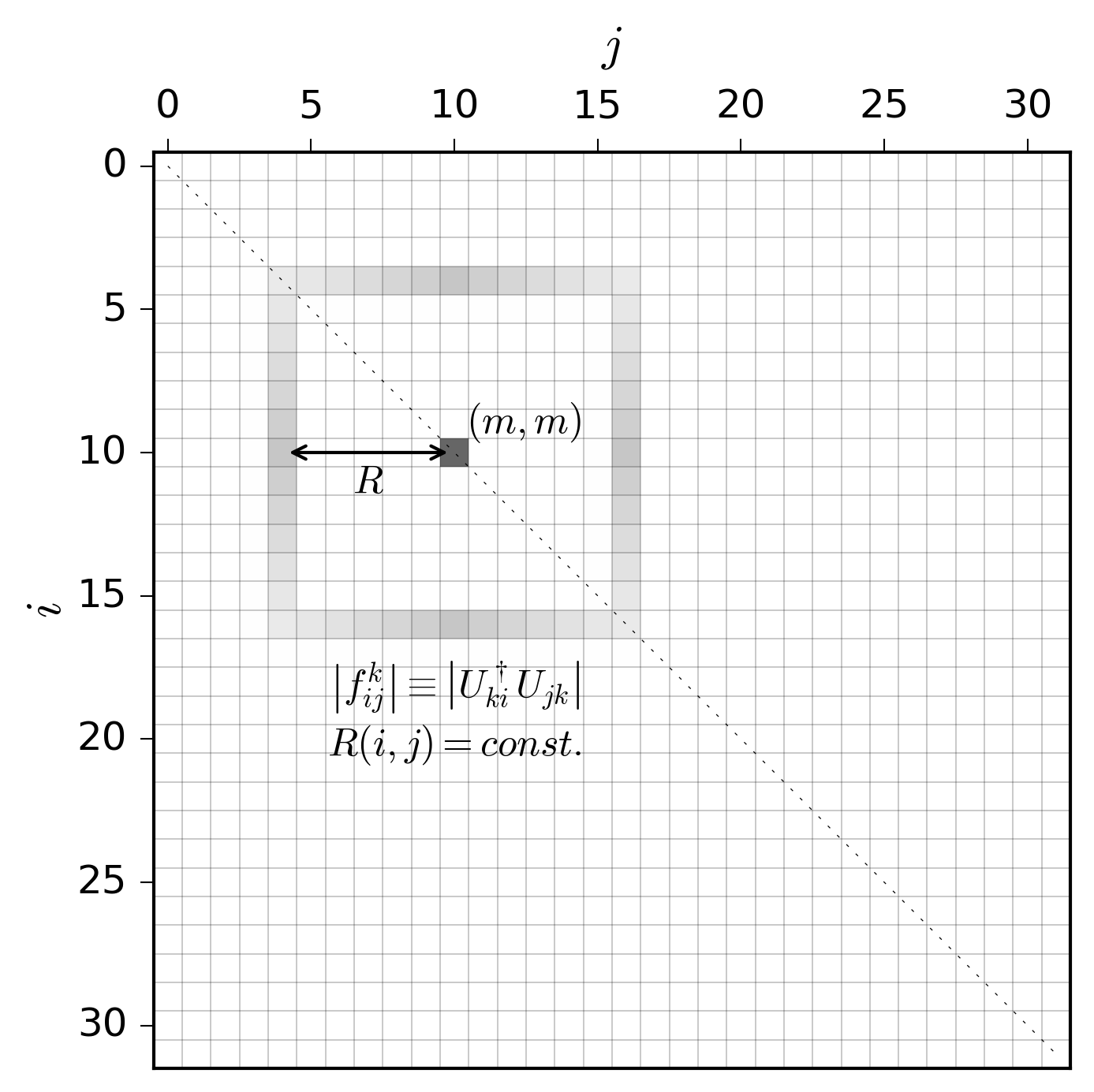}
\includegraphics[width=0.5\columnwidth]{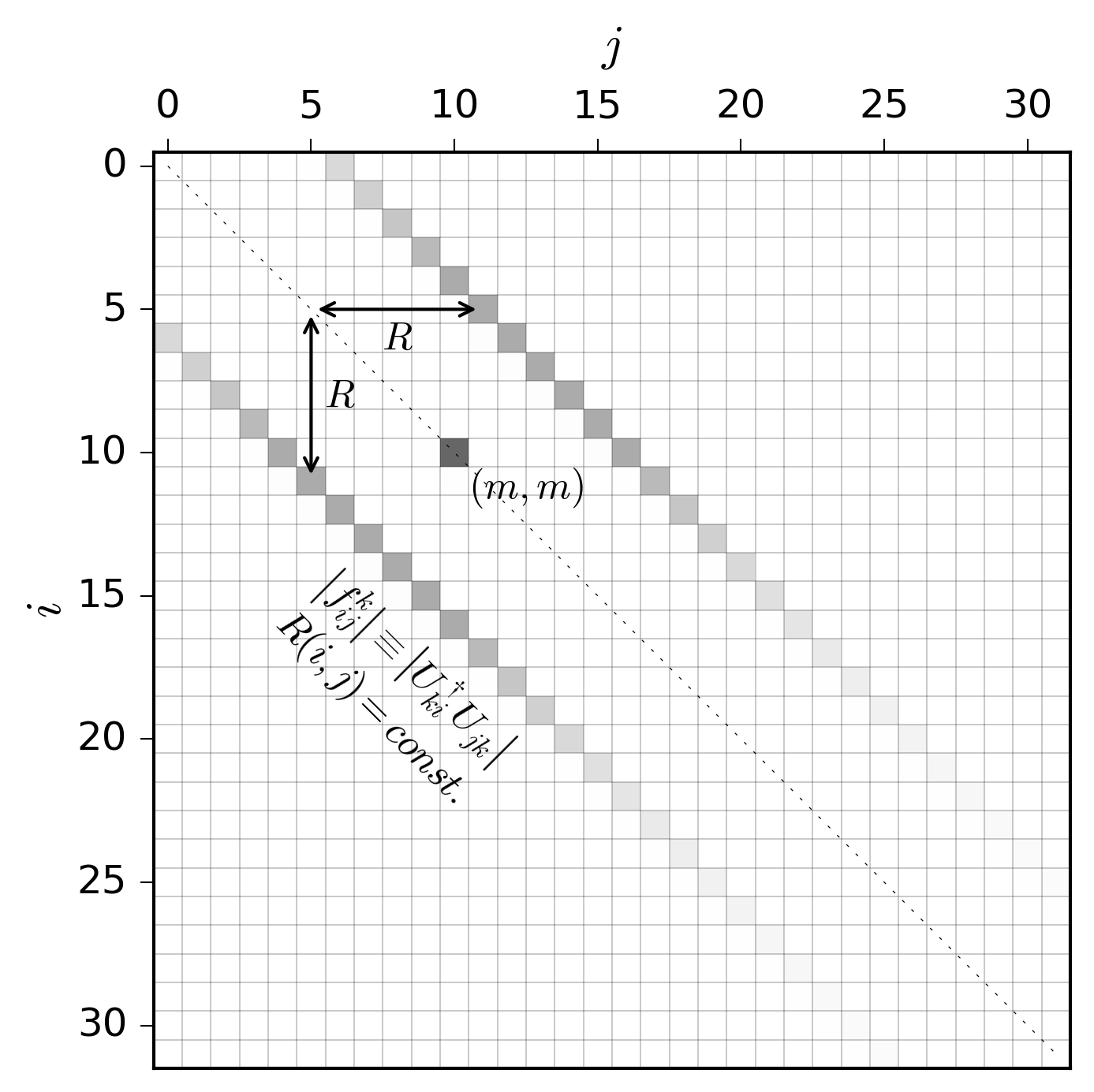}
\caption{\label{fig:scaling} Matrix $\left|f^k_{ij}\right|=\left|U^\dagger_{ki}U_{jk}\right|$ for a generic OPO with exponentially decaying $\left|U^\dagger_{ki}\right|\propto e^{-|i-m|/\xi}$ centered at site $m$.
Only the elements for which the range $R(i,j)=const.$ are represented, where $R(i,j)\equiv\max\left(\left|i-m\right|,\left|j-m\right|\right)$ (centered) applies to the \emph{top} panel, and $R(i,j)\equiv\left|i-j\right|$ (uncentered) applies to the \emph{bottom} panel.
In both example, $m=10$ and $R=6$ for a system of $L=32$.
}
\end{figure}

\begin{figure}[t]
\includegraphics[width=1.00\columnwidth]{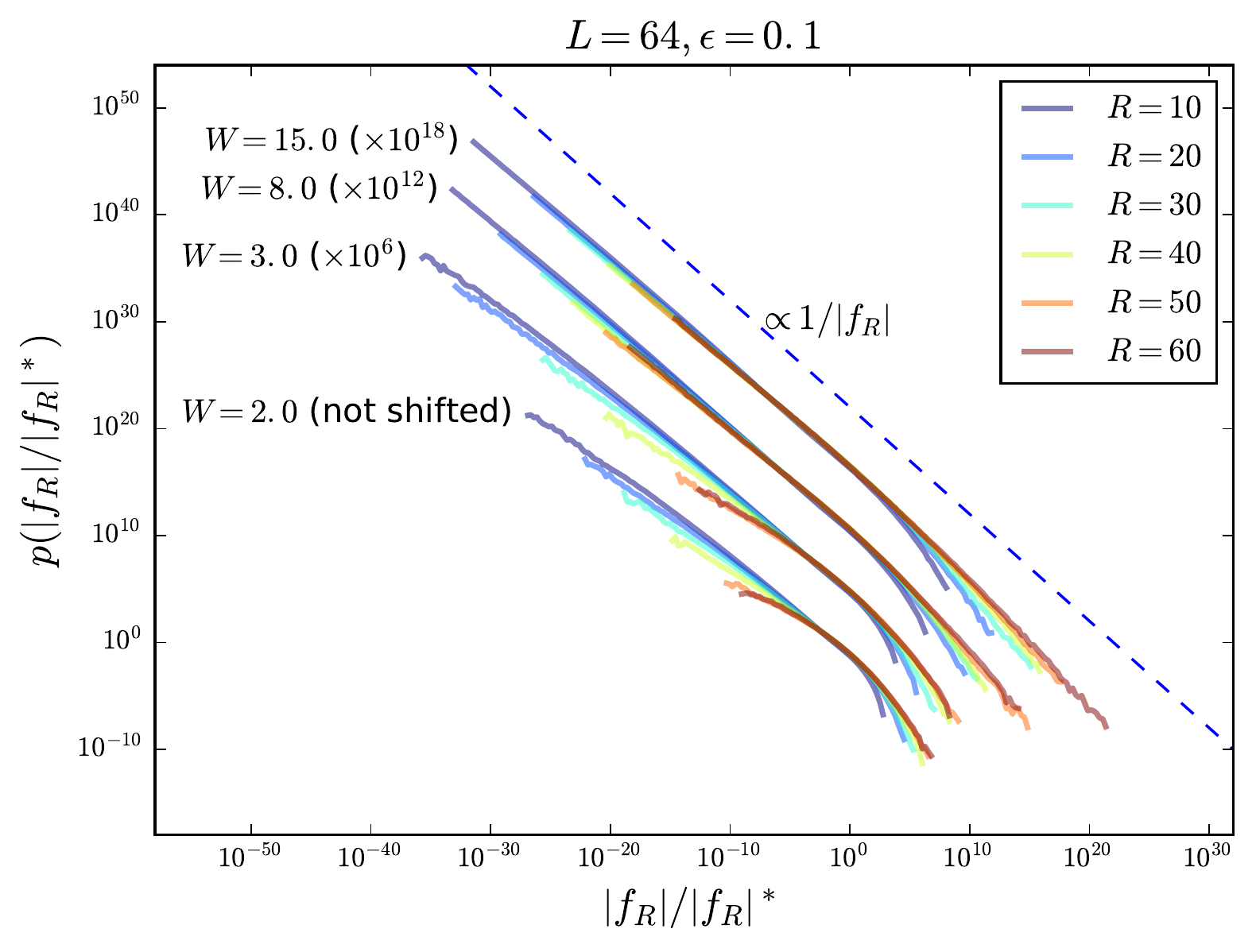}
\caption{\label{fig:pfR_vs_R_no_center} Equivalent to Fig.~\ref{fig:pfR_vs_R}, but computed using the uncentered range. Contrary to Fig.~\ref{fig:pfR_vs_R}, the distributions of $|f_R|$ are broader and flatter at small ranges, as discussed in Appendix~\ref{sec:range}.
}
\end{figure}

\begin{figure}[t]
\includegraphics[width=1.00\columnwidth]{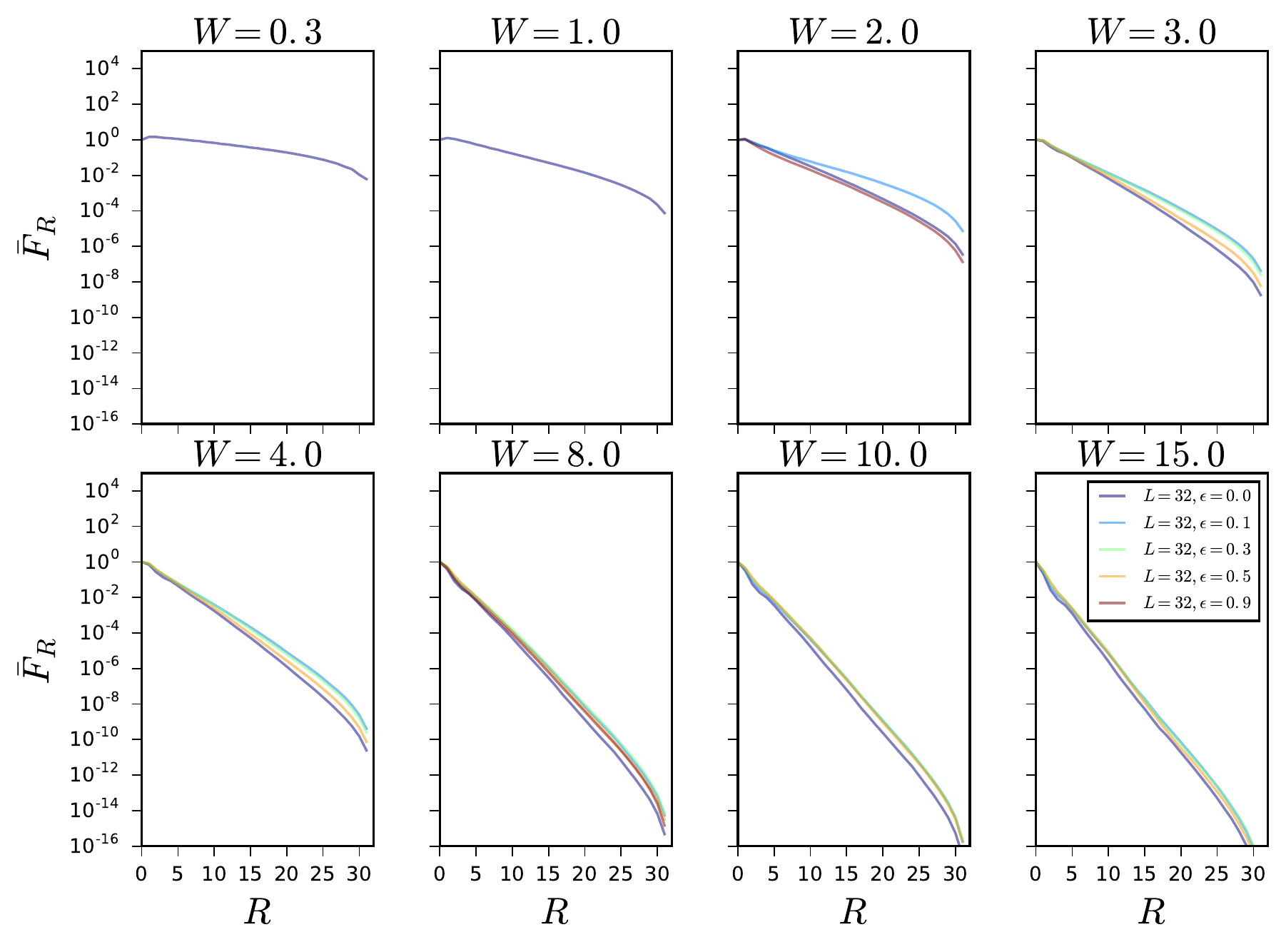}
\includegraphics[width=1.00\columnwidth]{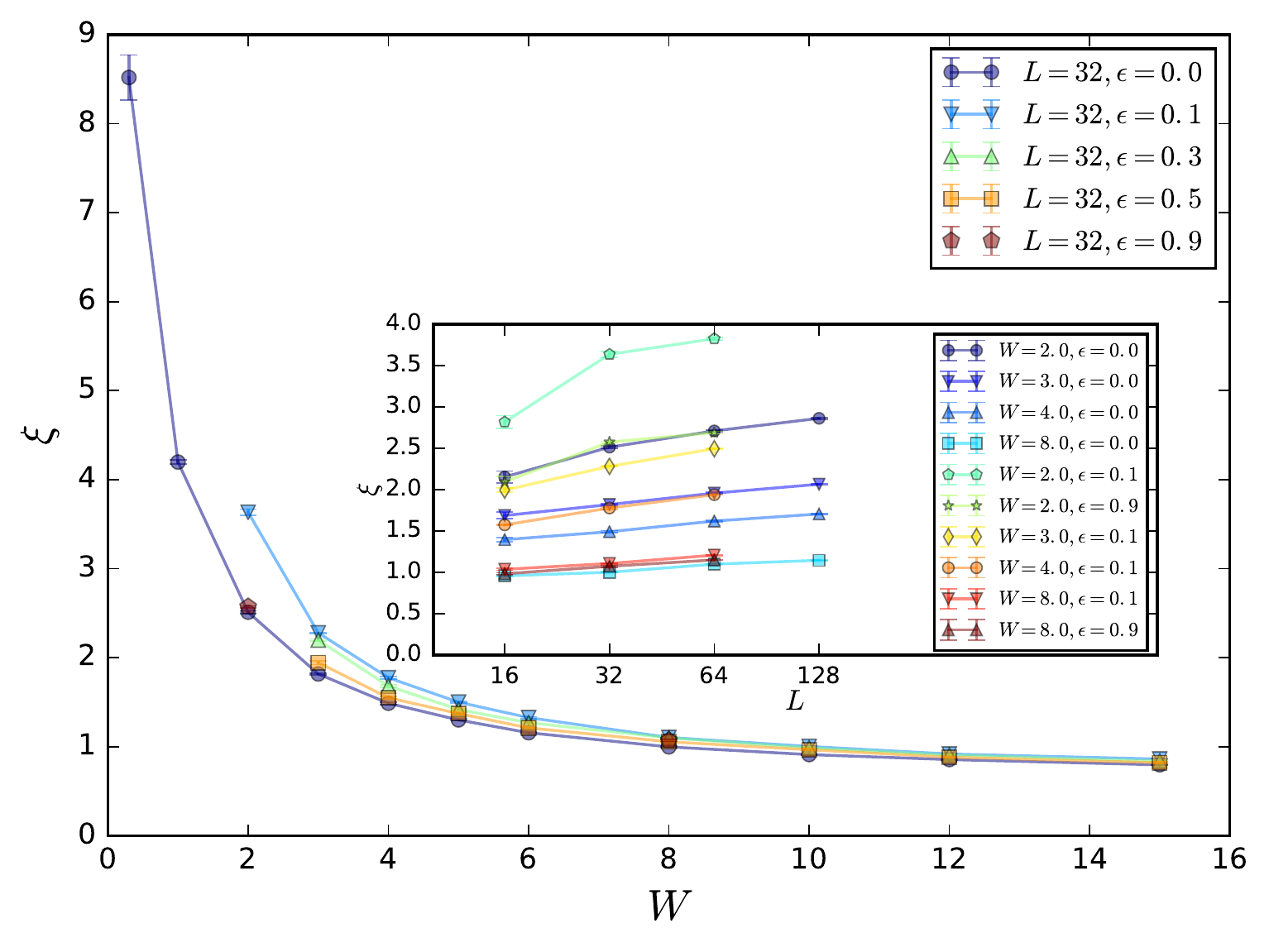}
\caption{\label{fig:FR_vs_R_no_center} Equivalent to Fig.~\ref{fig:FR_vs_R},
although the uncentered definition of the range $R$ is used.}
\end{figure}

\begin{figure}[t]
\includegraphics[width=1.00\columnwidth]{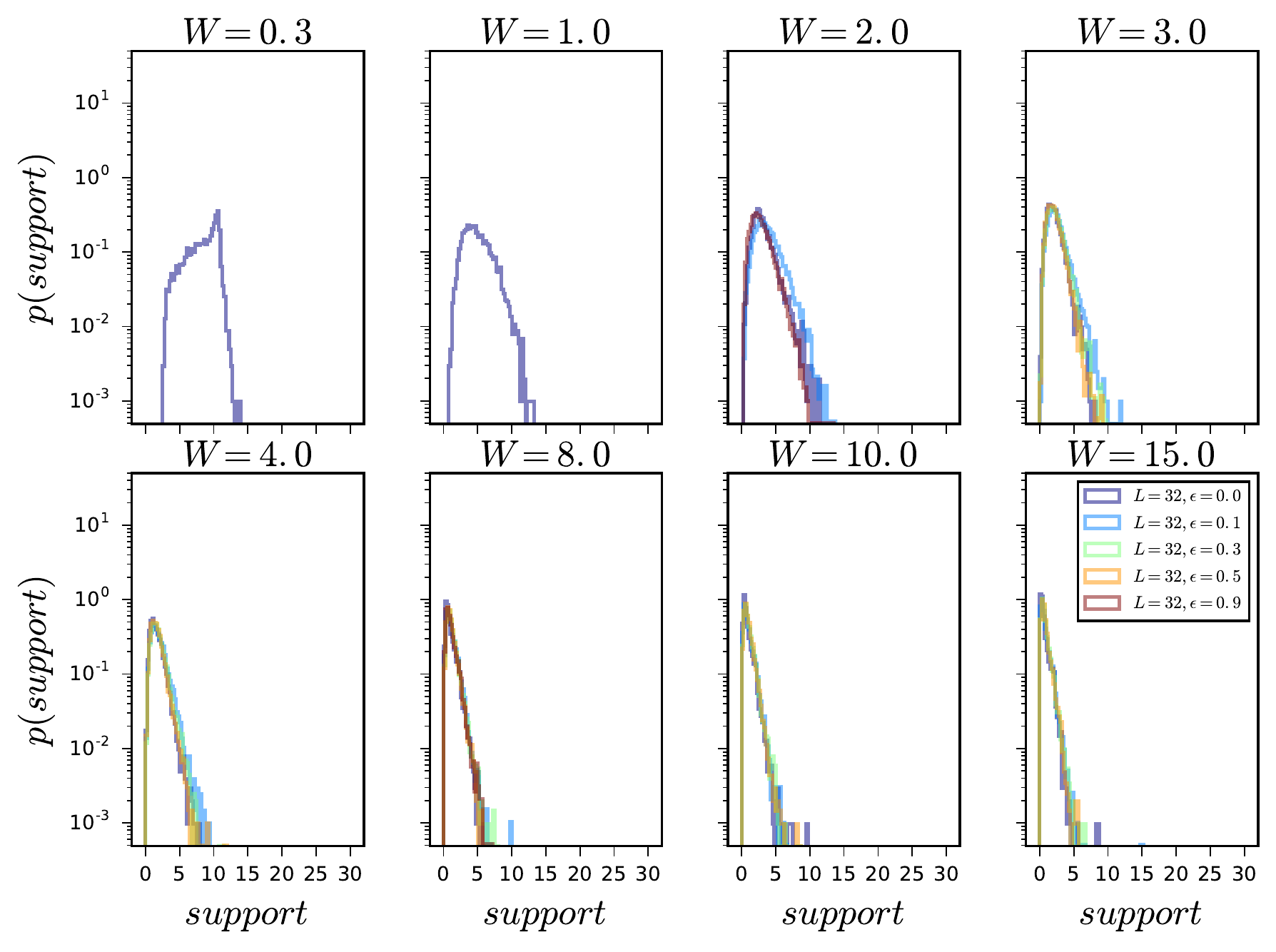}
\includegraphics[width=1.00\columnwidth]{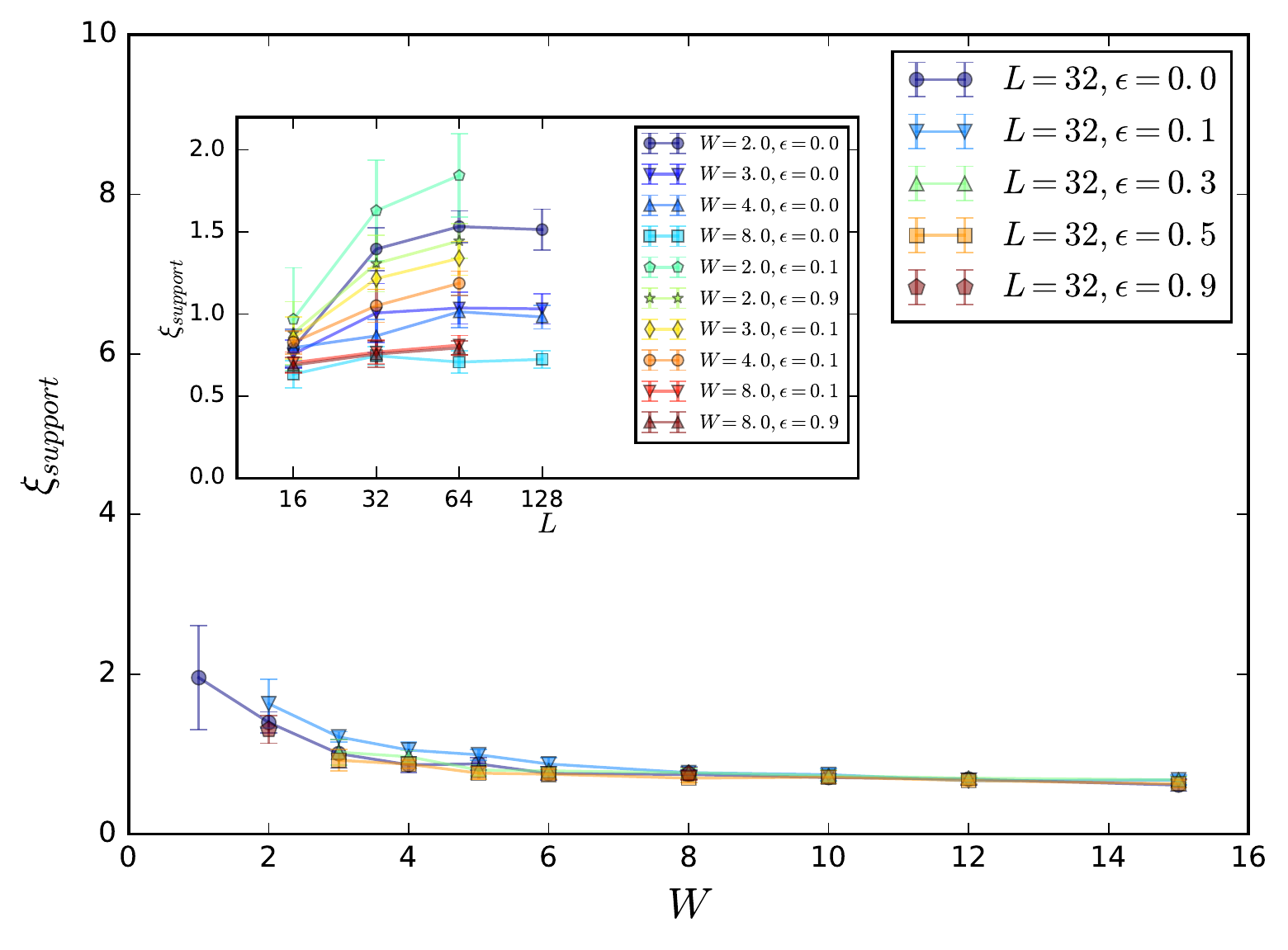}
\caption{\label{fig:hist_support_no_center} Equivalent to Fig.~\ref{fig:hist_support}, although using the uncentered definition for the range $R$.}
\end{figure}

\begin{figure}[t]
\includegraphics[width=1.00\columnwidth]{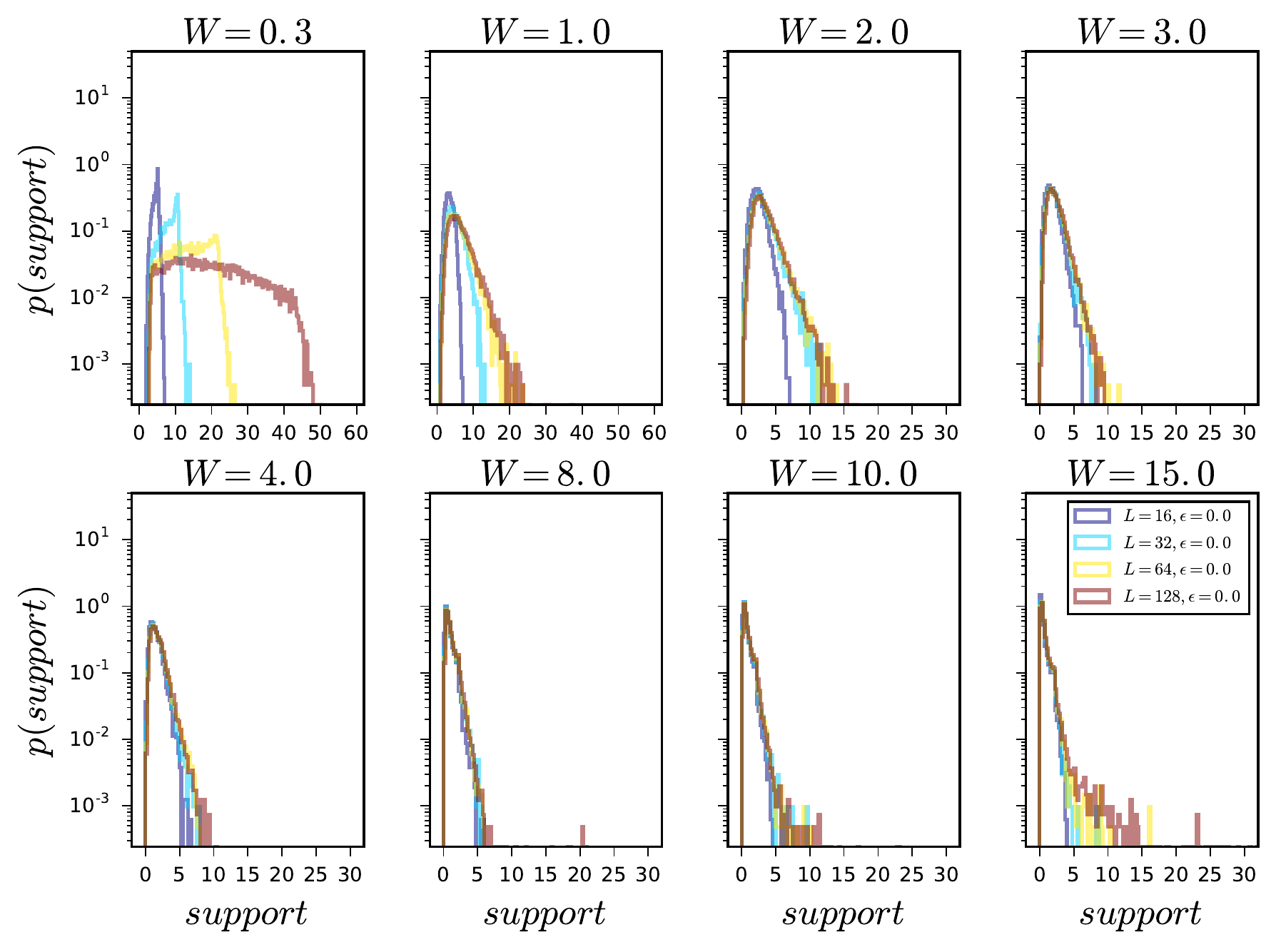}
\includegraphics[width=1.00\columnwidth]{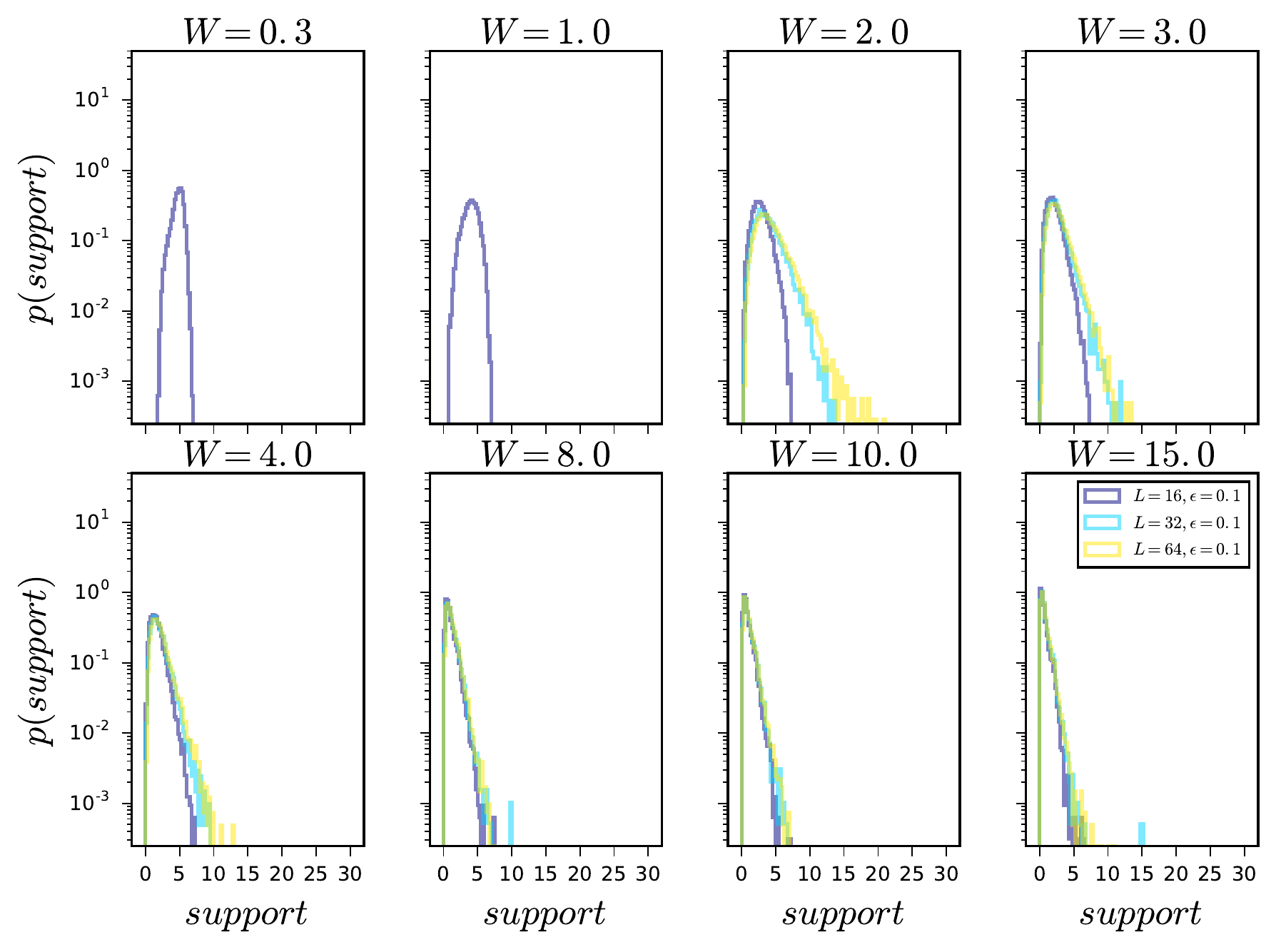}
\caption{\label{fig:hist_support_no_center_fixed_E} Equivalent to Fig.~\ref{fig:hist_support_fixed_E}, but using the uncentered definition of the range $R$.}
\end{figure}

The exponential decay of the total contribution $F_R$ from the string operators of range $R$ to the number operator of the OPOs is not only seen in average ($\bar{F}_R$), but also in the 2D histogram of $F_R$ vs. $R$.
We demonstrate in Fig.~\ref{fig:hist_FR_vs_R} for $\epsilon=0.0, 0.1$ and systems of size $L=64$ that at all disorder strengths $F_R$ presents a signal that decays exponentially with $R$ away from long ranges, for which the finite size effects (and possibly noise due to the numerics) are stronger.

The correlation length $\xi$ increases monotonically with $L$, as mentioned in Section~\ref{sec:support} (see Fig.~\ref{fig:FR_vs_R}).
The precise functional form of the scaling is not determined due the few data points available, but at $W\lessapprox W_c$ the points align suggesting a logarithmic scaling of the form $\xi = \log(\beta \cdot L^\alpha)$.
Assuming this form is correct,
we can estimate the value of $\alpha$ as a function of $W$ (see Fig.~\ref{fig:alphas}).
It is interesting to see that $\alpha$ increases as the disorder is lowered, implying a faster increase in $\xi$ with system size as $W$ gets smaller.

\section{Exponential decay of the OPOs}
\label{sec:exponential}

Assuming that the asymptotic exponential decay of the averaged $\bar{F}_R$ is representative of a typical OPO, we have $F_R \propto e^{-R/\xi}$ for a particular OPO $k$, where $R\equiv \max\left(|i-m|, |j-m|\right)$ and $m$ is the localization center, \emph{i.e.} the site with the maximum amplitude $\left|U^\dagger_{km}\right|$.
$F_R$ is defined as the total contribution from range $R$ to the definition of $a^\dagger_k a_k$:
\begin{align}
\label{F_R_appendix}
F_R \equiv \sum_{\max\left( |i-m|, |j-m| \right)=R} \left|f^k_{ij}\right| \text{,}
\end{align}
where $f^k_{ij}\equiv U^\dagger_{ki}U_{jk}$.
Using the fact that $f^k_{ij}$ is a hermitian matrix in indices $i$ and $j$, Eq.~\eqref{F_R_appendix} can be rewritten as:
\begin{align}
\label{F_R_symmetric}
F_R &= 2 \times \sum_{i\in(m-R, m+R)} \left|f^k_{m-R, i}\right| \nonumber \\
&\ \ + 2 \times \sum_{i\in(m-R, m+R)} \left|f^k_{m+R, i}\right| \nonumber \\
&\ \ + 2 \times \left|f^k_{m-R, m+R}\right| \nonumber \\
&\ \ + \left|f^k_{m-R, m-R}\right| + \left|f^k_{m+R, m+R}\right| \text{.}
\end{align}
Furthermore, if we assume that the decay of the OPO is symmetric to both sides of site $m$, Eq.~\eqref{F_R_symmetric} becomes:
\begin{align}
\label{F_R_space_symmetry}
F_R &= 4 \times \sum_{i\in(m-R, m+R)} \left|f^k_{m-R, i}\right| \nonumber \\
&\ \ + 4 \times \left|f^k_{m-R, m-R}\right| \text{,}
\end{align}
which is expressed in terms of $U^\dagger$ (note that in our case $U^\dagger = U^T$, since U has only real coefficients) as:
\begin{align}
\label{F_R_in_U}
F_R &= 4 \times \sum_{i\in(m-R, m+R)} \left|U^\dagger_{k, m-R} U^\dagger_{k, i}\right| \nonumber \\
&\ \ + 4 \times \left|U^\dagger_{k, m-R} U^\dagger_{k, m-R}\right| \nonumber \\
&= 4 \left|U^\dagger_{k, m-R} \right| \times \sum_{i\in[m-R, m+R)} \left|U^\dagger_{k, i}\right| \nonumber \\
&= 4 \left|U^\dagger_{k, m-R} \right| \times \nonumber \\
&\ \ \left( \left|U^\dagger_{k, m-R}\right| + \left|U^\dagger_{k, m}\right| + 2\times\sum_{i\in(m-R, 0)} \left|U^\dagger_{k, i}\right| \right) \text{,}
\end{align}
which by assumption has to decay as $e^{-R/\xi}$.
Solving for the decay of $\left|U^\dagger_{(k, m-R)}\right|$ as $m-R$ gets away from $m$ we get:
\begin{align}
\label{F_R_solve}
&\left|U^\dagger_{k, m-R} \right| \propto \nonumber \\
&e^{-R/\xi} \left( \left|U^\dagger_{k, m-R}\right| + \left|U^\dagger_{k, m}\right| + 2\sum_{i\in(m-R, 0)} \left|U^\dagger_{k, i}\right| \right)^{-1} \text{.}
\end{align}
It is clear from Eq.~\eqref{F_R_in_U} (third line) that $\left|U^\dagger_{k, m-R}\right|$ decays at least as fast as $e^{-R/\xi}$ as a function of $R$, and so the sum $2\sum_{i\in(m-R, 0)} \left|U^\dagger_{k, i} \right| + \left|U^\dagger_{k, m-R}\right|$
is convergent (in the limit $R\rightarrow \infty$).
We end up with the functional form:
\begin{align}
\label{F_R_functional}
&\left|U^\dagger_{k, m-R} \right| \propto \frac{e^{-R/\xi}}{A + B\cdot g(R)} \text{,}
\end{align}
where $A$ and $B$ are positive constants and $g(R)$ is a monotonically increasing function with the limits $g(0) = 0$ and $g(\infty) = 1$.
Therefore, the weight of the number operators of the OPOs ($F_R$) and the OPOs themselves (as one particle wave functions) have the same asymptotic exponential behavior, with the same correlation length $\xi$.

As we can see in Fig.~\ref{fig:tail}, the (logarithmically) averaged decay of the tails of the OPOs
is extremely similar, and equal asymptotically, to the one of $\bar{F}_R$ presented in Fig.~\ref{fig:FR_vs_R}.

\section{Supplementary data on the support of the OPOs}
\label{sec:sup_support}

Here we consider the distribution of the $support_{90}$ of the OPOs at different system sizes (see Fig.~\ref{fig:hist_support_90_fixed_E}).
While at strong disorder the distributions are pretty much system size independent and decay exponentially with $support_{90}$, at small disorder they clearly suffer from finite size effects and collapse to the system size.
Also, in the weak disorder limit the exponential decay seems to be lost, although it might be masked by the finite size effects on the distributions.

The definition of the support ($support_{90}$) involves the arbitrary choice of a region containing $90\%$ of the norm of the OPO.
An alternative way of defining the support of an OPO, which is less intuitive but does not depend on an arbitrary choice of some sort of threshold, is by considering its number operator $a_k^\dagger a_k$.
We define its $support$ as the average range $R$ weighted by $F_R$:
\begin{align}
\label{support}
	support \equiv \frac{\sum_R F_R R}{\sum_R F_R} \text{,}
\end{align}
which is equivalent to the average range of the string operators that define the number operator $a_k^\dagger a_k$ (see Eq.~\eqref{number_operator}) weighted by their amplitude $\left| f^k_{ij}\right|$; this is analogous to the definition for l-bits from Ref.~\onlinecite{huse_phenomenology_2014}, but our range $R$ always includes the distance to the center, as is presented in Ref.~\onlinecite{abanin_recent_2017} (see Appendix~\ref{sec:range} for more details).
Both Figs.~\ref{fig:hist_support}~and~\ref{fig:hist_support_fixed_E} show that the already discussed phenomenology captured by the $support_{90}$ is extremely similar to the one captured by the $support$ of Eq.~\eqref{support}.
For a given OPO, its $support$ is usually smaller than its $support_{90}$ due to the fact that the average over ranges will make the $support$ take roughly half of the value of the $support_{90}$; we can easily see this trend in the figures.

\section{System size independence of the distribution of the IPR}
\label{sec:size_independence}

In Fig.~\ref{fig:bimodality_fixed_E} we see that the distribution of the IPR is system size independent for $W \gtrapprox W_c$ and very slightly system size dependent at small disorder, where the OPOs delocalize and are affected by finite size effects, with a slight drift towards larger IPR for smaller systems.

\section{Supplementary data on the OPOs' overlaps}
\label{sec:sup_overlaps}

The high overlap between OPOs at different energy densities could be due to the localized form of the OPOs, which might match trivially with one another at their center.
However, we show in this appendix that their overlap is benefited from the particular shape of OPOs' tails, and is not only due to the overlap coming from two OPOs centered at the same site.
To study this we define the leading one site contribution to the overlap between two OPOs at different energy densities $\epsilon_1$ and $\epsilon_2$ as:
\begin{align}
\label{one_site}
&\left<\phi_k(\epsilon_1)|\psi_l(\epsilon_2)\right>_{1site} \nonumber\\
&\quad\equiv \max\left\{\left|U(\epsilon_1)_{ik} U^\dagger(\epsilon_2)_{li}\right|\right\}_{i\in[0,L-1]}
\end{align}
where $\left|\phi_{k}(\epsilon_1)\right>=\sum_i U^\dagger(\epsilon_1)_{ki}\left|i\right>$ and $\left|\psi_{l}(\epsilon_2)\right>=\sum_i U^\dagger(\epsilon_2)_{li}\left|i\right>$.
We can see in the top panel of Fig.~\ref{fig:hist_one_site_overlaps} that the distribution of the main one site contribution to the overlaps between corresponding OPOs ($k\leftrightarrow l(k)$) is always substantially lower than the total overlap over the entire chain (compare with Fig.~\ref{fig:hist_overlaps} in Section~\ref{sec:correspondence}).
The pairs of OPOs match therefore both at their center and throughout their tails in a non-trivial way.
The bottom panel of Fig.~\ref{fig:hist_one_site_overlaps} shows the mode of the distribution of overlaps as a function of $W$.
We see that for all pairs of energy densities depicted, and for all disorder strengths, the typical overlap is higher than or equal to the typical best one site contribution.
This is particularly noticeable at low disorder, where the mode of the one site overlaps drops substantially below $100\%$ at $W=3$, to $40-50\%$ at $W=2$ (as opposed to about $70\%$ when the tails are considered for $\epsilon_1$ far from $\epsilon_2$) and below $30\%$ at $W=1$ and $L=16$ (as opposed to $60-70\%$).

\section{The different definitions of the range of the string operators and the relation between the ``$1/f$'' distribution and the exponential decay of the OPOs}
\label{sec:range}

In this appendix we will discuss two different definitions of the range $R$. In either case, we will show how the ``$1/f$'' distribution of the coupling constants of the number operators of the OPOs is a consequence of the exponential decay of the OPOs in real space. We will also show the robustness to the two definitions of $R$ of our results from Section~\ref{sec:support} on the correlation length and the support of the OPOs.

In Section~\ref{sec:support} we defined the range associated to the string of operators $c_i^\dagger
c_j$ that contributes to the definition of the number operator of an OPO with its maximum amplitude
at site $m$ (see Eq.~\eqref{number_operator}) as $\max(|i-m|, |j-m|)$, in the spirit of the one for l-bits of Ref.~\onlinecite{abanin_recent_2017}; we will call this the ``centered'' definition of the range.
An alternative definition of the range is $R\equiv|i-j|$, which is considered in Refs.~\onlinecite{huse_phenomenology_2014}~and~\onlinecite{pekker_fixed_2017} for l-bits; we will call this the ``uncentered'' definition.
The centered range takes into account the notion of an l-bit being localized around a site $m$ in real space, and acting non-trivially mainly in a small region around this site.
The uncentered range ignores this notion of a center, and relates the concept of localization to the idea of an l-bit acting non-trivially between sites contained in small regions in real space, but these regions can be many and lay anywhere on the chain.
Both definitions are interesting in slightly different ways due to their different points of emphasis, but in practice they give rise to a very similar phenomenology.
Their relation with the matrix $f^k_{ij}$ of coupling constants of the number operator of an OPO is better understood graphically, with the aid of Fig.~\ref{fig:scaling}, where only the elements of a range $R(i,j)=const.$ are shown (top panel for centered range and bottom panel for uncentered range).
Let's first focus on the centered range, and leave the discussion of the uncentered range for later.
It is easy to see that the elements of a constant range $R$ correspond to squares of side $2R$ centered at $(m,m)$.
In addition, the elements within a particular square decay exponentially on each one of its four sides as either $\left|f^k_{ij}\right|\propto e^{-|i-m|/\xi}$ or $\left|f^k_{ij}\right|\propto e^{-|j-m|/\xi}$.
As a consequence, the elements of constant $R$ (that we denote by $|f_R|$) follow a distribution $p\left(\log|f_R|\right)=const.$, but that implies $p(|f_R|)=const./|f_R|$ due to $d\left(\log|f_R|\right)/d\left(p(f_R)\right)=1/|f_R|$.
If we consider an ensemble of exponentially decaying OPOs, the combined $p(|f_R|)$ will drop towards the ends, due to the individual distributions spanning different regions of the $|f_R|$ axis; we can see this in Fig.~\ref{fig:FR_vs_R}.
We can see that the ``$1/f$'' distribution of the coupling constants of $a^\dagger_k a_k$ can be derived from the exponential decay of the OPOs in real space.
Finally, at small disorder the distributions $p(|f_R|)$ get narrower as a natural consequence of the slower decay of the OPOs in this limit (see Fig.~\ref{fig:FR_vs_R}).

Let's now focus on the uncentered range (bottom panel of Fig.~\ref{fig:scaling}).
The elements within the secondary diagonals of the matrix are now those with a constant $R$ and decay exponentially as $e^{-\left(|i-m|+|j-m|\right)/\xi}$, which drops twice as fast as the OPO's amplitudes due to the simultaneous change of $i$ and $j$ along the diagonal.
The distribution $p(|f_R|)\propto 1/|f_R|$ for a fixed $R$ still holds (see Fig.~\ref{fig:pfR_vs_R_no_center}) due to the same argument discussed for the centered range case, although now the tails of the $|f_R|$ diagonals get shorter as $R$ is increased, due to the finite size of the system (as opposed to the squares of the centered range, which did grow in size with $R$).
This causes the distributions $p(|f_R|)$ to become narrower as the range $R$ is increased, contrary to the expectations for the centered range.

It is easy to see that the total contribution $F_R$ of a particular (uncentered) range to $a^\dagger_k a_k$ (see Eq.~\eqref{F_R}), \emph{i.e.} the sum of all elements in the diagonals shown in the bottom panel of Fig.~\ref{fig:scaling}, decays exponentially with $R$ as $e^{-R/\xi}$ for big enough systems, as was the case with the centered range.
This is demonstrated for the (logarithmic) average $\bar{F}_R$ in Fig.~\ref{fig:FR_vs_R_no_center}, where little difference is found as compared to Fig.~\ref{fig:FR_vs_R} of Section~\ref{sec:support} (where the centered range is used).
The $\bar{F}_R$ curves are slightly concave at large $R$, which we think is due to the shortening of the tails of the  $\left|f_R\right|$  diagonals with $R$.

The $support$ (see Eq.~\eqref{support}) is also robust to the change in the definition of the range.
We demonstrate in
Figs.~\ref{fig:hist_support_no_center}~and~\ref{fig:hist_support_no_center_fixed_E} that the
phenomenology (using the uncentered range) is similar to the one found with the centered definition
of the range in Figs.~\ref{fig:hist_support}~and~\ref{fig:hist_support_fixed_E}, and hence to the one discussed in Section~\ref{sec:support} for the simpler $support_{90}$ (see Fig.~\ref{fig:hist_support_90} and Fig.~\ref{fig:hist_support_90_fixed_E} of Appendix.~\ref{sec:sup_support}).

\bibliography{opdm_mbl}

\end{document}